\documentclass[11pt,a4paper]{article}
\DeclareMathAlphabet{\scr}{U}{rsfs}{m}{n}

\pdfoutput=1
\usepackage[utf8]{inputenc}
\usepackage{latexsym}
\usepackage{epsfig}
\usepackage[mathscr]{eucal}
\usepackage{amsfonts}
\usepackage{amscd}
\usepackage{cite}
\usepackage{array}   
\usepackage{amssymb}
\usepackage{colordvi}
\usepackage[centertags]{amsmath}
\usepackage{enumerate}
\usepackage{graphicx}
\usepackage{booktabs}
\usepackage{hyperref}
\usepackage{enumitem}
\setlist[description]{leftmargin=2\parindent,labelindent=\parindent}
\usepackage{theorem}
\usepackage{listings}
\usepackage{multirow}
\usepackage[framemethod=TikZ]{mdframed}
\usepackage{makecell}
\usepackage[footnotesize]{caption}
\usepackage{bbm}
\usepackage[labelfont=bf,textfont={sl,bf}]{subfig}
 \usepackage{slashed}
\usepackage{xcolor}
\xdefinecolor{mygray}{gray}{0.85}
\usepackage{ulem}
\usepackage{bm}
\usepackage{arydshln}
\usepackage{pifont}% http://ctan.org/pkg/pifont
\usepackage{mathtools}

\mdfsetup{%
   middlelinecolor=black,
   middlelinewidth=0.5pt,
   roundcorner=0pt,
   innertopmargin=-3pt,
   innerbottommargin=10pt
}

\newcommand{\newc}{\newcommand}
\newc{\be}{\begin{equation}}
\newc{\ee}{\end{equation}}
\newc{\bea}{\begin{eqnarray}}
\newc{\eea}{\end{eqnarray}}
\newc{\ol}{\overline}
\newc{\bs}{\boldsymbol}
\newc{\m}{\mathcal}
\newc{\lan}{\langle}
\newc{\ra}{\rangle}
\newc{\pa}{\partial}

\newcommand{\beq}{\begin{eqnarray}} 
\newcommand{\eeq}{\end{eqnarray}} 
\newcommand{\bpmatrix}{\begin{pmatrix}}
\newcommand{\epmatrix}{\end{pmatrix}}
\newcommand{\ba}{\begin{array}}
\newcommand{\ea}{\end{array}}

\renewcommand{\eqref}[1]{Eq.~(\ref{#1})}

\newcommand{\bc}{\begin{center}}
\newcommand{\ec}{\end{center}}

\renewcommand{\ol}{\text{1l}}

\newcommand{\s}{\newline \vspace*{-3.5mm}}

\begin{document}
\title{
\vspace*{-3cm}
\phantom{h} \hfill\mbox{\small KA-TP-25-2019}
%\vspace*{1cm}
\\[2cm]
%\vspace{13mm}   
\textbf{Impact of Electroweak Corrections on Neutral Higgs Boson
  Decays in Extended Higgs Sectors}}

\date{}
\author{Marcel Krause$^{1\,}$\footnote{E-mail:
  \texttt{marcel.krause@kit.edu}}, Margarete M\"{u}hlleitner$^{1\,}$\footnote{E-mail:
  \texttt{margarete.muehlleitner@kit.edu}}
\\[9mm]
{\small\it
$^1$Institute for Theoretical Physics, Karlsruhe Institute of Technology,} \\
{\small\it Wolfgang-Gaede-Str. 1, 76131 Karlsruhe, Germany.}\\[3mm]
}
\maketitle

\begin{abstract}
\noindent 
Precision predictions play an important role in the search for
indirect New Physics effects in the Higgs sector itself. For the
electroweak (EW) corrections of the Higgs bosons in extended Higgs sectors
several renormalization schemes have been worked out that provide
gauge-parameter-inde\-pen\-dent relations between the input parameters
and the computed observables. Our recently published program codes
{\tt 2HDECAY} and {\tt ewN2HDECAY} allow for the computation of the EW
corrections to the Higgs decay widths and branching ratios of the
Two-Higgs-Doublet Model (2HDM) and the Next-to-Minimal-2HDM (N2HDM) for
different renormalization schemes of the scalar mixing angles. In this paper, we present a
comprehensive and complete overview over the relative size of the EW
corrections to the branching ratios of the 2HDM and N2HDM neutral Higgs
bosons for different applied renormalization schemes. We
quantify the size of the EW corrections of Standard Model(SM)- and non-SM-like Higgs
bosons and moreover also identify renormalization schemes that are well-behaved
and do not induce unnaturally large corrections. We furthermore
pin down decays and parameter regions that feature large EW
corrections and need further treatment in order to improve the predictions. Our
study sets the scene for future work in the computation of higher-order corrections to
the decays of non-minimal Higgs sectors.
\end{abstract}
\thispagestyle{empty}
\vfill
\newpage

%%%%%%%%%%%%%%%%%%%%%%%%%%%%%%%%%%%%%%%%%%%%%%%%%%%%%%%
\section{Introduction}
\label{sec:Introduction}
%%%%%%%%%%%%%%%%%%%%%%%%%%%%%%%%%%%%%%%%%%%%%%%%%%%%%%%
While the discovery of the Higgs boson in 2012 by the LHC experiments
ATLAS \cite{Aad:2012tfa} and CMS \cite{Chatrchyan:2012xdj} marked a
milestone for particle physics, there are still 
many questions left open. The Higgs boson turned out to
behave very Standard-Model-like. The Standard Model (SM), however,
cannot solve open problems like {\it e.g.}~the generation of the
observed baryon-antibaryon asymmetry or provide an appropriate Dark
Matter candidate. This calls for New Physics extensions that usually
come along with an extended Higgs sector. So far no direct sign of any 
New Physics manifestation has been discovered by experiments, so that
the Higgs sector itself has moved into the focus of our search for New
Physics. There, it might reveal itself indirectly in deviations of the
Higgs properties from the SM expectations. Since experiments have pushed
the exclusion limits on new particles to high mass scales, these
effects are expected to be small, unless triggered by new light
particles in the spectrum. These could be {\it e.g.}~the additional light
Higgs bosons of extended Higgs sectors. Still, precision is required
in order to detect the indirect signs of New Physics, so that additionally the
nature of the underlying model can be revealed. From the theory
side this requires the inclusion of higher-order corrections to the
Higgs boson observables. \s

Higgs sector extensions have to ensure compatibility with 
experimental and theoretical constraints. The extensions may be based on a
weakly or strongly interacting model. Among the weakly interacting
models, both supersymmetric (SUSY) and non-SUSY Higgs sectors are possible. While SUSY
extensions are very well-motivated by their symmetry, non-SUSY models
allow for more freedom, in particular in the Higgs boson self-couplings. In
contrast to SUSY, where these coupling constants are given in terms of the gauge boson
couplings, they can be sizeable in non-SUSY models, modulo the constraints stemming from the
requirement of perturbativity. This may induce interesting effects not
only in Higgs pair production but also in the electroweak corrections to Higgs
boson observables. \s

In this paper, we investigate the impact of electroweak corrections on
the neutral Higgs boson decays of two non-SUSY extension of the Higgs
sector. These are the Two-Higgs-Doublet Model (2HDM)
\cite{Lee:1973iz,Branco:2011iw} and the Next-to-Minimal
Two-Higgs-Doublet-Model (N2HDM) 
\cite{Chen:2013jvg,Muhlleitner:2016mzt}. Both models feature an
extended Higgs sector with at least 
three neutral Higgs bosons, inducing interesting phenomenology stemming
from Higgs self-interactions. Both models allow for a strong first
order phase transition and can in principle provide a Dark
Matter candidate depending on the applied symmetries. While the 2HDM
is the simplest Higgs doublet extension of the SM, the more complex
structure of the N2HDM Higgs sector allows for more 
freedom in the parameter space, thus inducing additional interesting effects
in the phenomenology. In both models, light Higgs bosons are still
allowed in the spectrum and have an impact 
not only on Higgs observables but moreover, as we will show, on the size of
the electroweak corrections. \s

The aim of this paper is to demonstrate that electroweak (EW) corrections can be
important although the Higgs measurements push New Physics extensions
to be very close to the SM.\footnote{Other recent works
  including EW corrections to non-SM Higgs decays are \cite{Bojarski:2015kra}
where EW corrections to Higgs-to-Higgs decays in a singlet extension of
the SM have been computed, or \cite{Goodsell:2017pdq} which provides a 
generic calculation of the two-body partial decays widths at full
one-loop level in the $\overline{\mbox{DR}}$ renormalization scheme. 
Next-to-leading order corrections to Higgs boson decays in models with
non-minimal sectors have been computed in
\cite{Kanemura:2018yai,Kanemura:2019kjg}. 
For the Next-to-Minimal Supersymmetric extension of the SM (NMSSM),
the full one-loop renormalization and one-loop corrected two-body 
Higgs decay widths in the on-shell (OS) renormalization scheme have
been given in \cite{Belanger:2016tqb,Belanger:2017rgu}. Recent
calculations of the EW corrections to the NMSSM Higgs boson decays in
the CP-conserving and/or CP-violating NMSSM can be found in
\cite{Nhung:2013lpa,Muhlleitner:2015dua,Baglio:2015noa,Domingo:2018uim,
  Baglio:2019nlc,Domingo:2019vit,Dao:2019nxi}. They can be computed with the public code {\tt NMSSMCALCEW}
\cite{Baglio:2019nlc} (based on the extension of {\tt NMSSMCALC}
\cite{Baglio:2013iia}).} 
 We furthermore aim to show
the impact of different renormalization schemes. In previous papers it
has been shown that care has to be taken to choose a renormalization
scheme for the Higgs mixing angles that is not gauge-dependent \cite{Krause:2016oke,Krause:2016xku,Denner:2016etu,Altenkamp:2017ldc,Altenkamp:2017kxk,Krause:2017mal,Kanemura:2017wtm,Denner2018,Denner:2019xti}.\footnote{The gauge-independent renormalization of multi-Higgs
models has been discussed in \cite{Fox:2017hbw,Grimus:2018rte}.}
Additionally, it should not lead to unnaturally large loop 
corrections \cite{Kanemura:2002vm,Kanemura:2004mg,Krause:2016xku,Krause:2017mal,Belanger:2017rgu}. We want to compare the results for different
renormalization schemes that take into account these
considerations. Lastly, we want to identify those decay channels where
the relative corrections do not blow up due to an unsuitable renormalization
scheme but because they have small leading-order (LO) branching
ratios or because the corrections are parametrically enhanced. This
serves as a starting point for future work on the reduction of these
corrections by going beyond next-to-leading order (NLO) and applying
resummation techniques. \s 

The results for the loop corrected branching ratios (BRs)
that we will show in this paper have been generated with the
codes {\tt 2HDECAY} \cite{Krause:2018wmo}
and {\tt N2HDECAY} \cite{Krause:2019oar}
that are based on extensions of the {\tt FORTRAN} codes {\tt 
  HDECAY} \cite{DJOUADI199856,Djouadi:2018xqq} and {\tt N2HDECAY}
\cite{Muhlleitner:2016mzt}, respectively, to include also
electroweak corrections.\footnote{A public code for
  the computation of loop-corrected decay widths in extended Higgs
  sectors, {\tt H-COUP}, has also been provided in
  \cite{Kanemura:2017gbi,Kanemura:2019slf}. The Higgs
decay into four fermions has been implemented in the Monte Carlo
generator {\tt PROPHECY4F} 3.0 for the singlet
extended SM and the 2HDM including the full QCD and EW NLO
corrections \cite{Denner:2019fcr}.} Based on {\tt   
  HDECAY}, the decay widths and BRs already include the
state-of-the-art higher-order QCD corrections, so that the presented
numbers are the most precise predictions that can be provided at
present. \s

In Section~\ref{sec:modelintro} we introduce the 2HDM and
N2HDM and set our notation. In Section~\ref{sec:renormalization}, we
briefly present the applied renormalization schemes. The results for
the EW- and QCD-corrected BRs are given in 
Section~\ref{sec:numerics}. For the 2HDM they are presented and
discussed in
Subsections~\ref{sec:2HDMHiggsDecays}--\ref{sec:pseudoscalar2hdm}, and
for the N2HDM in
Subsections~\ref{sec:N2HDMHiggsDecays}--\ref{sec:pseudoscalarn2hdm}. For
both models we 
investigate the SM-like Higgs boson decays, the non-SM-like CP-even
Higgs decays and the pseudoscalar decays. We consider different mass
hierarchies, where either the lighter or the heavier CP-even Higgs state
represents the SM-like Higgs boson. In Section~\ref{sec:conclusions},
we conclude with a short summary.

%%%%%%%%%%%%%%%%%%%%%%%%%%%%%%%%%%%%%%%%%%%%%%%%%%%%%%%%%%%%%%
\section{Model Introduction}
\label{sec:modelintro}
%%%%%%%%%%%%%%%%%%%%%%%%%%%%%%%%%%%%%%%%%%%%%%%%%%%%%%%%%%%%%%
In the following we introduce the two models that we consider in this
work, namely the 2HDM and the N2HDM, and we set our notation.

%%%%%%%%%%%%%%%%%%%%%%%%%%%%%%%%%%%%%%%%%%%%%%%%%%%%%%%%%%%%%%
\subsection{Introduction of the 2HDM}
%%%%%%%%%%%%%%%%%%%%%%%%%%%%%%%%%%%%%%%%%%%%%%%%%%%%%%%%%%%%%%
\label{sec:introOf2HDM}
In terms of the two complex $SU(2)_L$ Higgs doublets $\Phi_1$ and $\Phi_2$
with hypercharge $Y=+1$,
\beq
\Phi_1 = \left( \begin{array}{c}  \phi_1^+ \\ \phi_1^0 \end{array}
\right) \quad \mbox{and} \quad
\Phi_2 = \left( \begin{array}{c}  \phi_2^+ \\ \phi_2^0 \end{array}
\right) \;,
\eeq
the tree-level potential of a general CP-conserving 2HDM
\cite{Lee:1973iz,Branco:2011iw} reads 
\begin{equation}
\begin{split}
V_{\text{2HDM}} =&~ m_{11}^2 \left| \Phi _1 \right| ^2 + m_{22}^2 \left|
  \Phi _2 \right| ^2 - m_{12}^2 \left( \Phi _1 ^\dagger \Phi _2 +
  \textit{h.c.} \right) + \frac{\lambda _1}{2} \left( \Phi _1^\dagger
  \Phi _1 \right) ^2 + \frac{\lambda _2}{2} \left( \Phi _2^\dagger
  \Phi _2 \right) ^2 \\ 
&+ \lambda _3 \left( \Phi _1^\dagger \Phi _1 \right) \left( \Phi
  _2^\dagger \Phi _2 \right) + \lambda _4 \left( \Phi _1^\dagger \Phi
  _2 \right) \left( \Phi _2^\dagger \Phi _1 \right) + \frac{\lambda
  _5}{2} \left[ \left( \Phi _1^\dagger \Phi _2 \right) ^2 +
  \textit{h.c.} \right] ~. 
\end{split}
\label{eq:scalarPotential}
\end{equation}
It is parametrized by three real mass parameters $m_{11}$,
$m_{22}$ and $m_{12}$ and five dimensionless real couplings
$\lambda_{1...5}$. The term proportional $m_{12}^2$ softly breaks the
global discrete $\mathbb{Z}_2$ symmetry under which the doublets
transform as $\Phi_1 \to \Phi_1$ and $\Phi_2 \to -\Phi_2$. After
electroweak symmetry breaking 
(EWSB), the neutral components of the Higgs doublets acquire vacuum
expectation values (VEVs) $v_1$ and $v_2$ which in the
CP-conserving case are real. The Higgs doublets can be expanded around their
VEVs in terms of the charged complex fields $\omega_i^\pm$ and the
real neutral CP-even and CP-odd fields $\rho_i$ and $\eta_i$
($i=1,2$), respectively, as 
\begin{equation}
\Phi _1 = \begin{pmatrix} \omega ^+ _1 \\ \frac{v_1 + \rho _1 + i \eta
    _1 }{\sqrt{2}} \end{pmatrix} ~~~\text{and}~~~~ \Phi _2 = \begin{pmatrix}
  \omega ^+ _2 \\ \frac{v_2 + \rho _2 + i \eta
    _2}{\sqrt{2}} \end{pmatrix} 
\label{eq:vevexpansion}
\end{equation} 
where 
\begin{equation}
v^2 = v_1^2 + v_2^2 \approx (246.22~\text{GeV})^2
\label{eq:vevRelations} 
\end{equation}
is the squared SM VEV obtained from the Fermi constant $G_F$, $v=1/\sqrt{\sqrt{2}G_F}$. We introduce the
mixing angle $\beta$ through
\beq
\tan\beta = \frac{v_2}{v_1} 
\eeq
so that
\beq 
v_1 = v \cos \beta \quad \mbox{and} \quad v_2 = v \sin \beta~.
\eeq
Inserting Eq.~(\ref{eq:vevexpansion}) in the scalar potential in
\eqref{eq:scalarPotential} yields
\begin{equation}
\begin{split}
V_{\text{2HDM}} =&~ \frac{1}{2} \left( \rho _1 ~~ \rho _2 \right) M_\rho ^2 \begin{pmatrix} \rho _1 \\ \rho _2 \end{pmatrix} + \frac{1}{2} \left( \eta _1 ~~ \eta _2 \right) M_\eta ^2 \begin{pmatrix} \eta _1 \\ \eta _2 \end{pmatrix} + \frac{1}{2} \left( \omega ^\pm _1 ~~ \omega ^\pm _2 \right) M_\omega ^2 \begin{pmatrix} \omega ^\pm _1 \\ \omega ^\pm _2 \end{pmatrix} \\
& + T_1 \rho _1 + T_2 \rho _2 + ~~ \cdots
\end{split}
\label{eq:scalarPotentialMultilinearFields}
\end{equation}
where $T_1$ and $T_2$ denote the tadpole terms and $M_\omega ^2$,
$M_\rho ^2$ and $M_\eta ^2$ the mass matrices of the charged, neutral
CP-even and CP-odd fields, respectively. Requiring the
VEVs of \eqref{eq:vevexpansion} to represent the minimum of the
potential, {\it i.e.}
\begin{equation}
\frac{\partial V_\text{2HDM}}{\partial \Phi _i} \Bigg| _{\left\langle \Phi _j \right\rangle} = 0~,
\end{equation}
yields the tree-level tadpole conditions 
\begin{align}
\frac{T_1}{v_1} &\equiv m_{11}^2 - m_{12}^2 \frac{v_2}{v_1} + \frac{\lambda _1}{2} v_1^2 + \frac{\lambda_{345}}{2} v_2 ^2 = 0 
\label{eq:tadpoleCondition1}  \\
\frac{T_2}{v_2} &\equiv m_{22}^2 - m_{12}^2 \frac{v_1}{v_2} +
\frac{\lambda _2}{2} v_2^2 + \frac{\lambda_{345}}{2} v_1 ^2 = 0 
\label{eq:tadpoleCondition2} \;,
\end{align}
where we used the short-hand notation
\beq
\lambda_{345} \equiv \lambda_3 + \lambda_4 + \lambda_5 \;.
\eeq
The tadpole equations can be solved for $m_{11}^2$ and $m_{22}^2$ in
order to replace these two parameters by the tadpole parameters $T_1$
and $T_2$. The mass matrices are given by 
\begin{align}
M_\rho ^2 &\equiv \begin{pmatrix}
m_{12}^2 \frac{v_2}{v_1} + \lambda _1 v_1^2 & -m_{12}^2 + \lambda _{345} v_1 v_2 \\ -m_{12}^2 + \lambda _{345} v_1 v_2 & m_{12}^2 \frac{v_1}{v_2} + \lambda _2 v_2^2
\end{pmatrix} + \begin{pmatrix} \frac{T_1}{v_1} & 0 \\ 0 & \frac{T_2}{v_2} \end{pmatrix} \label{eq:massMatrices1} \\
M_\eta ^2 &\equiv \left( \frac{m_{12}^2}{v_1v_2} - \lambda _5 \right) \begin{pmatrix}
v_2^2 & -v_1 v_2 \\ -v_1 v_2 & v_1 ^2 
\end{pmatrix} + \begin{pmatrix} \frac{T_1}{v_1} & 0 \\ 0 & \frac{T_2}{v_2} \end{pmatrix} \\
M_\omega ^2 &\equiv \left( \frac{m_{12}^2}{v_1v_2} - \frac{\lambda _4 + \lambda _5}{2} \right) \begin{pmatrix}
v_2^2 & -v_1 v_2 \\ -v_1 v_2 & v_1 ^2 
\end{pmatrix} + \begin{pmatrix} \frac{T_1}{v_1} & 0 \\ 0 & \frac{T_2}{v_2} \end{pmatrix} \;.\label{eq:massMatrices3}
\end{align}
They are diagonalized by the rotation matrices\footnote{Here and in
  the following, we use the short-hand notation $s_x \equiv \sin (x)$,
  $c_x \equiv \cos (x)$, $t_x \equiv \tan (x)$ for convenience.}   
\begin{equation}
	R (x) \equiv \begin{pmatrix} c_x & - s_x \\ s_x & c_x \end{pmatrix} 
\end{equation}
in terms the mixing angles $\alpha$ and $\beta$ which rotate the field
$\omega_i^\pm$, $\rho_i$ and $\eta_i$ in the gauge basis to the mass
basis as
\begin{align}
	\begin{pmatrix} \rho _1 \\ \rho _2 \end{pmatrix} &= R(\alpha ) \begin{pmatrix} H \\ h \end{pmatrix}  \label{eq:rotationCPEven} \\
	\begin{pmatrix} \eta _1 \\ \eta _2 \end{pmatrix} &= R(\beta ) \begin{pmatrix} G^0 \\ A \end{pmatrix}  \\
	\begin{pmatrix} \omega ^\pm _1 \\ \omega ^\pm _2 \end{pmatrix} &= R(\beta ) \begin{pmatrix} G^\pm \\ H^\pm \label{eq:rotationCharged} \end{pmatrix} \;.
\end{align}
Here, $h$ and $H$ denote the CP-even Higgs bosons with masses $m_h <
m_H$, respectively, $A$ the CP-odd Higgs boson with mass $m_A$ and $H^\pm$ the
charged Higgs bosons with masses $m_{H^\pm}$. The massless neutral and
charged Goldstone bosons are denoted by $G^0$ and $G^\pm$,
respectively. In order to avoid flavor-changing neutral currents
(FCNCs) at tree level, the $\mathbb{Z}_2$ symmetry of the Higgs
potential is extended to the Yukawa sector so that each of the up-type
quarks, down-type quarks and charged leptons can only couple to one of
the Higgs doublets. Depending on the
$\mathbb{Z}_2$ charge assignments, there are four phenomenologically
different types of 2HDMs that are shown in Tab.~\ref{tab:yukawaDefinitions}.
\begin{table}[tb]
\centering
  \begin{tabular}{ c c c c }
    \hline
    & $u$-type & $d$-type & leptons \\ \hline
    I &  $\Phi _2$ & $\Phi _2$ & $\Phi _2$ \\
    II & $\Phi _2$ & $\Phi _1$ & $\Phi _1$ \\
    lepton-specific & $\Phi _2$ & $\Phi _2$ & $\Phi _1$ \\
    flipped & $\Phi _2$ & $\Phi _1$ & $\Phi _2$ \\
    \hline
  \end{tabular}
    \caption{The four Yukawa types of the $\mathbb{Z}_2$-symmetric
      2HDM defined by the Higgs doublet that couples to each kind of fermions.}
   \label{tab:yukawaDefinitions}
\end{table}
We conclude by giving the set of independent parameters that
parametrize the tree-level potential of the CP-conserving
2HDM. By exploiting the minimum conditions from
Eqs.~(\ref{eq:tadpoleCondition1}) and (\ref{eq:tadpoleCondition2}), 
the set of independent input parameters in the mass basis is given by
\beq
m_h,\; m_H,\;  m_A,\;  m_{H^\pm},\;  m_{12}^2,\;  \alpha,\;  \tan\beta,\;  v \;.
\eeq
Alternatively, the original parametrization of the scalar
potential in the interaction basis can be
used so that the set of independent parameters is given by 
\beq
\lambda _1 , \; \lambda _2 , \; \lambda _3,\; \lambda _4 ,\; \lambda
_5 ,\; m_{12}^2,\; \tan\beta,\; v \;. 
\label{eq:inputSetInteractionBase}
\eeq

%%%%%%%%%%%%%%%%%%%%%%%%%%%%%%%%%%%%%%%%%%%%%%%%%%%%%%%%%%%%%%
\subsection{Introduction of the N2HDM}
%%%%%%%%%%%%%%%%%%%%%%%%%%%%%%%%%%%%%%%%%%%%%%%%%%%%%%%%%%%%%%
\label{sec:introOfN2HDM}
We consider a general CP-conserving N2HDM that is obtained from the
CP-conserving 2HDM with a softly broken $\mathbb{Z}_2$ symmetry by
adding a real singlet field $\Phi_S$. The phenomenology of this
version of the N2HDM has been extensively discussed in
\cite{Muhlleitner:2016mzt,Muhlleitner:2017dkd,Azevedo:2018llq}.\footnote{For
  the N2HDM in different phases, cf. \cite{Engeln:2018mbg}. A recent
  discussion of its vacuum instabilities can be found in
  \cite{Ferreira:2019iqb}.} In terms of the $SU(2)_L$ 
Higgs doublets $\Phi_{1,2}$ and the singlet field $\Phi_S$, the N2HDM
potential reads
\begin{align}
V_{\text{N2HDM}} =&~ m_{11}^2 \left| \Phi _1 \right| ^2 + m_{22}^2 \left|
  \Phi _2 \right| ^2 - m_{12}^2 \left( \Phi _1 ^\dagger \Phi _2 +
  \textit{h.c.} \right) + \frac{\lambda _1}{2} \left( \Phi _1^\dagger
  \Phi _1 \right) ^2 + \frac{\lambda _2}{2} \left( \Phi _2^\dagger
  \Phi _2 \right) ^2 \nonumber \\ 
&+ \lambda _3 \left( \Phi _1^\dagger \Phi _1 \right) \left( \Phi
  _2^\dagger \Phi _2 \right) + \lambda _4 \left( \Phi _1^\dagger \Phi
  _2 \right) \left( \Phi _2^\dagger \Phi _1 \right) + \frac{\lambda
  _5}{2} \left[ \left( \Phi _1^\dagger \Phi _2 \right) ^2 +
  \textit{h.c.} \right]   \nonumber \\
&~ +\frac{1}{2} m_S^2 \Phi _S^2 + \frac{1}{8} \lambda _6 \Phi _S^4 +
  \frac{1}{2} \lambda _7 \left( \Phi _1 ^\dagger \Phi _1 \right) \Phi
  _S^2 + \frac{1}{2} \lambda _8 \left( \Phi _2 ^\dagger \Phi _2
  \right) \Phi _S^2~, \label{eq:vpotn2hdm}
\end{align}
where the first two lines describe the 2HDM part of the N2HDM and the
last line contains the contribution of the singlet field $\Phi_S$. The
potential is obtained by imposing two $\mathbb{Z}_2$ symmetries. The first one, called $\mathbb{Z}_2$, under which the doublets and singlet transform as
\beq  
\Phi_1 \to \Phi_1 \;, \quad \Phi_2 \to -\Phi_2 \;, \quad \Phi_S \to
\Phi_S \;,
\label{eq:z2one}
\eeq 
is the trivial
generalization of the usual 2HDM $\mathbb{Z}_2$ symmetry to the N2HDM. It is softly broken by the term involving $m_{12}^2$ and will be
extended to the Yukawa sector to avoid FCNCs at tree level. The second
symmetry, called $\mathbb{Z}_2^\prime$, under which the doublets and singlet transform as
\beq 
\Phi_1 \to \Phi_1 \;, \quad \Phi_2 \to \Phi_2 \;, \quad \Phi_S \to -\Phi_S \;,
\eeq 
is not explicitly broken. The N2HDM potential depends on four real
mass parameters $m_{11}$, $m_{22}$, $m_{12}$, and $m_S$ and eight
dimensionless real coupling constants $\lambda_i$
($i=1,...,8$). After EWSB the doublet and singlet fields can be
expanded around their non-negative real VEVs $v_1$, $v_2$ and $v_S$ as 
\begin{equation}
\Phi _1 = \begin{pmatrix} \omega ^\pm _1 \\ \frac{v_1 + \rho _1 + i \eta
    _1 }{\sqrt{2}} \end{pmatrix} ~,~~~ \Phi _2 = \begin{pmatrix}
  \omega ^\pm _2 \\ \frac{v_2 + \rho _2 + i \eta
    _2}{\sqrt{2}} \end{pmatrix} ~,~~~ \Phi _S = v_S + \rho _S ~, 
\label{eq:vevexpansion1}
\end{equation} 
where $\rho _i$ and $\rho _S$ represent three CP-even fields, $\eta
_i$ two CP-odd fields and $\omega ^\pm _i$ are two charged fields
($i=1,2$). The two VEVs of the doublets are again related to the SM VEV $v$ as 
$v^2 = v_1^2 + v_2^2$, and the mixing angle $\beta$ is introduced as
before through $\tan\beta =
v_2/v_1$. Inserting Eq.~(\ref{eq:vevexpansion1}) into the N2HDM potential
Eq.~(\ref{eq:vpotn2hdm}) yields
\begin{equation}
	V_\text{N2HDM} = \frac{1}{2} \left( \rho _1 ~~ \rho _2 ~~ \rho
          _S \right) M_\rho ^2 \begin{pmatrix} \rho _1 \\ \rho _2 \\
          \rho _S \end{pmatrix} + T_1 \rho _1 + T_2 \rho _2 + T_S \rho
        _S + ~~ \cdots ~, \label{eq:vpotentialn2hdm}
\end{equation}
where $M_\rho$ is the $3\times 3$ mass matrix in the CP-even scalar
sector and $T_k$ ($k=1,2,S$) are the three tadpole terms. We demand
the three VEVs to represent the minimum of the potential by requiring ($k,l=1,2,S$)
\begin{equation}
\frac{\partial V_\text{N2HDM}}{\partial \Phi _l} \Bigg| _{\left\langle \Phi _k \right\rangle} = 0~,
\end{equation}
which at tree level leads to the three tadpole conditions
\begin{align}
\frac{T_1}{v_1} &\equiv m_{11}^2 - m_{12}^2 \frac{v_2}{v_1} +
                  \frac{v_1^2 \lambda _1}{2} + \frac{v_2 ^2 \lambda
                  _{345}}{2} + \frac{v_S ^2 \lambda _7}{2} = 0  \label{eq:tadpoleCondition1x} \\
\frac{T_2}{v_2} &\equiv m_{22}^2 - m_{12}^2 \frac{v_1}{v_2} +
                  \frac{v_2^2 \lambda _2}{2} + \frac{v_1 ^2 \lambda
                  _{345}}{2} + \frac{v_S ^2 \lambda _8}{2}  = 0 \label{eq:tadpoleCondition2x} \\
\frac{T_S}{v_S} &\equiv m_S^2 + \frac{v_1 ^2 \lambda _7}{2} +
                  \frac{v_2 ^2 \lambda _8}{2} + \frac{v_S ^2 \lambda
                  _6}{2} = 0~. \label{eq:tadpoleCondition3x}
\end{align}
These conditions can be used to replace $m_{11}^2$, $m_{22}^2$ and $m_S^2$
in favor of the three tadpole terms. The mass matrix of the CP-even scalar fields reads
\begin{align}
M_\rho ^2 &\equiv \begin{pmatrix}
m_{12}^2 \frac{v_2}{v_1} + \lambda _1 v_1^2 & -m_{12}^2 + \lambda _{345} v_1 v_2 & \lambda _7 v_1 v_S \\ -m_{12}^2 + \lambda _{345} v_1 v_2 & m_{12}^2 \frac{v_1}{v_2} + \lambda _2 v_2^2 & \lambda _8 v_2 v_S \\ \lambda _7 v_1 v_S & \lambda _8 v_2 v_S & \lambda _6 v_S^2
\end{pmatrix} + \begin{pmatrix} \frac{T_1}{v_1} & 0 & 0 \\ 0 & \frac{T_2}{v_2} & 0 \\ 0 & 0 & \frac{T_S}{v_S} \end{pmatrix} ~. \label{eq:massMatricesrho}
\end{align}
We introduce three mixing angles $\alpha _i$ ($i=1,2,3$) defined in the range
\begin{equation}
	- \frac{\pi}{2} \leq \alpha _i < \frac{\pi}{2} 
\end{equation}
in order to diagonalize the mass matrix by means of the the orthogonal matrix
$R$, parametrized as
\begin{equation}
	R = \begin{pmatrix}
	c_{\alpha _1} c_{\alpha _2} & s_{\alpha _1} c_{\alpha _2} & s_{\alpha _2} \\
	- \left( c_{\alpha _1} s_{\alpha _2} s_{\alpha _3} + s_{\alpha _1} c_{\alpha _3} \right) & c_{\alpha _1} c_{\alpha _3} - s_{\alpha _1} s_{\alpha _2} s_{\alpha _3} & c_{\alpha _2} s_{\alpha _3} \\
	-c_{\alpha _1} s_{\alpha _2} c_{\alpha _3} + s_{\alpha _1} s_{\alpha _3} & - \left( c_{\alpha _1} s_{\alpha _3} + s_{\alpha _1} s_{\alpha _2} c_{\alpha _3} \right) & c_{\alpha _2} c_{\alpha _3}
	\end{pmatrix} ~.
\end{equation}
This matrix transforms the CP-even interaction fields $\rho_i$
($i=1,2,3$, $\rho_3 \equiv \rho_S$) into the mass eigenstates $H_i$,
\begin{equation}
	\begin{pmatrix} H_1 \\ H_2 \\ H_3 \end{pmatrix} =  R \begin{pmatrix} \rho _1 \\ \rho _2 \\ \rho _3 \end{pmatrix} ~. \label{eq:rotationCPeven}
\end{equation}
This transformation yields the diagonalized mass matrix
\begin{equation}
	D ^2 _\rho \equiv R M^2 _\rho R^T \equiv \text{diag} \left( m_{H_1}^2 , m_{H_2}^2 , m_{H_3}^2 \right) ~, \label{eq:diagonalizationMassMatrix}
\end{equation}
where we demand the three CP-even Higgs bosons $H_i$ to be
ordered by ascending mass,
\begin{equation}
	m_{H_1} < m_{H_2} < m_{H_3} ~.
\end{equation}
We do not show the CP-odd and charged mass matrices explicitly in
Eq.~(\ref{eq:vpotentialn2hdm}) as they do not change with respect to
the 2HDM. At tree level they are diagonalized by the mixing angle $\beta$, yielding the CP-odd and charged
Higgs bosons $A$ and $H^\pm$ with masses $m_A$ and $m_{H^\pm}$,
respectively, as well as the massless neutral and charged Goldstone
bosons $G^0$ and $G^\pm$. \s

As mentioned before, the softly broken $\mathbb{Z}_2$ symmetry
transformation shown in Eq.~(\ref{eq:z2one}) is extended to the Yukawa
sector in order to avoid FCNCs at tree level. This leads to the same
four types of doublet couplings to the fermion fields as in the 2HDM. 
At tree level, the N2HDM potential is parametrized by twelve
parameters. By exploiting the minimum conditions in
Eqs.~(\ref{eq:tadpoleCondition1x}) to (\ref{eq:tadpoleCondition3x}), the set of independent parameters in the
mass basis is given by 
\beq
m_{H_{1}},\;  m_{H_{2}},\;  m_{H_{3}},\;  m_A,\;  m_{H^\pm},\;  m_{12}^2,\;  \alpha_{1},\; \alpha_{2},\; \alpha_{3},\; \tan\beta,\;  v,\; v_S \;.
\eeq

%%%%%%%%%%%%%%%%%%%%%%%%%%%%%%%%%%%%%%%%%%%%%%%%%%%%%%%%%%%%%%
\section{Renormalization}
\label{sec:renormalization}
The ultraviolet (UV) divergences that appear in the computation of the
electroweak corrections to the Higgs boson decays in the 2HDM and N2HDM require the
renormalization of the parameters which are involved in the calculations. The
renormalization schemes that we apply are chosen such that they
fulfil the following requirements as far as it is possible: 
\begin{itemize}
\item We require physical OS renormalization conditions, where possible.
\item The renormalization schemes are chosen such that they preserve gauge-parameter-inde\-pen\-dent relations between the input parameters and the computed observables.
\item The NLO corrections should not become unnaturally large. Throughout the paper, we refer to this behavior of a renormalization scheme as being 'numerically stable'.
\item Finally, conditions that depend on a physical process, \textit{i.e.} so-called process-dependent renormalization schemes, are avoided, if possible.
\end{itemize}
The renormalization conditions
respecting these requirements, in particular the gauge-inde\-pen\-dent
renormalization, have been introduced and extensively discussed by us
for the 2HDM in \cite{Krause:2016oke,Krause:2016xku} and for the N2HDM
in \cite{Krause:2017mal}. Moreover, renormalization schemes respecting some or all of these criteria were also discussed in \cite{Denner:2016etu,Altenkamp:2017ldc,Altenkamp:2017kxk,Denner2018} for the 2HDM and in \cite{Fox:2017hbw,Grimus:2018rte} for a generic
multi-Higgs sector.
In this paper we apply our renormalization conditions of
Refs.~\cite{Krause:2016oke,Krause:2016xku} for the 2HDM and of
Ref.~\cite{Krause:2017mal} for the N2HDM. Additionally, we apply the
renormalization conditions developed for the 2HDM in \cite{Denner:2016etu,Altenkamp:2017ldc,Altenkamp:2017kxk,Denner2018}. All these conditions have
been implemented in the codes {\tt 2HDECAY} \cite{Krause:2018wmo} and
{\tt N2HDECAY} \cite{Krause:2019oar} for the 2HDM and N2HDM, respectively, which we use to calculate the
presented loop-corrected branching 
ratios. We refer to the given literature for the detailed introduction
and description of the renormalization schemes. Here, we list them only
very briefly for a convenient overview. 

%%%%%%%%%%%%%%%%%%%%%%%%%%%%&%%%%%%%%%%%%%%%%%%%%%%%%%%%%%%%%%
\subsection{Renormalization of the 2HDM}
\label{sec:renormalization2HDM}
The renormalization schemes for the scalar mixing angles $\alpha$ and $\beta$ implemented in {\texttt{2HDECAY}} \cite{Krause:2018wmo} for the 2HDM are summarized in Tab.~\ref{tab:renSchemes2HDM}. For
details on these schemes and for the complete EW one-loop renormalization of the 2HDM, we refer to \cite{Krause:2016oke,Krause:2016xku,Krause:2018wmo}. The first
column in the table gives the identifier with which the user selects the
renormalization scheme in the input file of {\tt 2HDECAY}. The second 
column describes the scheme and the third one the abbreviation that
will be used for it in the presentation of the results. The fourth
column refers to the tadpole scheme that is used and the last column
lists the references where the respective renormalization scheme is
introduced and described. \s
\begin{table*}
\centering
{\renewcommand{\arraystretch}{1.2}\begin{tabular}{ccccc} \toprule
{\tt 2HDECAY} ID & Scheme & Abbreviation & Tadpole scheme & Reference \rule{0pt}{2.9ex}\rule[-1.2ex]{0pt}{0pt} \\[1.6mm] \hline 
1/2 & KOSY & $\text{KOSY}^{o/c}\text{(std)}$ & standard & \cite{Kanemura:2004mg} \\
3/4 & KOSY & $\text{KOSY}^{o/c}$ & alternative FJ & \cite{Kanemura:2004mg,krauseMaster2016,Krause:2016oke} \\
5/6 & $p_{*}$-pinched & $\text{p}_{*}^{o/c}$ & alternative FJ & \cite{krauseMaster2016,Krause:2016oke} \\
7/8 & OS-pinched & $\text{pOS}^{o/c}$ & alternative FJ & \cite{krauseMaster2016,Krause:2016oke} \\
9 & Process-dependent 1 & $\text{proc1}$ & alternative FJ & \cite{krauseMaster2016,Krause:2016oke,krausePhD2019} \\
10 & Process-dependent 2 & $\text{proc2}$ & alternative FJ & \cite{krausePhD2019} \\
11 & Process-dependent 3 & $\text{proc3}$ & alternative FJ & \cite{krausePhD2019} \\
12 & Physical OS 1 & $\text{OS1}$ & alternative FJ & \cite{Denner2018} \\
13 & Physical OS 2 & $\text{OS2}$ & alternative FJ & \cite{Denner2018} \\
14 & Physical OS 12 & $\text{OS12}$ & alternative FJ & \cite{Denner2018} \\
15 & Rigid symmetry (BFMS) & $\text{BFMS}$ & alternative FJ & \cite{Denner2018} \\
16 & $\overline{\text{MS}}$ & $\overline{\text{MS}}\text{(std)}$ & standard & \cite{krauseMaster2016,Krause:2016oke,Altenkamp:2017ldc} \\
17 & $\overline{\text{MS}}$ & $\overline{\text{MS}}$ & alternative FJ & \cite{krauseMaster2016,Krause:2016oke,Altenkamp:2017ldc} \\
\bottomrule
\end{tabular}}
\caption{Renormalization schemes of {\tt 2HDECAY}
  \cite{Krause:2018wmo}. For further explanations, we refer to the text.}\label{tab:renSchemes2HDM}
\end{table*}

The standard tadpole scheme is a commonly used renormalization schemes
for the tadpoles, {\it cf.}~{\it e.g.}~\cite{Denner:1991kt}
for the SM and \cite{Kanemura:2004mg, Kanemura:2015mxa} for the
2HDM. They are renormalized such that the ground state of the
potential represents the minimum also at higher orders. In the
standard tadpole scheme this condition is imposed on the
loop-corrected potential. As the latter is in general gauge dependent,
the counterterms (CTs) defined through this minimum, {\it e.g.}~the
scalar mass (matrix) CTs, become manifestly
gauge-dependent themselves. In the alternative tadpole scheme,
proposed by Fleischer and Jegerlehner (FJ) for the SM in \cite{Fleischer:1980ub}
and applied to the 2HDM for the 
first time in Refs.\,\cite{krauseMaster2016,Krause:2016oke}, the VEVs are
defined through the gauge-independent tree-level potential so that
they become manifestly gauge-independent quantities and hence, also
the mass (matrix) CTs become manifestly gauge-independent. \s

The KOSY scheme, introduced in \cite{Kanemura:2004mg} and named by us
after the authors' initials, defines the CTs for the mixing angles
$\alpha$ and $\beta$ through 
off-diagonal wave-function renormalization constants. As shown in
\cite{krauseMaster2016,Krause:2016oke}, this not only implies a
gauge-dependent definition of the mixing angle CTs but also leads to
explicitly gauge-dependent decay amplitudes and partial decay
widths. Nevertheless, in {\tt 2HDECAY} it has been implemented as a
benchmark scheme for comparison with other schemes, both in the
standard and alternative tadpole scheme. Due to its intricate gauge
dependence, we do not recommend to use the KOSY scheme for actual
computations, however. \s 

On the other hand, the pinched schemes lead to manifestly
gauge-independent mixing angle CTs 
\cite{krauseMaster2016,Krause:2016oke}. In these schemes, the OS-based
definition of the mixing angle counterterms for $\alpha$ and $\beta$
of the KOSY scheme is kept, but instead of using the usual
gauge-dependent off-diagonal wave function renormalization
counterterms (WFRCs), the WFRCs are defined through pinched
self-energies in the alternative tadpole scheme by applying the pinch
technique (PT) \cite{Binosi:2004qe,
  Binosi:2009qm, Cornwall:1989gv, Papavassiliou:1989zd,
  Degrassi:1992ue, Papavassiliou:1994pr, Watson:1994tn,
  Papavassiliou:1995fq} We employ two different definitions that
differ solely by the scale at which the self-energies are
evaluated. In the $p_\star$-pinched scheme they are evaluated at the
arithmetic average of the squared masses of the external particles. In
the OS-pinched scheme, OS-motivated scales are chosen for the
self-energies. \s

The upper indices in the KOSY and pinched schemes refer to the part of
the Higgs sector that is applied in the renormalization of
$\beta$. The angle appears both in the charged and in the CP-odd mass
matrix and its renormalization can be defined either through the
charged ($'c'$) or the CP-odd ($'o'$) sector. \s

Process-dependent renormalization schemes have the advantage to lead to manifestly
gauge-independent mixing angle CTs. However, they make the renormalization conditions
dependent on a specific physical processes. In {\texttt{2HDECAY}}, three different
processes were implemented for the renormalization of the scalar mixing angles \cite{krauseMaster2016,Krause:2016oke,krausePhD2019}. In the
first process-dependent scheme ('proc1'), the CT for $\beta$, 
$\delta\beta$ is defined through the decay of the pseudoscalar into a
$\tau$-lepton pair, $A\to \tau^+\tau^-$, and the CT for
$\alpha$, $\delta \alpha$, through the decay of the heavy scalar into
a $\tau$-lepton pair, $H\to \tau^+ \tau^-$. In the second
process-dependent scheme ('proc2'), $\delta \beta$ is again defined
through $A\to \tau^+\tau^-$, but $\delta \alpha$ is defined through
the decay of the lighter CP-even Higgs boson into a $\tau$-lepton
pair, $h\to \tau^+ \tau^-$. In the third process-dependent scheme
('proc3'), $\delta\beta$ and $\delta\alpha$ are simultaneously defined
through $H\to \tau^+ \tau^-$ and $h\to \tau^+ \tau^-$. \s

In the physical OS scheme proposed in \cite{Denner2018}, the mixing angles CTs
are defined through ratios of processes such that they are manifestly
gauge-independent, while at the same time avoiding potentially large NLO
corrections that are usually present in process-dependent schemes. Solely for the
purpose of renormalization, two right-handed fermion singlets
$\nu_{1R}$ and $\nu_{2R}$ are introduced which transform under an
additionally introduced $\mathbb{Z}_2$ symmetry transformation as $\nu_{1R} \to -
\nu_{1R}$ and $\nu_{2R} \to \nu_{2R}$. The singlets are coupled via
Yukawa couplings to left-handed lepton doublets of the 2HDM, giving
rise to two massive Dirac neutrinos $\nu_1$ and $\nu_2$. Denoting by
${\cal A}$ the decay amplitude, the three implemented OS schemes are
defined as follows. In 'OS1' the ratio ${\cal A}_{H\to
    \nu_1 \bar{\nu}_1}/{\cal A}_{h\to \nu_1 \bar{\nu}_1}$ is used to
define $\delta \alpha$ while ${\cal A}_{A\to
    \nu_1 \bar{\nu}_1}/{\cal A}_{H\to \nu_1 \bar{\nu}_1}$ is used to define
$\delta\beta$. In 'OS2' we have ${\cal A}_{H\to
    \nu_2 \bar{\nu}_2}/{\cal A}_{h\to \nu_2 \bar{\nu}_2}$ for $\delta
\alpha$ and ${\cal A}_{A\to 
    \nu_2 \bar{\nu}_2}/{\cal A}_{H\to \nu_2 \bar{\nu}_2}$ for
$\delta\beta$. In 'OS12', again the ratio ${\cal A}_{H\to
    \nu_1 \bar{\nu}_1}/{\cal A}_{h\to \nu_1 \bar{\nu}_1}$ is used for $\delta
\alpha$ and a specific combination of all possible decay amplitudes
$\mathcal{A} _{h/H \rightarrow \nu _j \bar{\nu }_j}$ and $\mathcal{A}
_{A \rightarrow \nu _j \bar{\nu }_j}$ ($i,j=1,2$) for $\delta \beta$.
\s

In the rigid symmetry scheme (BFMS), the rigid symmetry of the Lagrangian is
used to connect the renormalization of $\alpha$ and $\beta$ to the
renormalization of the WFRCs. 
In \cite{{Denner2018}}, this scheme was worked out and applied to derive
gauge-independent counterterms for the scalar mixing
angles of the 2HDM within the framework of the background field method
(BFM) \cite{KlubergStern:1974xv,KlubergStern:1975hc,Boulware:1980av,Abbott:1980hw,Abbott:1981ke,Hart:1984jy,Denner:1994xt}. \s

We also apply the $\overline{\mbox{MS}}$ scheme for $\delta \alpha$
and $\delta \beta$
\cite{krauseMaster2016,Krause:2016oke,Denner:2016etu,Altenkamp:2017ldc} for
reference, both in the standard and the alternative tadpole scheme,
although it can typically lead to very large corrections at NLO for many two-body decay
processes \cite{krauseMaster2016,LorenzMaster2015}. Note that the
$\overline{\mbox{MS}}$ scheme induces gauge-dependent $\delta\alpha$
and $\delta \beta$ and hence gauge-dependent partial decay widths unless the alternative tadpole scheme is applied.

%%%%%%%%%%%%%%%%%%%%%%%%%%%%&%%%%%%%%%%%%%%%%%%%%%%%%%%%%%%%%%
\subsection{Renormalization of the N2HDM}
In Tab.~\ref{tab:renSchemesN2HDM}, we summarize the renormalization schemes
that we apply for the four scalar mixing angles in the computation of the
EW-corrected NLO widths of the neutral N2HDM Higgs bosons. For details
on these schemes and for the complete electroweak one-loop renormalization of the N2HDM, we refer to \cite{Krause:2017mal,krausePhD2019}. \s

\label{sec:renormalizationN2HDM}
\begin{table*}[t]
\centering
{\renewcommand{\arraystretch}{1.2}\begin{tabular}{ccccc} \toprule
{\tt ewN2HDECAY} ID & Scheme & Abbreviation & Tadpole scheme & Reference \rule{0pt}{2.9ex}\rule[-1.2ex]{0pt}{0pt} \\[1.6mm] \hline 
1/2 & Adapted KOSY & $\text{KOSY}^{o/c}\text{(std)}$ & standard & \cite{Kanemura:2004mg,Krause:2017mal,krausePhD2019} \\
3/4 & Adapted KOSY & $\text{KOSY}^{o/c}$ & alternative FJ & \cite{Kanemura:2004mg,Krause:2017mal,krausePhD2019} \\
5/6 & $p_{*}$-pinched & $\text{p}_{*}^{o/c}$ & alternative FJ & \cite{Krause:2017mal,krausePhD2019} \\
7/8 & OS-pinched & $\text{pOS}^{o/c}$ & alternative FJ & \cite{Krause:2017mal,krausePhD2019} \\
9 & $\overline{\text{MS}}$ & $\overline{\text{MS}}\text{(std)}$ & standard & \cite{Krause:2017mal,krausePhD2019} \\
10 & $\overline{\text{MS}}$ & $\overline{\text{MS}}$ & alternative FJ & \cite{Krause:2017mal,krausePhD2019} \\
\bottomrule
\end{tabular}}
\caption{Renormalization schemes of {\tt ewN2HDECAY}
  \cite{Krause:2019oar}. For further explanations, we refer to the text.}\label{tab:renSchemesN2HDM}
\end{table*}

In contrast to the 2HDM, in the N2HDM three mixing angles $\alpha_i$ ($i=1,2,3$) in the
CP-even neutral Higgs sector require renormalization. Together with the mixing angle $\beta$ from the CP-odd and charged Higgs sectors, in total four mixing angle
CTs are required, namely $\delta\alpha_i$ and $\delta\beta$. In
\cite{Krause:2017mal}, we generalized the renormalization schemes
for the mixing angles of the 2HDM developed in \cite{Krause:2016oke} to the more intricate Higgs sector of the N2HDM. The schemes are defined in complete analogy to the 2HDM and
have been implemented in {\tt ewN2HDECAY} \cite{Krause:2019oar}. 
The first column in Tab.~\ref{tab:renSchemesN2HDM} again gives 
the identifier with which the user selects the renormalization scheme
in the input file of {\tt ewN2HDECAY}. The corresponding scheme is
named in the second column along with its abbreviation in the third
column. The fourth column refers to the tadpole scheme that is used
and the last column cites the references where the respective
renormalization scheme is introduced and described. \s

The adapted KOSY scheme is the generalization of the KOSY scheme
\cite{Kanemura:2004mg} to the N2HDM, both by applying the standard and
the alternative tadpole scheme. Accordingly, the $p_\star$-pinched and
OS-pinched schemes are the extensions of the corresponding renormalization schemes introduced in the 2HDM to the case of the N2HDM. All four schemes provide the renormalization of the mixing angle $\beta$ both via the charged ($'c'$) and the CP-odd ($'o'$) Higgs sector. For
reference, also the renormalization of the mixing angles via the
$\overline{\mbox{MS}}$ scheme has been implemented in both the standard and the alternative tadpole scheme. 

%%%%%%%%%%%%%%%%%%%%%%%%%%%%%%%%%%%%%%%%%%%%%%%%%%%%%%%%%%%%%%
\section{Numerical Results}
\label{sec:numerics}
%%%%%%%%%%%%%%%%%%%%%%%%%%%%%%%%%%%%%%%%%%%%%%%%%%%%%%%%%%%%%%
In the following subsections, we give in tabular format the relative sizes of the
EW corrections to the BRs for the various Higgs
decays in the 2HDM and in the N2HDM. In order to define them, we first
have to explain what is meant when we talk about the inclusion of QCD
and/or EW corrections in the BRs. Higher-order
corrected BRs means that we include higher-order
corrections in the partial Higgs decay widths and hence also the total
widths as follows\footnote{For further details, {\it
    cf.}~Refs.~\cite{Krause:2018wmo,Krause:2019oar}}. 
\begin{itemize}
\item All decays include the state-of-the-art higher-order QCD
  corrections to the OS and off-shell Higgs boson decays where
  appropriate. Also the loop-induced decays into gluonic
  and photonic final states include higher-order QCD corrections. The
  exact description of the 
  implemented loop-order and applied approximation in the various
  decays can be found in \cite{Djouadi:2018xqq,Spira:2016ztx}. The
  results for the SM and MSSM, respectively, have been translated to
  the 2HDM \cite{Harlander:2013qxa} and subsequently included in {\tt
    2HDECAY}. The generalizaton to the N2HDM
  has been performed in {\tt N2HDECAY} \cite{Muhlleitner:2016mzt} and
  subsequently included in {\tt ewN2HDECAY}. 
\item The codes include off-shell decays into heavy-quark pairs,
  massive gauge boson pairs, neutral Higgs pairs as well as Higgs and gauge
  boson final states. The EW corrections are included only for decays into OS
  final states, however. Off-shell decays are computed at LO or, where
  appropriate and available, with the inclusion of higher-order QCD corrections. 
\item The loop-induced decays into gluon and photon final states as
  well as into $Z\gamma$ do not include any EW corrections as they are
  of two-loop order. 
\item For the combination of the QCD and EW corrections, we assume
  that these corrections factorize.  
\end{itemize}

The relative size $\Delta \mbox{BR}$ of the higher-order corrected
BRs, including QCD and EW corrections as described
above, is defined as 
\beq
\Delta \mbox{BR} =
\frac{\mbox{BR}^{\text{QCD\&EW}}-\mbox{BR}^{\text{QCD}}}{\mbox{BR}^{\text{QCD}}}
\;. 
\eeq
The superscript 'QCD\&EW' means that both QCD and EW corrections have
been implemented in the various partial decay widths where appropriate
and/or possible as described above. The superscript 'QCD'
refers to the BR where only QCD corrections are taken
into account in the various Higgs decays, where appropriate. The
$\Delta \mbox{BR}$ hence quantifies the relative importance of the EW
corrections in the BRs with respect to the QCD corrected BRs. 

%%%%%%%%%%%%%%%%%%%%%%%%%%%%%%%%%%%%%%%%%%%%%%%%%%%%%%%%%%%%%%
\subsection{2HDM Higgs Decays}
\label{sec:2HDMHiggsDecays}
%%%%%%%%%%%%%%%%%%%%%%%%%%%%%%%%%%%%%%%%%%%%%%%%%%%%%%%%%%%%%%
We start with the results for the 2HDM. We consider two different kinds of
parameter sets, one where the lighter of the CP-even Higgs bosons, $h$, is
the SM-like Higgs, and another set where the heavier one, $H$, is SM-like. For both kinds of parameter sets, we consider all four 2HDM types. For the case that $h$ is SM-like, the total amount of points used for the numerical analyses is as follows,
\beq
\begin{array}{lllll}
\underline{h \mbox{ is SM-like:}} \qquad \qquad &
\mbox{Type I:} & \; 373\,517 & \qquad \mbox{Type II:} & \; 413\,377 \\[0.2cm]
& \mbox{Type LS:} & \; 373\,000 & \qquad \mbox{Type FL:} & \; 431\,540 ~,
\end{array}
\label{eq:numberslighter}
\eeq 
where we introduced the abbreviations LS for the lepton-specific 2HDM and FL
for the flipped 2HDM. In case that $H$ is SM-like, we found less valid parameter
points in our scans that could be used for our numerical analyses,
\beq
\begin{array}{lllll}
\underline{H \mbox{ is SM-like:}} \qquad \qquad &
\mbox{Type I:} & \; 747 & \qquad \mbox{Type II:} & \; 39 \\[0.2cm]
& \mbox{Type LS:} & \; 132 & \qquad \mbox{Type FL:} & \; 124 ~.
\end{array}
\label{eq:numbersheavier}
\eeq 
As can be inferred from these numbers, this case is strongly
disfavored in comparison to the case that $h$ is the SM-like
Higgs. This also means that the statistics for our analyses in these
scenarios is very low.

%%%%%%%%%%%%%%%%%%%%%%%%%%%%%%%%%%%%%%%%%%%%%%%%%%%%%%%%%%%%%%
\subsection{SM-Like 2HDM Higgs Decays}
\label{sec:SMlikeHiggsDecays}
%%%%%%%%%%%%%%%%%%%%%%%%%%%%%%%%%%%%%%%%%%%%%%%%%%%%%%%%%%%%%%
We start by giving our results for the 2HDM decays of the SM-like
Higgs boson before moving on to the non-SM-like Higgs decays.

%%%%%%%%%%%%%%%%%%%%%%%%%%%%%%%%%%%%%%%%%%%%%%%%%%%%%%%%%%%%%%
\subsubsection{$h$ is the SM-like Higgs Boson}
%%%%%%%%%%%%%%%%%%%%%%%%%%%%%%%%%%%%%%%%%%%%%%%%%%%%%%%%%%%%%%
In the tables of this section we list the relative size of the EW
corrections $\Delta\mbox{BR}$ to the BRs of the SM-like Higgs boson
$h$ for the four 2HDM types I, II, LS and FL. We use subscripts for
$\Delta\mbox{BR}$ in order to refer to the corresponding decay
channel. In the 
computation of the EW corrections we apply various renormalization
schemes presented in Tab.~\ref{tab:renSchemes2HDM} of
Section~\ref{sec:renormalization2HDM}. The 
comparison of the results in the different renormalization schemes
gives an estimate of the theoretical error on the BR due
to the missing higher-order corrections that are not included. Moreover, it
also shows which renormalization schemes are less suitable because
they lead to very large NLO corrections. In order to keep the
presentation of the higher-order EW effects computed in the various
renormalization schemes clear, we group the results into bins
quantifying the size of the relative corrections to the BRs and into subsets of
renormalization schemes that lead to similar results. The subsets of
schemes are given as superscript of $\Delta\mbox{BR}$. The following
scheme sets could be defined here:
\begin{align}
\pmb{S_1} &\equiv \big\{ ~ \text{KOSY}^o ~,~ \text{KOSY}^c ~,~ \text{p}_*^o ~,~ \text{p}_*^c ~,~ \text{pOS}^o ~,~ \text{pOS}^c ~,~ \text{BFMS} ~ \big\} \nonumber\\
\pmb{S_2} &\equiv \big\{ ~ \text{proc1} ~,~ \text{OS1} ~ \big\} \nonumber\\
\pmb{S_3} &\equiv \big\{ ~ \text{proc2} ~,~ \text{proc3} ~,~
            \text{OS12} ~ \big\} \nonumber \\
\pmb{\text{OS2}} &\equiv \big\{ ~ \text{OS2} ~ \big\} \nonumber \\
\pmb{\overline{\text{MS}}} &\equiv \big\{ ~ \overline{\text{MS}} ~
                             \big\} \nonumber
\end{align}
Let us clarify that in the generation of the numbers for a
specific scheme the respective set of input parameters is understood
to be given in this scheme. This means in particular that we did not
convert an input parameter set given in one scheme to the input
parameter set for a different renormalization scheme. Therefore the
numbers calculated in two different schemes cannot be directly compared to each
other in order to derive {\it e.g.}~the theoretical error due to
missing higher-order corrections. Our approach is instead a
large-scale analysis for each scheme separately. This way, we cannot
estimate the remaining theoretical errors, but we can judge if schemes
or sets of schemes behave similarly with respect to their relative
size of corrections. \s 

In the tables of this section, the relative size of the corrections
to the BRs are quantified via a binning as
\begin{align}
	\lesssim / \gtrsim a\,\% ~(b\,\%) \;. \label{eq:percentageValues}
\end{align}
This is to be read as ``$b\,\%$ of all used input parameter sets lead
to corrections $\Delta \text{BR}$ below/above or approximately equal
to $a\,\%$''. For each renormalization scheme set, we show two pairs
of percentage values ``$a\,\%$'' according to the format described in
Eq.~(\ref{eq:percentageValues}): a lower value where the peak of
all corrections is found and a larger value where the majority of
all results lies in. An exception to this is the case when all
corrections are situated in the lowest bin ($2.5\,\%$) in which case
we only show one pair of numbers. For the $\overline{\text{MS}}$
scheme (and for some other schemes and decay channels) where the corrections
are typically huge, we show the following two pairs
of values: one where half of the results lies in (``half'' only if half
of the results are actually below $\pm 100\,\%$, otherwise a lower
appropriate value) and a second pair of values that  
corresponds to the largest bin (\textit{i.e.} above $\pm$100\,\%). \s

\begin{figure}[ht!]
\centering
\includegraphics[width=9cm]{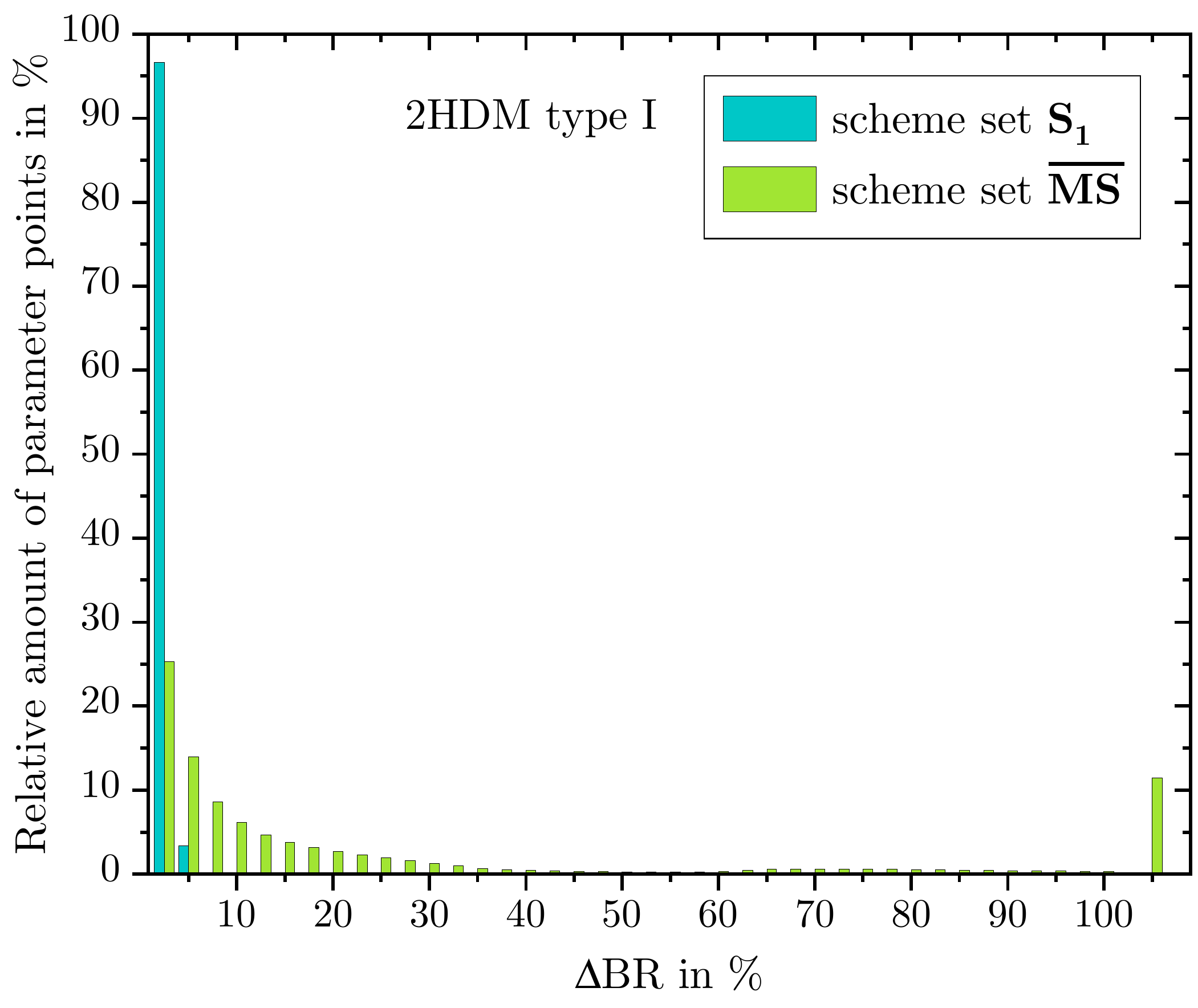}
\caption{2HDM type I: Binned amount of parameter points in \% for the scheme sets
  $\pmb{S_1}$ (blue) and $\pmb{\overline{\text{MS}}}$ (green) as a
  function of $\Delta \text{BR}$ in \% for the decay $h\to b\bar{b}$.}
\label{fig:numerical}
\end{figure}
This means for example for the first upper left entry of
Tab.~\ref{tab:hlight2hdmbr} that in the 2HDM type I, the relative
corrections $\Delta \mbox{BR}_{hb\bar{b}}^{\pmb{S_1}}$ to 
the BR of the SM-like Higgs boson $h$ into the $b\bar{b}$ 
final state amount to less than about 2.5\% for
96\% of the valid parameter sets and to less than or roughly equal to
5\% for all of them. This is exemplified in
Fig.~\ref{fig:numerical}. In the plot, we present the binned
percentage of parameter points for $\Delta \text{BR}_{hb\bar{b}}$ of
the electroweak corrections for the scheme sets $\pmb{S_1}$ and
$\pmb{\overline{\text{MS}}}$. To that  
end, we computed the arithmetic average\footnote{We point out again that 
the numerical results computed for the schemes within one set are close 
to each other such that the arithmetic average of the results for the 
scheme set does not deviate much from the results computed for each 
scheme individually.} of the numerical results obtained for all 
renormalization schemes included in scheme set $\pmb{S_1}$ and
$\pmb{\overline{\text{MS}}}$, respectively. As 
can be inferred from the plot, for the scheme set $\pmb{S_1}$
96\% of all input parameter sets lead  
to corrections of 2.5\% or less and 100\% of all input parameter 
sets yield corrections less than 5.0\%, as stated above. For the 
$\overline{\text{MS}}$ scheme on the other hand, roughly 50\% of all 
parameter points yield relative corrections below 10.0\% while 12\% 
of all points lead to corrections above 100.0\%. 
The same methodology was applied for all 2HDM types and for all decays and translated into values in the tables. For example, the lower right entry of Tab.~\ref{tab:hlight2hdmbr} tells us that in the flipped
2HDM, the relative corrections $\Delta \mbox{BR}_{h\tau^+
  \tau^-}^{\pmb{\overline{\text{MS}}}}$ to 
the BR of $h$ into $\tau^+ \tau^-$ for 38\% of the
parameter points is as large as about 90\%, with 58\% leading to
corrections even beyond 100\%. The remaining 4\% of parameter points
lead to corrections between 90\% and 100\% and form the complement of
the two preceding pairs of numbers in the table. \s

\begin{table}[t!]
\begin{center}
\renewcommand{\arraystretch}{1.2}
\scriptsize
 \begin{tabular}{|c|c c c c c|}
  \hline
  Type & $\Delta \text{BR}_{hb\bar{b}}^{\pmb{S_1}}$ & $\Delta \text{BR}_{hb\bar{b}}^{\pmb{S_2}}$ & $\Delta \text{BR}_{hb\bar{b}}^{\pmb{S_3}}$ & $\Delta \text{BR}_{hb\bar{b}}^{\pmb{\text{OS2}}}$ & $\Delta \text{BR}_{hb\bar{b}}^{\pmb{\overline{\text{MS}}}}$ \\ \hline
  I  & $\lesssim 2.5\,\%$ $(96\,\%)$  & $\lesssim 5.0\,\%$ $(98\,\%)$ & $\lesssim 2.5\,\%$ $(90\,\%)$ & $\lesssim 2.5\,\%$ $(94\,\%)$ & $\lesssim 10.0\,\%$ $(50\,\%)$  \\  
     & $\lesssim 5.0\,\%$ $(100\,\%)$ & $\lesssim 7.5\,\%$ $(99\,\%)$ & $\lesssim 5.0\,\%$ $(99\,\%)$ & $\lesssim 5.0\,\%$ $(99\,\%)$ & $\gtrsim 100.0\,\%$ $(12\,\%)$ \\ \hdashline
  II & $\lesssim 2.5\,\%$ $(99\,\%)$  & $\lesssim 2.5\,\%$ $(54\,\%)$ & $\lesssim 2.5\,\%$ $(98\,\%)$ & $\lesssim 2.5\,\%$ $(81\,\%)$ & $\lesssim 40.0\,\%$ $(50\,\%)$ \\
     & $\lesssim 5.0\,\%$ $(100\,\%)$ & $\lesssim 7.5\,\%$ $(96\,\%)$
                                                                                                 & $\lesssim 5.0\,\%$ $(99\,\%)$ & $\lesssim 5.0\,\%$ $(99\,\%)$ & $\gtrsim 100.0\,\%$ $(36\,\%)$ \\ \hdashline
  LS & $\lesssim 2.5\,\%$ $(96\,\%)$ & $\lesssim 2.5\,\%$ $(54\,\%)$ & $\lesssim 2.5\,\%$ $(75\,\%)$ & $\lesssim 2.5\,\%$ $(94\,\%)$ & $\lesssim 17.5\,\%$ $(50\,\%)$ \\
     & $\lesssim 5.0\,\%$ $(99\,\%)$ & $\lesssim 5.0\,\%$ $(97\,\%)$ & $\lesssim 5.0\,\%$ $(99\,\%)$ & $\lesssim 5.0\,\%$ $(99\,\%)$ & $\gtrsim 100.0\,\%$ $(14\,\%)$ \\ \hdashline
  FL & $\lesssim 2.5\,\%$ $(96\,\%)$ & $\lesssim 2.5\,\%$ $(54\,\%)$ & $\lesssim 2.5\,\%$ $(75\,\%)$ & $\lesssim 2.5\,\%$ $(94\,\%)$ & $\lesssim 17.5\,\%$ $(50\,\%)$ \\  
     & $\lesssim 5.0\,\%$ $(99\,\%)$ & $\lesssim 5.0\,\%$ $(97\,\%)$ & $\lesssim 5.0\,\%$ $(99\,\%)$ & $\lesssim 5.0\,\%$ $(99\,\%)$ & $\gtrsim 100.0\,\%$ $(12\,\%)$ \\ \hline\hline
  Type & $\Delta \text{BR}_{h\gamma \gamma / hZZ}^{\pmb{S_1}}$ & $\Delta \text{BR}_{h\gamma \gamma / hZZ}^{\pmb{S_2}}$ & $\Delta \text{BR}_{h\gamma \gamma / hZZ}^{\pmb{S_3}}$ & $\Delta \text{BR}_{h\gamma \gamma / hZZ}^{\pmb{\text{OS2}}}$ & $\Delta \text{BR}_{h\gamma \gamma / hZZ}^{\pmb{\overline{\text{MS}}}}$ \\ \hline
  I  & $\lesssim 5.0\,\%$ $(97\,\%)$ & $\lesssim 5.0\,\%$ $(90\,\%)$  & $\lesssim 5.0\,\%$ $(90\,\%)$ & $\lesssim 5.0\,\%$ $(94\,\%)$ & $\lesssim 20.0\,\%$ $(50\,\%)$ \\
     & $\lesssim 7.5\,\%$ $(99\,\%)$ & $\lesssim 10.0\,\%$ $(98\,\%)$ & $\lesssim 7.5\,\%$ $(99\,\%)$ & $\lesssim 7.5\,\%$ $(99\,\%)$ & $\gtrsim 100.0\,\%$ $(21\,\%)$ \\ \hdashline
  II & $\lesssim 5.0\,\%$ $(99\,\%)$ & $\lesssim 5.0\,\%$ $(60\,\%)$  & $\lesssim 2.5\,\%$ $(96\,\%)$ & $\lesssim 5.0\,\%$ $(82\,\%)$ & $\lesssim 62.0\,\%$ $(50\,\%)$ \\
     & $\lesssim 7.5\,\%$ $(99\,\%)$ & $\lesssim 12.5\,\%$ $(96\,\%)$ & $\lesssim 5.0\,\%$ $(99\,\%)$ & $\lesssim 7.5\,\%$ $(97\,\%)$ & $\gtrsim 100.0\,\%$ $(47\,\%)$ \\ \hdashline
  LS & $\lesssim 5.0\,\%$ $(97\,\%)$ & $\lesssim 5.0\,\%$ $(75\,\%)$  & $\lesssim 2.5\,\%$ $(88\,\%)$ & $\lesssim 5.0\,\%$ $(95\,\%)$ & $\lesssim 12.5\,\%$ $(50\,\%)$ \\
     & $\lesssim 7.5\,\%$ $(99\,\%)$ & $\lesssim 10.0\,\%$ $(99\,\%)$ & $\lesssim 5.0\,\%$ $(99\,\%)$ & $\lesssim 7.5\,\%$ $(99\,\%)$ & $\gtrsim 100.0\,\%$ $(13\,\%)$ \\ \hdashline
  FL & $\lesssim 5.0\,\%$ $(97\,\%)$ & $\lesssim 5.0\,\%$ $(75\,\%)$ & $\lesssim 2.5\,\%$ $(88\,\%)$ & $\lesssim 5.0\,\%$ $(95\,\%)$ & $\lesssim 15.0\,\%$ $(50\,\%)$ \\
     & $\lesssim 7.5\,\%$ $(99\,\%)$ & $\lesssim 10.0\,\%$ $(99\,\%)$ & $\lesssim 5.0\,\%$ $(99\,\%)$ & $\lesssim 7.5\,\%$ $(99\,\%)$ & $\gtrsim 100.0\,\%$ $(11\,\%)$ \\ \hline\hline
  Type & $\Delta \text{BR}_{h\tau ^+ \tau ^-}^{\pmb{S_1}}$ & $\Delta \text{BR}_{h\tau ^+ \tau ^-}^{\pmb{S_2}}$ & $\Delta \text{BR}_{h\tau ^+ \tau ^-}^{\pmb{S_3}}$ & $\Delta \text{BR}_{h\tau ^+ \tau ^-}^{\pmb{\text{OS2}}}$ & $\Delta \text{BR}_{h\tau ^+ \tau ^-}^{\pmb{\overline{\text{MS}}}}$ \\ \hline
  I  & $\lesssim 2.5\,\%$ $(98\,\%)$ & $\lesssim 2.5\,\%$ $(88\,\%)$ & $\lesssim 2.5\,\%$ $(97\,\%)$ & $\lesssim 2.5\,\%$ $(98\,\%)$ & $\lesssim 7.5\,\%$ $(50\,\%)$ \\
     & $\lesssim 5.0\,\%$ $(99\,\%)$ & $\lesssim 5.0\,\%$ $(99\,\%)$ & $\lesssim 5.0\,\%$ $(99\,\%)$ & $\lesssim 5.0\,\%$ $(99\,\%)$ & $\gtrsim 100.0\,\%$ $(12\,\%)$ \\ \hdashline
  II & $\lesssim 5.0\,\%$ $(98\,\%)$ & $\lesssim 5.0\,\%$ $(85\,\%)$ & $\lesssim 5.0\,\%$ $(96\,\%)$ & $\lesssim 2.5\,\%$ $(57\,\%)$ & $\lesssim 35.0\,\%$ $(50\,\%)$ \\
     & $\lesssim 7.5\,\%$ $(99\,\%)$ & $\lesssim 10.0\,\%$ $(97\,\%)$ & $\lesssim 10.0\,\%$ $(99\,\%)$ & $\lesssim 5.0\,\%$ $(96\,\%)$ & $\gtrsim 100.0\,\%$ $(34\,\%)$ \\ \hdashline
  LS & $\lesssim 5.0\,\%$ $(94\,\%)$ & $\lesssim 7.5\,\%$ $(42\,\%)$ & $\lesssim 5.0\,\%$ $(95\,\%)$ & $\lesssim 5.0\,\%$ $(70\,\%)$ & $\lesssim 90.0\,\%$ $(37\,\%)$ \\
     & $\lesssim 7.5\,\%$ $(99\,\%)$ & $\lesssim 25.0\,\%$ $(96\,\%)$ & $\lesssim 7.5\,\%$ $(99\,\%)$ & $\lesssim 10.0\,\%$ $(98\,\%)$ & $\gtrsim 100.0\,\%$ $(58\,\%)$ \\ \hdashline
  FL & $\lesssim 5.0\,\%$ $(94\,\%)$ & $\lesssim 7.5\,\%$ $(42\,\%)$ & $\lesssim 5.0\,\%$ $(95\,\%)$ & $\lesssim 5.0\,\%$ $(70\,\%)$ & $\lesssim 90.0\,\%$ $(38\,\%)$ \\
     & $\lesssim 7.5\,\%$ $(99\,\%)$ & $\lesssim 25.0\,\%$ $(96\,\%)$ & $\lesssim 7.5\,\%$ $(99\,\%)$ & $\lesssim 10.0\,\%$ $(98\,\%)$ & $\gtrsim 100.0\,\%$ $(58\,\%)$ \\ \hline
 \end{tabular}
\caption{Relative size of the EW corrections to the BRs of the 2HDM
  SM-like Higgs boson $h$ in the four 2HDM types I, II, LS and FL. For details,
  we refer to the text. \label{tab:hlight2hdmbr}}
\end{center}
\vspace*{-0.5cm}
\end{table}
In Tab.~\ref{tab:hlight2hdmbr}, we show the results for the main discovery channels and
the dominant down-type fermion final states, {\it i.e.}~for the decays
into $\gamma\gamma$, $ZZ$, $b\bar{b}$ and $\tau^+ \tau^-$. We have
calculated also the corrections for all other final states and provide them on demand. In order to keep the presentation of the results clear, we stick to the most important
final states here, however. Note that the decay into $Z$ boson pairs is off-shell
for the 125~GeV SM-like 
Higgs boson, whereas the decay into photons is loop-induced already at
tree level. Therefore, as described above, these decays do not include
any EW corrections so that the relative corrections of their branching
ratios are the same. Inspecting the numbers of the table, we make the
following observations: 
\begin{itemize}
\item[--] The relative sizes $\Delta \text{BR}$ of the EW corrections
  to the BRs for the scheme set $\pmb{S_1}$ for all four
  2HDM types are typically very small, with the bulk of the parameter
  points resulting in corrections below $2.5\,\%$ and $5.0\,\%$,
  indicating both small EW corrections as well as numerical
  stability of the schemes. 
\item[--] For $\pmb{S_3}$, the corrections are slightly increased for
  the decay $h\rightarrow \tau ^+ \tau ^-$ in the 2HDM types II, LS
  and FL, but for the other decays in all four types, the results are
  comparable to the ones in set $\pmb{S_1}$. 
\item[--] For $\pmb{S_2}$, the corrections for the decays in all 2HDM
  types are increased compared to $\pmb{S_1}$ and $\pmb{S_3}$ and in
  particular in the 2HDM types LS and FL, the corrections to the decay
  $h\rightarrow \tau ^+ \tau ^-$ can be considerably larger. 
\item[--] The results for the OS2 scheme are similar to the ones for
  the scheme set $\pmb{S_1}$, but depending on the decay and 2HDM
  type, the majority of corrections are sometimes shifted towards
  lower/higher bins. Nevertheless, the corrections are moderate and
  the scheme is numerically stable. 
\item[--] The results for the $\overline{\mbox{MS}}$ scheme
  are typically very large for all decays and 2HDM
  types and hence clearly demonstrate numerical
  instability of the renormalization scheme. 
We find that this is the case throughout all decay channels and
mass hierarchies as presented in the following subsections, both for the 2HDM and
N2HDM and for all types of the two models. We do not remark this explicitly any more when
discussing the various decays and models but only present the
corresponding results in the tables.
\end{itemize}

Overall, with the exception of the $\overline{\mbox{MS}}$ scheme, the
corrections to the SM-like decays are of the typical size 
of EW corrections with values ranging between below 2.5\% up to about
25\% depending on the decay, the renormalization scheme and the type
of the 2HDM. 

%%%%%%%%%%%%%%%%%%%%%%%%%%%%%%%%%%%%%%%%%%%%%%%%%%%%%%%%%%%%%%
\subsubsection{$H$ is the SM-like Higgs Boson}
%%%%%%%%%%%%%%%%%%%%%%%%%%%%%%%%%%%%%%%%%%%%%%%%%%%%%%%%%%%%%%
\begin{table}[t]
\begin{center}
\renewcommand{\arraystretch}{1.2}
\scriptsize
 \begin{tabular}{|c|c c c c c|}
  \hline
  Type & $\Delta \text{BR}_{Hb\bar{b}}^{\pmb{S_1}}$ & $\Delta \text{BR}_{Hb\bar{b}}^{\pmb{S_2}}$ & $\Delta \text{BR}_{Hb\bar{b}}^{\pmb{S_3}}$ & $\Delta \text{BR}_{Hb\bar{b}}^{\pmb{S_4}}$ & $\Delta \text{BR}_{Hb\bar{b}}^{\pmb{\overline{\text{MS}}}}$ \\ \hline
  I  & $\lesssim 2.5\,\%$ $(95\,\%)$  & $\lesssim 2.5\,\%$ $(100\,\%)$ & $\lesssim 2.5\,\%$ $(87\,\%)$ & $\lesssim 2.5\,\%$ $(97\,\%)$ & $\lesssim 7.5\,\%$ $(50\,\%)$ \\  
     & $\lesssim 5.0\,\%$ $(100\,\%)$  & & $\lesssim 5.0\,\%$ $(98\,\%)$ & $\lesssim 5.0\,\%$ $(100\,\%)$ & $\gtrsim 100.0\,\%$ $(4\,\%)$ \\ \hdashline
  II & $\lesssim 2.5\,\%$ $(75\,\%)$  & $\lesssim 2.5\,\%$ $(100\,\%)$ & $\lesssim 15.0\,\%$ $(51\,\%)$ & $\lesssim 5.0\,\%$ $(74\,\%)$ & $\lesssim 75.0\,\%$ $(50\,\%)$ \\  
     & $\lesssim 5.0\,\%$ $(99\,\%)$  & & $\lesssim 30.0\,\%$ $(89\,\%)$ & $\lesssim 10.0\,\%$ $(99\,\%)$ & $\gtrsim 100.0\,\%$ $(13\,\%)$ \\ \hdashline
  LS & $\lesssim 2.5\,\%$ $(94\,\%)$  & $\lesssim 2.5\,\%$ $(62\,\%)$ & $\lesssim 7.5\,\%$ $(57\,\%)$ & $\lesssim 2.5\,\%$ $(96\,\%)$ & $\lesssim 90.0\,\%$ $(34\,\%)$ \\  
     & $\lesssim 5.0\,\%$ $(100\,\%)$  & $\lesssim 5.0\,\%$ $(100\,\%)$ & $\lesssim 20.0\,\%$ $(93\,\%)$ & $\lesssim 5.0\,\%$ $(100\,\%)$ & $\gtrsim 100.0\,\%$ $(63\,\%)$ \\ \hdashline
  FL & $\lesssim 2.5\,\%$ $(62\,\%)$  & $\lesssim 40.0\,\%$ $(50\,\%)$ & $\lesssim 10.0\,\%$ $(50\,\%)$ & $\lesssim 5.0\,\%$ $(50\,\%)$ & $\lesssim 90.0\,\%$ $(17\,\%)$ \\  
     & $\lesssim 7.5\,\%$ $(98\,\%)$  & $\gtrsim 100.0\,\%$ $(23\,\%)$ & $\lesssim 35.0\,\%$ $(91\,\%)$ & $\lesssim 15.0\,\%$ $(97\,\%)$ & $\gtrsim 100.0\,\%$ $(58\,\%)$ \\ \hline\hline
  Type & $\Delta \text{BR}_{H\gamma \gamma / HZZ}^{\pmb{S_1}}$ & $\Delta \text{BR}_{H\gamma \gamma / HZZ}^{\pmb{S_2}}$ & $\Delta \text{BR}_{H\gamma \gamma / HZZ}^{\pmb{S_3}}$ & $\Delta \text{BR}_{H\gamma \gamma / HZZ}^{\pmb{S_4}}$ & $\Delta \text{BR}_{H\gamma \gamma / HZZ}^{\pmb{\overline{\text{MS}}}}$ \\ \hline
  I  & $\lesssim 5.0\,\%$ $(95\,\%)$  & $\lesssim 2.5\,\%$ $(100\,\%)$ & $\lesssim 5.0\,\%$ $(86\,\%)$ & $\lesssim 5.0\,\%$ $(97\,\%)$ & $\lesssim 17.5\,\%$ $(50\,\%)$ \\  
     & $\lesssim 100.0\,\%$ $(100\,\%)$  & & $\lesssim 10.0\,\%$ $(98\,\%)$ & $\lesssim 7.5\,\%$ $(100\,\%)$ & $\gtrsim 100.0\,\%$ $(6\,\%)$ \\ \hdashline
  II & $\lesssim 5.0\,\%$ $(82\,\%)$  & $\lesssim 2.5\,\%$ $(100\,\%)$ & $\lesssim 22.5\,\%$ $(55\,\%)$ & $\lesssim 7.5\,\%$ $(56\,\%)$ & $\lesssim 90.0\,\%$ $(3\,\%)$ \\  
     & $\lesssim 7.5\,\%$ $(99\,\%)$  & & $\lesssim 50.0\,\%$ $(92\,\%)$ & $\lesssim 15.0\,\%$ $(99\,\%)$ & $\gtrsim 100.0\,\%$ $(96\,\%)$ \\ \hdashline
  LS & $\lesssim 5.0\,\%$ $(77\,\%)$  & $\lesssim 5.0\,\%$ $(80\,\%)$ & $\lesssim 5.0\,\%$ $(48\,\%)$ & $\lesssim 5.0\,\%$ $(66\,\%)$ & $\lesssim 90.0\,\%$ $(48\,\%)$ \\  
     & $\lesssim 7.5\,\%$ $(100\,\%)$  & $\lesssim 7.5\,\%$ $(100\,\%)$ & $\lesssim 20.0\,\%$ $(94\,\%)$ & $\lesssim 7.5\,\%$ $(99\,\%)$ & $\gtrsim 100.0\,\%$ $(51\,\%)$ \\ \hdashline
  FL & $\lesssim 5.0\,\%$ $(79\,\%)$  & $\lesssim 50.0\,\%$ $(50\,\%)$ & $\lesssim 10.0\,\%$ $(44\,\%)$ & $\lesssim 7.5\,\%$ $(52\,\%)$ & $\lesssim 90.0\,\%$ $(1\,\%)$ \\  
     & $\lesssim 7.5\,\%$ $(98\,\%)$  & $\gtrsim 100.0\,\%$ $(26\,\%)$ & $\lesssim 40.0\,\%$ $(90\,\%)$ & $\lesssim 15.0\,\%$ $(98\,\%)$ & $\gtrsim 100.0\,\%$ $(98\,\%)$ \\ \hline\hline
  Type & $\Delta \text{BR}_{H\tau ^+ \tau ^-}^{\pmb{S_1}}$ & $\Delta \text{BR}_{H\tau ^+ \tau ^-}^{\pmb{S_2}}$ & $\Delta \text{BR}_{H\tau ^+ \tau ^-}^{\pmb{S_3}}$ & $\Delta \text{BR}_{H\tau ^+ \tau ^-}^{\pmb{S_4}}$ & $\Delta \text{BR}_{H\tau ^+ \tau ^-}^{\pmb{\overline{\text{MS}}}}$ \\ \hline
  I  & $\lesssim 2.5\,\%$ $(99\,\%)$  & $\lesssim 2.5\,\%$ $(100\,\%)$ & $\lesssim 2.5\,\%$ $(89\,\%)$ & $\lesssim 2.5\,\%$ $(99\,\%)$ & $\lesssim 7.5\,\%$ $(50\,\%)$ \\  
     & $\lesssim 5.0\,\%$ $(100\,\%)$  & & $\lesssim 5.0\,\%$ $(99\,\%)$ & $\lesssim 5.0\,\%$ $(100\,\%)$ & $\gtrsim 100.0\,\%$ $(4\,\%)$ \\ \hdashline
  II & $\lesssim 2.5\,\%$ $(64\,\%)$  & $\lesssim 2.5\,\%$ $(100\,\%)$ & $\lesssim 10.0\,\%$ $(32\,\%)$ & $\lesssim 7.5\,\%$ $(59\,\%)$ & $\lesssim 77.5\,\%$ $(50\,\%)$ \\  
     & $\lesssim 7.5\,\%$ $(99\,\%)$  & & $\lesssim 30.0\,\%$ $(90\,\%)$ & $\lesssim 12.5\,\%$ $(99\,\%)$ & $\gtrsim 100.0\,\%$ $(13\,\%)$ \\ \hdashline
  LS & $\lesssim 5.0\,\%$ $(72\,\%)$  & $\lesssim 2.5\,\%$ $(88\,\%)$ & $\lesssim 62.5\,\%$ $(50\,\%)$ & $\lesssim 10.0\,\%$ $(43\,\%)$ & $\lesssim 90.0\,\%$ $(1\,\%)$ \\  
     & $\lesssim 15.0\,\%$ $(98\,\%)$  & $\lesssim 5.0\,\%$ $(100\,\%)$ & $\gtrsim 100.0\,\%$ $(36\,\%)$ & $\lesssim 25.0\,\%$ $(98\,\%)$ & $\gtrsim 100.0\,\%$ $(98\,\%)$ \\ \hdashline
  FL & $\lesssim 5.0\,\%$ $(99\,\%)$  & $\lesssim 45.0\,\%$ $(50\,\%)$ & $\lesssim 5.0\,\%$ $(48\,\%)$ & $\lesssim 5.0\,\%$ $(95\,\%)$ & $\lesssim 90.0\,\%$ $(1\,\%)$ \\  & $\lesssim 7.5\,\%$ $(100\,\%)$  & $\gtrsim 100.0\,\%$
   $(25\,\%)$ & $\lesssim 50.0\,\%$ $(92\,\%)$ &
$\lesssim 10.0\,\%$ $(98\,\%)$ & $\gtrsim 100.0\,\%$ $(98\,\%)$ \\ \hline 
\end{tabular}
\caption{Relative size of the EW corrections to the BRs of the 2HDM
  SM-like $H$ in the four 2HDM types I, II, LS and FL. For details,
  we refer to the text. \label{tab:hheavy2hdmbr}}
\end{center}
\vspace*{-0.5cm}
\end{table}
In Tab.~\ref{tab:hheavy2hdmbr} we show the relative corrections $\Delta \text{BR}$ for the
2HDM parameter sets where the heavier of the two CP-even Higgs
bosons, $H$, is the SM-like Higgs boson. We want to mention again that in contrast to the case where $h$ is SM-like, we found much less valid parameter points in the case that $H$ is SM-like and hence, the statistics of the analysis is rather low. For a clear presentation of the results, we grouped renormalization schemes to the following sets: \s
\begin{align}
\pmb{S_1} &\equiv \big\{ ~ \text{KOSY}^o ~,~ \text{KOSY}^c ~,~
            \text{p}_*^o ~,~ \text{p}_*^c ~,~ \text{pOS}^o ~,~
            \text{pOS}^c ~,~ \text{BFMS} ~ \big\} \nonumber \\
\pmb{S_2} &\equiv \big\{ ~ \text{proc1} ~,~ \text{proc3} ~ \big\} \nonumber\\
\pmb{S_3} &\equiv \big\{ ~ \text{proc2} ~,~ \text{OS1} ~ \big\} \nonumber\\
\pmb{S_4}  &\equiv \big\{ ~ \text{OS2} ~,~ \text{OS12} ~ \big\} \nonumber\\
\pmb{\overline{\text{MS}}} &\equiv \big\{ ~ \overline{\text{MS}} ~ \big\}\nonumber
\end{align}

As before, we show results for the decays into $\gamma\gamma$, $ZZ$,
$b\bar{b}$ and $\tau^+ \tau^-$. The data in the table allows us to make the following observations:
\begin{itemize}
\item[--] The scheme set $\pmb{S_1}$, comprising the KOSY, the pinched and
  the BFMS schemes, induces the smallest EW corrections for all decays and 2HDM types.
\item[--] The process-dependent schemes, collected in the scheme sets
  $\pmb{S_2}$ to $\pmb{S_4}$, induce moderate corrections but for some decays and 2HDM types, they can also lead to very large corrections. 
\end{itemize}

In principle, also decays $H\,\rightarrow h\,h$ are possible, with the
heavier Higgs boson $H$ being 
SM-like. Since we did not find enough valid parameter points to reach
significant statistics, we do not show results here for this final
state, however. In our scan, did not find any parameter points that
allow for the OS decay $H\rightarrow ZA$. 

%%%%%%%%%%%%%%%%%%%%%%%%%%%%%%%%%%%%%%%%%%%%%%%%%%%%%%%%%%%%%%
\subsubsection{Comparison with the SM}
%%%%%%%%%%%%%%%%%%%%%%%%%%%%%%%%%%%%%%%%%%%%%%%%%%%%%%%%%%%%%%
We compare our results for the 2HDM with the size of the EW
corrections to the decays of the SM Higgs boson $H_{\text{SM}}$ with mass
$m_{H_{\text{SM}}}=125.09$~GeV \cite{Aad:2015zhl}. The decays with and without the
EW corrections on top of the QCD corrections (where applicable) have
been calculated with {\tt HDECAY} version 6.52
\cite{DJOUADI199856,Djouadi:2018xqq}. By setting the flag 'OMIT ELW'
in the input file equal to 0 (1), the EW corrections are computed (omitted). They
have been implemented for the decay into gluons
\cite{Actis:2008ug,Actis:2008ts,Actis:2008uh}, into the 
fermion final states $b\bar{b}$, $\tau^+ \tau^-$, $\mu^+ \mu^-$,
$s\bar{s}$, and $c\bar{c}$ \cite{Spira:2016ztx,Spira:1997dg}, into
$\gamma\gamma$ \cite{Actis:2008ts,Djouadi:1997rj,Degrassi:2005mc,Passarino:2007fp} and into the massive gauge boson final states $W^+W^-$ and $ZZ$ \cite{Bredenstein:2006rh,Bredenstein:2006ha,Boselli:2015aha}. 
Hence, in the SM also the EW corrections to the loop-induced decays into
$gg$ and $\gamma\gamma$ and into the off-shell final states of massive
gauge bosons are included. For the decay
width into $Z\gamma$ no EW corrections are implemented. Still the
BR changes because of the EW-corrected total decay
width entering the BR. \s
\begin{table}[t!]
\begin{center}
\vspace*{0.2cm}
\renewcommand{\arraystretch}{1.2}
\scriptsize
\begin{tabular}{|c | c c c c c c c c c c|}
\hline
%$\Delta \text{BR}_{Hb\bar{b}}$ & $\Delta \text{BR}_{H\tau^+\tau^-}$
%& $\Delta \text{BR}_{H\mu^+\mu^-}$ & $\Delta \text{BR}_{Hs\bar{s}}$
%& $\Delta \text{BR}_{Hc\bar{c}}$ & $\Delta \text{BR}_{Hgg}$
%& $\Delta \text{BR}_{H\gamma\gamma}$ & $\Delta \text{BR}_{HZ\gamma}$ &
%$\Delta \text{BR}_{HWW}$ & $\Delta \text{BR}_{HZZ}$
$\Delta \text{BR}$ & $b\bar{b}$ & $\tau^+\tau^-$ & $\mu^+\mu^-$ &
$s\bar{s}$ & $c\bar{c}$ & $gg$ & $\gamma\gamma$ & $Z\gamma$ & $W^+W^-$ & $ZZ$
\\ 
   \hline
& -1.76\% & -1.59\% & -3.52\% & 2.24\% & -3.81\% & 4.34\% & -2.29\% &
-0.71\% &  3.68\% & 1.61\% 
\\ \hline
\end{tabular}
\caption{Relative size of the EW corrections to the BRs of the SM Higgs
boson $H_{\text{SM}}$ with mass $m_{H_{\text{SM}}}=125.09$~GeV. \label{tab:smdeltabrs}}
\vspace*{-0.4cm}
\end{center}
\end{table}

The results for the relative corrections $\Delta \text{BR}$ into the various final states are
given in Tab.~\ref{tab:smdeltabrs}. As can be inferred from the table
the relative size of the EW corrections is always small, ranging below
5\%, with the maximum value given by 4.34\% for the decay into a pair of
gluons. The comparison of these results with the relative EW corrections to the 2HDM
SM-like Higgs BRs (computed with the renormalization
scheme set $\pmb{S_1}$, which delivers the most moderate corrections)
shows that in the 2HDM the EW corrections are in general somewhat more important.

%%%%%%%%%%%%%%%%%%%%%%%%%%%%%%%%%%%%%%%%%%%%%%%%%%%%%%%%%%%%%%
\subsection{Non-SM-Like CP-even 2HDM Higgs Decays}
\label{sec:nonSMlikeHiggsDecays}
%%%%%%%%%%%%%%%%%%%%%%%%%%%%%%%%%%%%%%%%%%%%%%%%%%%%%%%%%%%%%%
Next, we consider the decays of the non-SM-like CP-even Higgs bosons,
again for the two possible Higgs hierarchies, with the total amount of
points that have been used for the analysis as given in Sec.~\ref{sec:2HDMHiggsDecays}.

%%%%%%%%%%%%%%%%%%%%%%%%%%%%%%%%%%%%%%%%%%%%%%%%%%%%%%%%%%%%%%
\subsubsection{$h$ is the SM-like Higgs boson}
%%%%%%%%%%%%%%%%%%%%%%%%%%%%%%%%%%%%%%%%%%%%%%%%%%%%%%%%%%%%%%
With $h$ being SM-like, the non-SM-like CP-even Higgs boson corresponds to the heavier Higgs boson $H$. Its important decay channels are
those into $t\bar{t}$, but also decays into mixed Higgs plus gauge boson final
states, {\it i.e.}~$ZA$ and $W^\pm H^\mp$, can become important if they are kinematically allowed. In contrast to the SM-like decays presented above, 
for the relative corrections $\Delta \text{BR}$ of individual final
states we now consider only the parameter
points for which the decays are OS and not loop-induced. This
reduces the number of points available for the analysis as shown in
Tab.~\ref{tab:points2hdmlightersmlike} where we list the numbers for the
individual final states that are left over with the additional
requirement of the decays being OS. The relative corrections $\Delta
\text{BR}$ for the decays into $b\bar{b}$, $t\bar{t}$, $\tau^+\tau^-$,
$ZA$ and $W^\pm H^\mp$ are given in
Tab.~\ref{tab:heaviernonsmlikedecay1}, and those for the decays into
$ZZ$ and into two light Higgs bosons $hh$ in
Tab.~\ref{tab:heaviernonsmlikedecay2}. For simplicity, we do not list
the relative corrections to the $WW$ final state separately here, as
they behave similarly to those of the $ZZ$ final state. \s
\begin{table}[t!]
\begin{center}
\renewcommand{\arraystretch}{1.2}
\scriptsize
 \begin{tabular}{|c|c c c c c|}
  \hline
Type & $H \to t\bar{t}$ & $H \to ZA$ & $H \to W^\pm H^\mp$ & $H\to ZZ$
   & $H\to hh$\\
   \hline
I & $166\,444$ & $10\,406$ & 7600 & $186\,474$ & $179\,777$ \\
II & $239\,747$ & $19\,464$ & $12\,569$ & $239\,759$ & $239\,755$ \\
LS & $365\,299$ & $14\,506$ & $10\,898$ & $369\,448$ & $168\,129$ \\
FL & $419\,079$ & $12\,388$ & 8472 & $419\,149$ & $289\,146$ \\
\hline
\end{tabular}
\caption{Number of parameter points available for the analysis in the
  individual OS decay channels of $H$ ($h$ is SM-like).}
\label{tab:points2hdmlightersmlike}
\vspace*{-0.4cm}
\end{center}
\end{table}

%The relative corrections of the BRs of the heavier Higgs bosons $H$ with the lighter one, $h$, being SM-like are given in Tab.s~\ref{tab:heaviernonsmlikedecay1} and \ref{tab:heaviernonsmlikedecay2}. 
The scheme sets with comparable sizes in the EW corrections, shown in the table, are defined as follows: 
\begin{table}
\begin{center}
\renewcommand{\arraystretch}{1.2}
\scriptsize
 \begin{tabular}{|c|c c c c c|}
  \hline
  Type & $\Delta \text{BR}_{Hb\bar{b}}^{\pmb{S_1}}$ & $\Delta \text{BR}_{Hb\bar{b}}^{\pmb{S_2}}$ & $\Delta \text{BR}_{Hb\bar{b}}^{\pmb{S_3}}$ & $\Delta \text{BR}_{Hb\bar{b}}^{\pmb{S_4}}$ & $\Delta \text{BR}_{Hb\bar{b}}^{\pmb{\overline{\text{MS}}}}$ \\ \hline
  I  & $\lesssim 15.0\,\%$ $( 48\,\%)$  & $\lesssim 12.5\,\%$ $( 52\,\%)$ & $\lesssim 12.5\,\%$ $( 48\,\%)$ & $\lesssim 10.0\,\%$ $( 45\,\%)$ & $\lesssim 60.0\,\%$ $( 50\,\%)$  \\  
     & $\lesssim 27.5\,\%$ $( 93\,\%)$  & $\lesssim 35.0\,\%$ $( 89\,\%)$ & $\lesssim 40.0\,\%$ $( 80\,\%)$ & $\lesssim 35.0\,\%$ $( 89\,\%)$ & $\gtrsim 100.0\,\%$ $( 40\,\%)$ \\ \hdashline
  II & $\lesssim 10.0\,\%$ $( 52\,\%)$  & $\lesssim 20.0\,\%$ $( 48\,\%)$ & $\lesssim 10.0\,\%$ $( 52\,\%)$ & $\lesssim 25.0\,\%$ $( 44\,\%)$ & $\lesssim 90.0\,\%$ $( 14\,\%)$ \\
     & $\lesssim 25.0\,\%$ $( 92\,\%)$  & $\lesssim 32.5\,\%$ $( 93\,\%)$ & $\lesssim 35.0\,\%$ $( 91\,\%)$ & $\lesssim 42.5\,\%$ $( 88\,\%)$ & $\gtrsim 100.0\,\%$ $( 85\,\%)$ \\ \hdashline
  LS & $\lesssim 10.0\,\%$ $( 52\,\%)$  & $\lesssim 10.0\,\%$ $( 52\,\%)$ & $\lesssim 10.0\,\%$ $( 46\,\%)$ & $\lesssim 7.5\,\%$ $( 42\,\%)$ & $\lesssim 45.0\,\%$ $( 50\,\%)$ \\
     & $\lesssim 25.0\,\%$ $( 92\,\%)$  & $\lesssim 25.0\,\%$ $( 92\,\%)$ & $\lesssim 30.0\,\%$ $( 90\,\%)$ & $\lesssim 22.5\,\%$ $( 90\,\%)$ & $\gtrsim 100.0\,\%$ $( 36\,\%)$ \\ \hdashline
  FL & $\lesssim 12.5\,\%$ $( 52\,\%)$  & $\lesssim 10.0\,\%$ $( 52\,\%)$ & $\lesssim 10.0\,\%$ $( 46\,\%)$ & $\lesssim 7.5\,\%$ $( 42\,\%)$ & $\lesssim 45.0\,\%$ $( 50\,\%)$ \\  
     & $\lesssim 32.5\,\%$ $( 88\,\%)$  & $\lesssim 25.0\,\%$ $( 92\,\%)$ & $\lesssim 30.0\,\%$ $( 90\,\%)$ & $\lesssim 22.5\,\%$ $( 90\,\%)$ & $\gtrsim 100.0\,\%$ $( 36\,\%)$ \\ \hline\hline
  Type & $\Delta \text{BR}_{Ht\bar{t}}^{\pmb{S_1}}$ & $\Delta \text{BR}_{Ht\bar{t}}^{\pmb{S_2}}$ & $\Delta \text{BR}_{Ht\bar{t}}^{\pmb{S_3}}$ & $\Delta \text{BR}_{Ht\bar{t}}^{\pmb{S_4}}$ & $\Delta \text{BR}_{Ht\bar{t}}^{\pmb{\overline{\text{MS}}}}$ \\ \hline
  I  & $\lesssim 5.0\,\%$ $( 48\,\%)$  & $\lesssim 7.5\,\%$ $( 58\,\%)$ & $\lesssim 7.5\,\%$ $( 54\,\%)$ & $\lesssim 5.0\,\%$ $( 50\,\%)$ & $\lesssim 52.5\,\%$ $( 50\,\%)$  \\  
     & $\lesssim 22.5\,\%$ $( 85\,\%)$  & $\lesssim 22.5\,\%$ $( 85\,\%)$ & $\lesssim 25.0\,\%$ $( 80\,\%)$ & $\lesssim 25.0\,\%$ $( 88\,\%)$ & $\gtrsim 100.0\,\%$ $( 36\,\%)$ \\ \hdashline
  II & $\lesssim 2.5\,\%$ $( 60\,\%)$  & $\lesssim 2.5\,\%$ $( 60\,\%)$ & $\lesssim 2.5\,\%$ $( 55\,\%)$ & $\lesssim 2.5\,\%$ $( 61\,\%)$ & $\lesssim 37.5\,\%$ $( 50\,\%)$ \\
     & $\lesssim 10.0\,\%$ $( 86\,\%)$  & $\lesssim 10.0\,\%$ $( 86\,\%)$ & $\lesssim 12.5\,\%$ $( 87\,\%)$ & $\lesssim 10.0\,\%$ $( 87\,\%)$ & $\gtrsim 100.0\,\%$ $( 33\,\%)$ \\ \hdashline
  LS & $\lesssim 5.0\,\%$ $( 61\,\%)$  & $\lesssim 2.5\,\%$ $( 46\,\%)$ & $\lesssim 5.0\,\%$ $( 54\,\%)$ & $\lesssim 2.5\,\%$ $( 47\,\%)$ & $\lesssim 45.0\,\%$ $( 50\,\%)$ \\
     & $\lesssim 15.0\,\%$ $( 88\,\%)$  & $\lesssim 15.0\,\%$ $( 87\,\%)$ & $\lesssim 20.0\,\%$ $( 88\,\%)$ & $\lesssim 15.0\,\%$ $( 88\,\%)$ & $\gtrsim 100.0\,\%$ $( 35\,\%)$ \\ \hdashline
  FL & $\lesssim 5.0\,\%$ $( 68\,\%)$  & $\lesssim 5.0\,\%$ $( 68\,\%)$ & $\lesssim 5.0\,\%$ $( 56\,\%)$ & $\lesssim 2.5\,\%$ $( 54\,\%)$ & $\lesssim 50.0\,\%$ $( 50\,\%)$ \\  
     & $\lesssim 12.5\,\%$ $( 87\,\%)$  & $\lesssim 12.5\,\%$ $( 87\,\%)$ & $\lesssim 20.0\,\%$ $( 87\,\%)$ & $\lesssim 12.5\,\%$ $( 87\,\%)$ & $\gtrsim 100.0\,\%$ $( 39\,\%)$ \\ \hline\hline
  Type & $\Delta \text{BR}_{H\tau ^+\tau ^-}^{\pmb{S_1}}$ & $\Delta \text{BR}_{H\tau ^+\tau ^-}^{\pmb{S_2}}$ & $\Delta \text{BR}_{H\tau ^+\tau ^-}^{\pmb{S_3}}$ & $\Delta \text{BR}_{H\tau ^+\tau ^-}^{\pmb{S_4}}$ & $\Delta \text{BR}_{H\tau ^+\tau ^-}^{\pmb{\overline{\text{MS}}}}$ \\ \hline
  I  & $\lesssim 15.0\,\%$ $( 49\,\%)$  & $\lesssim 15.0\,\%$ $( 51\,\%)$ & $\lesssim 15.0\,\%$ $( 48\,\%)$ & $\lesssim 15.0\,\%$ $( 55\,\%)$ & $\lesssim 60.0\,\%$ $( 50\,\%)$  \\  
     & $\lesssim 35.0\,\%$ $( 88\,\%)$  & $\lesssim 35.0\,\%$ $( 88\,\%)$ & $\lesssim 35.0\,\%$ $( 77\,\%)$ & $\lesssim 35.0\,\%$ $( 88\,\%)$ & $\gtrsim 100.0\,\%$ $( 40\,\%)$ \\ \hdashline
  II & $\lesssim 15.0\,\%$ $( 54\,\%)$  & $\lesssim 20.0\,\%$ $( 53\,\%)$ & $\lesssim 10.0\,\%$ $( 51\,\%)$ & $\lesssim 25.0\,\%$ $( 47\,\%)$ & $\lesssim 85.0\,\%$ $( 14\,\%)$ \\
     & $\lesssim 25.0\,\%$ $( 91\,\%)$  & $\lesssim 30.0\,\%$ $( 90\,\%)$ & $\lesssim 35.0\,\%$ $( 90\,\%)$ & $\lesssim 40.0\,\%$ $( 86\,\%)$ & $\gtrsim 100.0\,\%$ $( 84\,\%)$ \\ \hdashline
  LS & $\lesssim 15.0\,\%$ $( 54\,\%)$  & $\lesssim 17.5\,\%$ $( 48\,\%)$ & $\lesssim 7.5\,\%$ $( 46\,\%)$ & $\lesssim 25.0\,\%$ $( 46\,\%)$ & $\lesssim 77.5\,\%$ $( 15\,\%)$ \\
     & $\lesssim 27.5\,\%$ $( 90\,\%)$  & $\lesssim 30.0\,\%$ $( 88\,\%)$ & $\lesssim 30.0\,\%$ $( 88\,\%)$ & $\lesssim 40.0\,\%$ $( 85\,\%)$ & $\gtrsim 100.0\,\%$ $( 81\,\%)$ \\ \hdashline
  FL & $\lesssim 15.0\,\%$ $( 55\,\%)$  & $\lesssim 17.5\,\%$ $( 48\,\%)$ & $\lesssim 7.5\,\%$ $( 46\,\%)$ & $\lesssim 25.0\,\%$ $( 46\,\%)$ & $\lesssim 77.5\,\%$ $( 15\,\%)$ \\  
     & $\lesssim 27.5\,\%$ $( 90\,\%)$  & $\lesssim 30.0\,\%$ $( 88\,\%)$ & $\lesssim 30.0\,\%$ $( 88\,\%)$ & $\lesssim 40.0\,\%$ $( 85\,\%)$ & $\gtrsim 100.0\,\%$ $( 81\,\%)$ \\ \hline\hline
  Type & $\Delta \text{BR}_{HZA}^{\pmb{S_1}}$ & $\Delta \text{BR}_{HZA}^{\pmb{S_2}}$ & $\Delta \text{BR}_{HZA}^{\pmb{S_3}}$ & $\Delta \text{BR}_{HZA}^{\pmb{S_4}}$ & $\Delta \text{BR}_{HZA}^{\pmb{\overline{\text{MS}}}}$ \\ \hline
  I  & $\lesssim 5.0\,\%$ $( 51\,\%)$  & $\lesssim 5.0\,\%$ $( 51\,\%)$ & $\lesssim 10.0\,\%$ $( 46\,\%)$ & $\lesssim 10.0\,\%$ $( 53\,\%)$ & $\lesssim 80.0\,\%$ $( 26\,\%)$  \\  
     & $\lesssim 15.0\,\%$ $( 80\,\%)$  & $\lesssim 15.0\,\%$ $( 80\,\%)$ & $\lesssim 30.0\,\%$ $( 80\,\%)$ & $\lesssim 22.5\,\%$ $( 83\,\%)$ & $\gtrsim 100.0\,\%$ $( 52\,\%)$ \\ \hdashline
  II & $\lesssim 5.0\,\%$ $( 68\,\%)$  & $\lesssim 5.0\,\%$ $( 69\,\%)$ & $\lesssim 10.0\,\%$ $( 50\,\%)$ & $\lesssim 7.5\,\%$ $( 73\,\%)$ & $\lesssim 85.0\,\%$ $( 20\,\%)$ \\
     & $\lesssim 10.0\,\%$ $( 91\,\%)$  & $\lesssim 12.5\,\%$ $( 94\,\%)$ & $\lesssim 25.0\,\%$ $( 81\,\%)$ & $\lesssim 10.0\,\%$ $( 90\,\%)$ & $\gtrsim 100.0\,\%$ $( 56\,\%)$ \\ \hdashline
  LS & $\lesssim 5.0\,\%$ $( 65\,\%)$  & $\lesssim 5.0\,\%$ $( 65\,\%)$ & $\lesssim 10.0\,\%$ $( 48\,\%)$ & $\lesssim 7.5\,\%$ $( 41\,\%)$ & $\lesssim 85.0\,\%$ $( 29\,\%)$ \\
     & $\lesssim 10.0\,\%$ $( 86\,\%)$  & $\lesssim 10.0\,\%$ $( 86\,\%)$ & $\lesssim 27.5\,\%$ $( 80\,\%)$ & $\lesssim 15.0\,\%$ $( 90\,\%)$ & $\gtrsim 100.0\,\%$ $( 44\,\%)$ \\ \hdashline
  FL & $\lesssim 5.0\,\%$ $( 65\,\%)$  & $\lesssim 5.0\,\%$ $( 63\,\%)$ & $\lesssim 10.0\,\%$ $( 53\,\%)$ & $\lesssim 7.5\,\%$ $( 51\,\%)$ & $\lesssim 82.5\,\%$ $( 20\,\%)$ \\  
     & $\lesssim 10.0\,\%$ $( 88\,\%)$  & $\lesssim 10.0\,\%$ $( 88\,\%)$ & $\lesssim 15.0\,\%$ $( 83\,\%)$ & $\lesssim 10.0\,\%$ $( 84\,\%)$ & $\gtrsim 100.0\,\%$ $( 30\,\%)$ \\ \hline\hline
  Type & $\Delta \text{BR}_{HW^\pm H^\mp}^{\pmb{S_1}}$ & $\Delta \text{BR}_{HW^\pm H^\mp}^{\pmb{S_2}}$ & $\Delta \text{BR}_{HW^\pm H^\mp}^{\pmb{S_3}}$ & $\Delta \text{BR}_{HW^\pm H^\mp}^{\pmb{S_4}}$ & $\Delta \text{BR}_{HW^\pm H^\mp}^{\pmb{\overline{\text{MS}}}}$ \\ \hline
  I  & $\lesssim 5.0\,\%$ $( 56\,\%)$  & $\lesssim 5.0\,\%$ $( 55\,\%)$ & $\lesssim 10.0\,\%$ $( 49\,\%)$ & $\lesssim 10.0\,\%$ $( 57\,\%)$ & $\lesssim 70.0\,\%$ $( 25\,\%)$  \\  
     & $\lesssim 17.5\,\%$ $( 81\,\%)$  & $\lesssim 17.5\,\%$ $( 81\,\%)$ & $\lesssim 30.0\,\%$ $( 78\,\%)$ & $\lesssim 25.0\,\%$ $( 82\,\%)$ & $\gtrsim 100.0\,\%$ $( 52\,\%)$ \\ \hdashline
  II & $\lesssim 5.0\,\%$ $( 60\,\%)$  & $\lesssim 5.0\,\%$ $( 59\,\%)$ & $\lesssim 12.5\,\%$ $( 49\,\%)$ & $\lesssim 5.0\,\%$ $( 55\,\%)$ & $\lesssim 82.5\,\%$ $( 18\,\%)$ \\
     & $\lesssim 10.0\,\%$ $( 87\,\%)$  & $\lesssim 10.0\,\%$ $( 85\,\%)$ & $\lesssim 30.0\,\%$ $( 81\,\%)$ & $\lesssim 10.0\,\%$ $( 94\,\%)$ & $\gtrsim 100.0\,\%$ $( 50\,\%)$ \\ \hdashline
  LS & $\lesssim 5.0\,\%$ $( 71\,\%)$  & $\lesssim 5.0\,\%$ $( 70\,\%)$ & $\lesssim 12.5\,\%$ $( 52\,\%)$ & $\lesssim 7.5\,\%$ $( 57\,\%)$ & $\lesssim 75.0\,\%$ $( 26\,\%)$ \\
     & $\lesssim 7.5\,\%$ $( 84\,\%)$  & $\lesssim 7.5\,\%$ $( 84\,\%)$ & $\lesssim 27.5\,\%$ $( 81\,\%)$ & $\lesssim 12.5\,\%$ $( 85\,\%)$ & $\gtrsim 100.0\,\%$ $( 45\,\%)$ \\ \hdashline
  FL & $\lesssim 5.0\,\%$ $( 67\,\%)$  & $\lesssim 5.0\,\%$ $( 62\,\%)$ & $\lesssim 7.5\,\%$ $( 48\,\%)$ & $\lesssim 5.0\,\%$ $( 53\,\%)$ & $\lesssim 82.5\,\%$ $( 19\,\%)$ \\  
     & $\lesssim 7.5\,\%$ $( 85\,\%)$  & $\lesssim 7.5\,\%$ $( 84\,\%)$ & $\lesssim 15.0\,\%$ $( 87\,\%)$ & $\lesssim 10.0\,\%$ $( 95\,\%)$ & $\gtrsim 100.0\,\%$ $( 36\,\%)$ \\ \hline
 \end{tabular}
\caption{Relative size of the EW corrections to the BRs for the non-SM-like 2HDM Higgs boson $H$ decays into $b\bar{b}$, $t\bar{t}$, $\tau^+\tau^-$, $ZA$, $W^\pm H^\mp$, in the four 2HDM types I, II, LS and FL ($h$ is SM-like). \label{tab:heaviernonsmlikedecay1}}
\end{center}
\end{table}
\begin{table}[t]
\begin{center}
\renewcommand{\arraystretch}{1.2}
\scriptsize
 \begin{tabular}{|c|c c c c c|}
  \hline
 Type & $\Delta \text{BR}_{HZZ}^{\pmb{S_1}}$ & $\Delta \text{BR}_{HZZ}^{\pmb{S_2}}$ & $\Delta \text{BR}_{HZZ}^{\pmb{S_3}}$ & $\Delta \text{BR}_{HZZ}^{\pmb{S_4}}$ & $\Delta \text{BR}_{HZZ}^{\pmb{\overline{\text{MS}}}}$ \\ \hline
  I  & $\lesssim 47.5\,\%$ $( 50\,\%)$  & $\lesssim 45.0\,\%$ $( 50\,\%)$ & $\lesssim 90.0\,\%$ $( 48\,\%)$ & $\lesssim 52.5\,\%$ $( 50\,\%)$ & $\lesssim 80.0\,\%$ $( 44\,\%)$  \\  
     & $\gtrsim 100.0\,\%$ $( 29\,\%)$  & $\gtrsim 100.0\,\%$ $( 29\,\%)$ & $\gtrsim 100.0\,\%$ $( 49\,\%)$ & $\gtrsim 100.0\,\%$ $( 32\,\%)$ & $\gtrsim 100.0\,\%$ $( 50\,\%)$ \\ \hdashline
  II & $\lesssim 62.5\,\%$ $( 50\,\%)$  & $\lesssim 60.0\,\%$ $( 50\,\%)$ & $\lesssim 90.0\,\%$ $( 22\,\%)$ & $\lesssim 82.5\,\%$ $( 35\,\%)$ & $\lesssim 82.5\,\%$ $( 34\,\%)$ \\
     & $\gtrsim 100.0\,\%$ $( 39\,\%)$  & $\gtrsim 100.0\,\%$ $( 39\,\%)$ & $\gtrsim 100.0\,\%$ $( 76\,\%)$ & $\gtrsim 100.0\,\%$ $( 59\,\%)$ & $\gtrsim 100.0\,\%$ $( 61\,\%)$ \\ \hdashline
  LS & $\lesssim 67.5\,\%$ $( 50\,\%)$  & $\lesssim 65.0\,\%$ $( 50\,\%)$ & $\lesssim 90.0\,\%$ $( 30\,\%)$ & $\lesssim 80.0\,\%$ $( 50\,\%)$ & $\lesssim 80.0\,\%$ $( 38\,\%)$ \\
     & $\gtrsim 100.0\,\%$ $( 38\,\%)$  & $\gtrsim 100.0\,\%$ $( 37\,\%)$ & $\gtrsim 100.0\,\%$ $( 68\,\%)$ & $\gtrsim 100.0\,\%$ $( 43\,\%)$ & $\gtrsim 100.0\,\%$ $( 56\,\%)$ \\ \hdashline
  FL & $\lesssim 90.0\,\%$ $( 40\,\%)$  & $\lesssim 90.0\,\%$ $( 40\,\%)$ & $\lesssim 90.0\,\%$ $( 30\,\%)$ & $\lesssim 82.5\,\%$ $( 35\,\%)$ & $\lesssim 80.0\,\%$ $( 38\,\%)$ \\  
     & $\gtrsim 100.0\,\%$ $( 57\,\%)$  & $\gtrsim 100.0\,\%$ $( 57\,\%)$ & $\gtrsim 100.0\,\%$ $( 68\,\%)$ & $\gtrsim 100.0\,\%$ $( 60\,\%)$ & $\gtrsim 100.0\,\%$ $( 56\,\%)$ \\ \hline\hline
  Type & $\Delta \text{BR}_{Hhh}^{\pmb{S_1}}$ & $\Delta \text{BR}_{Hhh}^{\pmb{S_2}}$ & $\Delta \text{BR}_{Hhh}^{\pmb{S_3}}$ & $\Delta \text{BR}_{Hhh}^{\pmb{S_4}}$ & $\Delta \text{BR}_{Hhh}^{\pmb{\overline{\text{MS}}}}$ \\ \hline
  I  & $\lesssim 90.0\,\%$ $( 28\,\%)$  & $\lesssim 90.0\,\%$ $( 28\,\%)$ & $\lesssim 90.0\,\%$ $( 25\,\%)$ & $\lesssim 82.5\,\%$ $( 27\,\%)$ & $\lesssim 85.0\,\%$ $( 44\,\%)$  \\  
     & $\gtrsim 100.0\,\%$ $( 70\,\%)$  & $\gtrsim 100.0\,\%$ $( 70\,\%)$ & $\gtrsim 100.0\,\%$ $( 73\,\%)$ & $\gtrsim 100.0\,\%$ $( 69\,\%)$ & $\gtrsim 100.0\,\%$ $( 53\,\%)$ \\ \hdashline
  II & $\lesssim 90.0\,\%$ $( 10\,\%)$  & $\lesssim 90.0\,\%$ $( 10\,\%)$ & $\lesssim 90.0\,\%$ $( 8\,\%)$ & $\lesssim 85.0\,\%$ $( 10\,\%)$ & $\lesssim 87.5\,\%$ $( 35\,\%)$ \\
     & $\gtrsim 100.0\,\%$ $( 89\,\%)$  & $\gtrsim 100.0\,\%$ $( 89\,\%)$ & $\gtrsim 100.0\,\%$ $( 91\,\%)$ & $\gtrsim 100.0\,\%$ $( 88\,\%)$ & $\gtrsim 100.0\,\%$ $( 63\,\%)$ \\ \hdashline
  LS & $\lesssim 90.0\,\%$ $( 20\,\%)$  & $\lesssim 90.0\,\%$ $( 21\,\%)$ & $\lesssim 85.0\,\%$ $( 15\,\%)$ & $\lesssim 85.0\,\%$ $( 20\,\%)$ & $\lesssim 85.0\,\%$ $( 40\,\%)$ \\
     & $\gtrsim 100.0\,\%$ $( 78\,\%)$  & $\gtrsim 100.0\,\%$ $( 78\,\%)$ & $\gtrsim 100.0\,\%$ $( 83\,\%)$ & $\gtrsim 100.0\,\%$ $( 77\,\%)$ & $\gtrsim 100.0\,\%$ $( 57\,\%)$ \\ \hdashline
  FL & $\lesssim 90.0\,\%$ $( 14\,\%)$  & $\lesssim 90.0\,\%$ $( 14\,\%)$ & $\lesssim 85.0\,\%$ $( 8\,\%)$ & $\lesssim 90.0\,\%$ $( 14\,\%)$ & $\lesssim 85.0\,\%$ $( 38\,\%)$ \\  
     & $\gtrsim 100.0\,\%$ $( 84\,\%)$  & $\gtrsim 100.0\,\%$ $( 85\,\%)$ & $\gtrsim 100.0\,\%$ $( 90\,\%)$ & $\gtrsim 100.0\,\%$ $( 84\,\%)$ & $\gtrsim 100.0\,\%$ $( 59\,\%)$ \\ \hline
 \end{tabular}
\caption{Relative size of the EW corrections to the BRs for the non-SM-like
2HDM Higgs boson $H$ decays into $ZZ$, and $hh$, in
the four 2HDM types I, II, LS and FL ($h$ is
SM-like). \label{tab:heaviernonsmlikedecay2}} 
\end{center}
\vspace*{-0.5cm}
\end{table}
\begin{align}
\pmb{S_1} &\equiv \big\{ ~ \text{KOSY}^o ~,~ \text{KOSY}^c ~,~ \text{pOS}^o ~,~ \text{pOS}^c ~,~ \text{BFMS} \big\} \nonumber\\
\pmb{S_2} &\equiv \big\{ ~ \text{p}_*^o ~,~ \text{p}_*^c ~ \big\} \nonumber\\
\pmb{S_3} &\equiv \big\{ ~ \text{proc1} ~,~ \text{proc2} ~,~
                   \text{proc3} ~,~ \text{OS1} ~ \big\} \nonumber\\
\pmb{S_4} &\equiv \big\{ ~ \text{OS2} ~,~ \text{OS12} \big\} \nonumber\\
\pmb{\overline{\text{MS}}} &\equiv \big\{ ~ \overline{\text{MS}} ~ \big\}\nonumber
\end{align}
From the tables, we deduce the following:
\begin{itemize}
\item[--] For all renormalization schemes apart from the
  $\overline{\mbox{MS}}$ scheme, the corrections to the decays into the
  fermionic final states as well as to into $ZA$ and $H^\pm W^\mp$ are
  mostly of 
  moderate size. Depending on the scheme and 2HDM type they can become
  also significant, however, with relative corrections of up to
  40\%. The process-dependent schemes summarized in the sets
  $\pmb{S}_3$ and $\pmb{S}_4$ have larger maximal correction values
  than the schemes of $\pmb{S}_1$ and $\pmb{S}_2$ which exploit the
  symmetries of the model. 
\item[--] The corrections to the (typically dominant) decay
  $H\,\rightarrow \, t\,\bar{t}$ are the most moderate ones which is
  due to the relatively large BR already at tree level.
\item[--] The decays $H\,\rightarrow \, Z\,Z$ and $H\,\rightarrow \, h
  \, h$ typically feature huge EW corrections. For the decay
  into $ZZ$, this is mostly due to the smallness of the tree-level
  width which is proportional to $c_{\beta - \alpha }^2$. As $h$
  behaves SM-like, its coupling to massive gauge bosons, proportional
  to $s_{\beta - \alpha }^2$, is close to one, so that the branching ratio for the decay of $H$
  into $ZZ/WW$ is almost zero.\footnote{We note that in the computation of the
  EW-corrected decay width we only include one-loop terms proportional to the
  product of the tree-level and the one-loop amplitude. We do not include
  terms proportional to the squared one-loop amplitude, as might be
  considered if the tree-level decay width is small. These terms are
  formally of two-loop order so that we chose not to include them.} 
Therefore, the relative EW corrections become large. 
Another reason for the large corrections can be parametrically
enhanced counterterm contributions that arise due to small coupling constants
(due to the sum rules) in the denominator that are multiplied by the
counterterms, or the counterterms themselves become large.
\s

For the decay into $hh$, the dominant effect that
  enhances the EW corrections are non-decoupling effects, inducing parametrically enhanced EW corrections
to the Higgs-to-Higgs decays. This has been discussed in detail in
\cite{Krause:2016xku,Krause:2017mal,Kanemura:2004mg}. Such enhanced
corrections call for the resummation of all orders of perturbation
theory, which is beyond the phenomenological investigation performed
here. Recently, the two-loop corrections to the Higgs
  trilinear couplings in extended scalar sectors have been calculated in
\cite{Braathen:2019pxr,Braathen:2019zoh}. The authors find that they
remain smaller than the one-loop corrections so that the large
deviations predicted at one-loop level do not change significantly.
In \cite{Krause:2016xku} where we investigated in detail the
non-decoupling effects in the EW corrections to the Higgs-to-Higgs
decays we also provided parameter scenarios where we are truly
in the decoupling limit and find decent EW corrections. Finding such
scenarios required a dedicated scan. On the other hand, for this work we performed a
scan without demanding special parameter features and hence our sample of valid
points does not contain such specific parameter configurations. For an
exemplary analysis in the decoupling regime, we therefore refer to
Ref.~\cite{Krause:2016xku}. \s

Note, finally, that these huge corrections also indirectly affect the
$\Delta \mbox{BR}$ of the other decay channels through the total width, namely
the decays with smaller BRs. This also explains their larger values of
$\Delta BR$ in some cases.
\end{itemize}

%%%%%%%%%%%%%%%%%%%%%%%%%%%%%%%%%%%%%%%%%%%%%%%%%%%%%%%%%%%%%%
\subsubsection{$H$ is the SM-like Higgs boson}
%%%%%%%%%%%%%%%%%%%%%%%%%%%%%%%%%%%%%%%%%%%%%%%%%%%%%%%%%%%%%%
In case that the lighter CP-even Higgs boson $h$ is the non-SM-like Higgs
boson many decay channels are kinematically closed. In
Tab.~\ref{tab:lighternonsmlikedecay2} we show the relative corrections $\Delta \text{BR}$
for the OS decays into $b\bar{b}$ and $\tau^+ \tau^-$. The schemes
that are grouped together in scheme sets are as follows: \s 
\begin{table}[t]
\begin{center}
\renewcommand{\arraystretch}{1.2}
\scriptsize
 \begin{tabular}{|c|c c c c c|}
  \hline
  Type & $\Delta \text{BR}_{hb\bar{b}}^{\pmb{S_1}}$ & $\Delta \text{BR}_{hb\bar{b}}^{\pmb{S_2}}$ & $\Delta \text{BR}_{hb\bar{b}}^{\pmb{S_3}}$ & $\Delta \text{BR}_{hb\bar{b}}^{\pmb{S_4}}$ & $\Delta \text{BR}_{hb\bar{b}}^{\pmb{\overline{\text{MS}}}}$ \\ \hline
  I  & $\lesssim 2.5\,\%$ $(91\,\%)$  & $\lesssim 5.0\,\%$ $(61\,\%)$ & $\lesssim 2.5\,\%$ $(91\,\%)$ & $\lesssim 2.5\,\%$ $(91\,\%)$ & $\lesssim 10.0\,\%$ $(50\,\%)$ \\  
     & $\lesssim 10.0\,\%$ $(96\,\%)$  & $\lesssim 40.0\,\%$ $(90\,\%)$ & $\lesssim 15.0\,\%$ $(97\,\%)$ & $\lesssim 15.0\,\%$ $(98\,\%)$ & $\gtrsim 100.0\,\%$ $(12\,\%)$ \\ \hdashline
  II & $\lesssim 2.5\,\%$ $(100\,\%)$  & $\lesssim 2.5\,\%$ $(100\,\%)$ & $\lesssim 2.5\,\%$ $(100\,\%)$ & $\lesssim 2.5\,\%$ $(100\,\%)$ & $\lesssim 2.5\,\%$ $(77\,\%)$ \\  
     & & & & & $\gtrsim 100.0\,\%$ $(3\,\%)$ \\ \hdashline
  LS & $\lesssim 10.0\,\%$ $(67\,\%)$  & $\lesssim 7.5\,\%$ $(63\,\%)$ & $\lesssim 50.0\,\%$ $(50\,\%)$ & $\lesssim 17.5\,\%$ $(49\,\%)$ & $\lesssim 90.0\,\%$ $(18\,\%)$ \\  
     & $\lesssim 20.0\,\%$ $(99\,\%)$  & $\lesssim 20.0\,\%$ $(98\,\%)$ & $\lesssim 90.0\,\%$ $(69\,\%)$ & $\lesssim 30.0\,\%$ $(97\,\%)$ & $\gtrsim 100.0\,\%$ $(79\,\%)$ \\ \hdashline
  FL & $\lesssim 2.5\,\%$ $(100\,\%)$  & $\lesssim 2.5\,\%$ $(96\,\%)$ & $\lesssim 2.5\,\%$ $(100\,\%)$ & $\lesssim 2.5\,\%$ $(100\,\%)$ & $\lesssim 2.5\,\%$ $(73\,\%)$ \\  
     & & $\lesssim 7.5\,\%$ $(99\,\%)$ & & & $\gtrsim 100.0\,\%$
                                             $(3\,\%)$ \\ \hline\hline
  Type & $\Delta \text{BR}_{h\tau ^+ \tau ^-}^{\pmb{S_1}}$ & $\Delta \text{BR}_{h\tau ^+ \tau ^-}^{\pmb{S_2}}$ & $\Delta \text{BR}_{h\tau ^+ \tau ^-}^{\pmb{S_3}}$ & $\Delta \text{BR}_{h\tau ^+ \tau ^-}^{\pmb{S_4}}$ & $\Delta \text{BR}_{h\tau ^+ \tau ^-}^{\pmb{\overline{\text{MS}}}}$ \\ \hline
  I  & $\lesssim 7.5\,\%$ $(93\,\%)$  & $\lesssim 5.0\,\%$ $(51\,\%)$ & $\lesssim 5.0\,\%$ $(83\,\%)$ & $\lesssim 5.0\,\%$ $(86\,\%)$ & $\lesssim 12.5\,\%$ $(63\,\%)$ \\  
     & $\lesssim 20.0\,\%$ $(98\,\%)$  & $\lesssim 35.0\,\%$ $(87\,\%)$ & $\lesssim 15.0\,\%$ $(98\,\%)$ & $\lesssim 10.0\,\%$ $(97\,\%)$ & $\gtrsim 100.0\,\%$ $(12\,\%)$ \\ \hdashline
  II & $\lesssim 2.5\,\%$ $(100\,\%)$  & $\lesssim 2.5\,\%$ $(100\,\%)$ & $\lesssim 2.5\,\%$ $(100\,\%)$ & $\lesssim 2.5\,\%$ $(96\,\%)$ & $\lesssim 2.5\,\%$ $(74\,\%)$ \\  
     &                                 &                                &                      &      $\lesssim 5.0\,\%$ $(100\,\%)$ & $\gtrsim 100.0\,\%$ $(3\,\%)$ \\ \hdashline
  LS & $\lesssim 2.5\,\%$ $(94\,\%)$  & $\lesssim 2.5\,\%$ $(91\,\%)$ & $\lesssim 2.5\,\%$ $(83\,\%)$ & $\lesssim 2.5\,\%$ $(88\,\%)$ & $\lesssim 2.5\,\%$ $(64\,\%)$ \\  
     & $\lesssim 7.5\,\%$ $(100\,\%)$  & $\lesssim 10.0\,\%$ $(98\,\%)$ & $\lesssim 15.0\,\%$ $(97\,\%)$ & $\lesssim 10.0\,\%$ $(97\,\%)$ & $\gtrsim 100.0\,\%$ $(13\,\%)$ \\ \hdashline
  FL & $\lesssim 5.0\,\%$ $(96\,\%)$  & $\lesssim 77.5\,\%$ $(50\,\%)$ & $\lesssim 35.0\,\%$ $(52\,\%)$ & $\lesssim 15.0\,\%$ $(49\,\%)$ & $\lesssim 90.0\,\%$ $(24\,\%)$ \\  
     & $\lesssim 10.0\,\%$ $(98\,\%)$  & $\gtrsim 100.0\,\%$ $(43\,\%)$ & $\lesssim 60.0\,\%$ $(98\,\%)$ & $\lesssim 30.0\,\%$ $(100\,\%)$ & $\gtrsim 100.0\,\%$ $(73\,\%)$ \\ \hline
 \end{tabular}
\caption{Relative size of the EW corrections to the BRs for the non-SM-like
2HDM Higgs boson $h$ decays into $b\bar{b}$ and $\tau^+ \tau^-$, in
the four 2HDM types I, II, LS and 
FL ($H$ is SM-like). \label{tab:lighternonsmlikedecay2}} 
\vspace*{-0.4cm}
\end{center}
\end{table}
\begin{align}
\pmb{S_1} &\equiv \big\{ ~ \text{KOSY}^o ~,~ \text{KOSY}^c ~,~ \text{p}_*^o ~,~ \text{p}_*^c ~,~ \text{pOS}^o ~,~ \text{pOS}^c ~,~ \text{BFMS} ~ \big\} \nonumber\\
\pmb{S_2} &\equiv \big\{ ~ \text{proc1} ~,~ \text{proc3} ~ \big\} \nonumber\\
\pmb{S_3} &\equiv \big\{ ~ \text{proc2} ~,~ \text{OS1} ~ \big\} \nonumber\\
\pmb{S_4}  &\equiv \big\{ ~ \text{OS2} ~,~ \text{OS12} ~ \big\} \nonumber\\
\pmb{\overline{\text{MS}}} &\equiv \big\{ ~ \overline{\text{MS}} ~ \big\}
\nonumber
\end{align}

From the table, we read off:
\begin{itemize}
\item[--] For the renormalization schemes exploiting the symmetries of the
  Lagrangian, grouped together in $\pmb{S_1}$, the EW corrections are
  of moderate size, not exceeding 20\%. 
\item[--] The process-dependent renormalization schemes, {\it cf.}~sets
  $\pmb{S_2},$ $\pmb{S_3}$, and $\pmb{S_4}$, in general also lead to
  moderate corrections. For some decays and 2HDM types, however, the
  corrections become very large. Note, however, that these claims are
  based on the very low statistics stemming from the low amount of
  input parameter sets used. 
\end{itemize}

%%%%%%%%%%%%%%%%%%%%%%%%%%%%%%%%%%%%%%%%%%%%%%%%%%%%%%%%%%%%%%
\subsection{Pseudoscalar 2HDM Decays \label{sec:pseudoscalar2hdm}}
%%%%%%%%%%%%%%%%%%%%%%%%%%%%%%%%%%%%%%%%%%%%%%%%%%%%%%%%%%%%%%
\subsubsection{$h$ is the SM-like Higgs boson}
%%%%%%%%%%%%%%%%%%%%%%%%%%%%%%%%%%%%%%%%%%%%%%%%%%%%%%%%%%%%%%
\begin{table}[t!]
\begin{center}
\renewcommand{\arraystretch}{1.2}
\scriptsize
 \begin{tabular}{|c|c c c|}
  \hline
Type & $A \to t\bar{t}$ & $A \to Zh$ & $A \to ZH$ \\
   \hline
I & $176\,057$ & $183\,927$ & $101\,561$ \\
II & $239\,750$ & $239\,756$ & $87\,004$ \\
LS & $238\,520$ & $239\,122$ & $95\,863$ \\
FL & $219\,123$ & $219\,146$ & $108\,082$ \\
\hline
\end{tabular}
\caption{Number of parameter points available for the analysis in the
  individual OS decay channels of the pseudoscalar Higgs boson $A$ ($h$ is SM-like).}
\label{tab:pseudohlighter}
\vspace*{-0.4cm}
\end{center}
\end{table}
We turn to the decays of the pseudoscalar Higgs boson $A$ for the input parameter sets
where $h$ is SM-like. Besides the usual fermionic decays, $A$ can
also decay into a mixed gauge boson plus Higgs boson final
state. Electroweak corrections are computed only for OS decays so that
the number of available parameter points reduces as shown
in Tab.~\ref{tab:pseudohlighter}.
In Tab.~\ref{tab:lighterpseudoscalar}, the relative corrections $\Delta \text{BR}$ for the
pseudoscalar decays into $b\bar{b}$, $\tau^+ \tau^-$ and $t\bar{t}$ as
well as in the gauge boson plus Higgs boson final states $Zh$ and $ZH$ are
given for the case that $h$ is SM-like, for all four 2HDM types. 
The scheme sets shown in the table are defined as follows:
\begingroup
\allowdisplaybreaks
\begin{align}
\pmb{S_1} &\equiv \big\{ ~ \text{KOSY}^o ~,~ \text{KOSY}^c ~,~
                    \text{pOS}^o ~,~ \text{pOS}^c ~ \big\} \nonumber \nonumber\\
	\pmb{S_2} &\equiv \big\{ ~ \text{p}_*^o ~,~ \text{p}_*^c ~ \big\} \nonumber\\
	\pmb{S_3} &\equiv \big\{ ~ \text{proc1} ~,~ \text{proc2} ~,~ \text{proc3} ~,~ \text{OS1} ~,~ \text{OS2} ~ \big\} \nonumber\\
	\pmb{\text{OS12}} &\equiv \big\{ ~ \text{OS12} ~ \big\} \nonumber\\
	\pmb{\text{BFMS}} &\equiv \big\{ ~ \text{BFMS} ~ \big\} \nonumber\\
	\pmb{\overline{\text{MS}}} &\equiv \big\{ ~ \overline{\text{MS}} ~ \big\}
\nonumber
\end{align}
\endgroup

\begin{table}[t!]
\begin{center}
\renewcommand{\arraystretch}{1.2}
\scriptsize
 \begin{tabular}{|c|c c c c c c|}
  \hline
  Type & $\Delta \text{BR}_{Ab\bar{b}}^{\pmb{S_1}}$ & $\Delta \text{BR}_{Ab\bar{b}}^{\pmb{S_2}}$ & $\Delta \text{BR}_{Ab\bar{b}}^{\pmb{S_3}}$ & $\Delta \text{BR}_{Ab\bar{b}}^{\pmb{\text{OS12}}}$ & $\Delta \text{BR}_{Ab\bar{b}}^{\pmb{\text{BFMS}}}$ & $\Delta \text{BR}_{Ab\bar{b}}^{\pmb{\overline{\text{MS}}}}$ \\ \hline
  I  & $\lesssim 7.5\,\%$ $( 53\,\%)$  & $\lesssim 7.5\,\%$ $( 55\,\%)$ & $\lesssim 5.0\,\%$ $( 40\,\%)$ & $\lesssim 5.0\,\%$ $( 40\,\%)$ & $\lesssim 7.5\,\%$ $( 46\,\%)$ & $\lesssim 37.5\,\%$ $( 50\,\%)$  \\  
     & $\lesssim 17.5\,\%$ $( 95\,\%)$  & $\lesssim 17.5\,\%$ $( 94\,\%)$ & $\lesssim 17.5\,\%$ $( 86\,\%)$ & $\lesssim 17.5\,\%$ $( 86\,\%)$ & $\lesssim 20.0\,\%$ $( 96\,\%)$ & $\gtrsim 100.0\,\%$ $( 36\,\%)$ \\ \hdashline
  II & $\lesssim 10.0\,\%$ $( 45\,\%)$  & $\lesssim 17.5\,\%$ $( 50\,\%)$ & $\lesssim 12.5\,\%$ $( 47\,\%)$ & $\lesssim 15.0\,\%$ $( 44\,\%)$ & $\lesssim 7.5\,\%$ $( 37\,\%)$ & $\lesssim 50.0\,\%$ $( 2\,\%)$ \\
     & $\lesssim 22.5\,\%$ $( 96\,\%)$  & $\lesssim 30.0\,\%$ $( 94\,\%)$ & $\lesssim 30.0\,\%$ $( 82\,\%)$ & $\lesssim 27.5\,\%$ $( 89\,\%)$ & $\lesssim 25.0\,\%$ $( 97\,\%)$ & $\gtrsim 100.0\,\%$ $( 95\,\%)$ \\ \hdashline
  LS & $\lesssim 7.5\,\%$ $( 49\,\%)$  & $\lesssim 7.5\,\%$ $( 54\,\%)$ & $\lesssim 7.5\,\%$ $( 48\,\%)$ & $\lesssim 7.5\,\%$ $( 48\,\%)$ & $\lesssim 7.5\,\%$ $( 46\,\%)$ & $\lesssim 32.5\,\%$ $( 50\,\%)$ \\
     & $\lesssim 17.5\,\%$ $( 94\,\%)$  & $\lesssim 17.5\,\%$ $( 95\,\%)$ & $\lesssim 25.0\,\%$ $( 95\,\%)$ & $\lesssim 25.0\,\%$ $( 95\,\%)$ & $\lesssim 20.0\,\%$ $( 96\,\%)$ & $\gtrsim 100.0\,\%$ $( 30\,\%)$ \\ \hdashline
  FL & $\lesssim 10.0\,\%$ $( 45\,\%)$  & $\lesssim 17.5\,\%$ $( 50\,\%)$ & $\lesssim 30.0\,\%$ $( 50\,\%)$ & $\lesssim 17.5\,\%$ $( 44\,\%)$ & $\lesssim 7.5\,\%$ $( 50\,\%)$ & $\lesssim 55.0\,\%$ $( 2\,\%)$ \\  
     & $\lesssim 22.5\,\%$ $( 96\,\%)$  & $\lesssim 30.0\,\%$ $( 94\,\%)$ & $\lesssim 50.0\,\%$ $( 89\,\%)$ & $\lesssim 30.0\,\%$ $( 94\,\%)$ & $\lesssim 20.0\,\%$ $( 98\,\%)$ & $\gtrsim 100.0\,\%$ $( 94\,\%)$ \\ \hline\hline
  Type & $\Delta \text{BR}_{At\bar{t}}^{\pmb{S_1}}$ & $\Delta \text{BR}_{At\bar{t}}^{\pmb{S_2}}$ & $\Delta \text{BR}_{At\bar{t}}^{\pmb{S_3}}$ & $\Delta \text{BR}_{At\bar{t}}^{\pmb{\text{OS12}}}$ & $\Delta \text{BR}_{At\bar{t}}^{\pmb{\text{BFMS}}}$ & $\Delta \text{BR}_{At\bar{t}}^{\pmb{\overline{\text{MS}}}}$ \\ \hline
  I  & $\lesssim 2.5\,\%$ $( 87\,\%)$  & $\lesssim 2.5\,\%$ $( 90\,\%)$ & $\lesssim 2.5\,\%$ $( 60\,\%)$ & $\lesssim 2.5\,\%$ $( 83\,\%)$ & $\lesssim 2.5\,\%$ $( 83\,\%)$ & $\lesssim 45.0\,\%$ $( 50\,\%)$  \\  
     & $\lesssim 5.0\,\%$ $( 95\,\%)$  & $\lesssim 5.0\,\%$ $( 95\,\%)$ & $\lesssim 10.0\,\%$ $( 85\,\%)$ & $\lesssim 5.0\,\%$ $( 92\,\%)$ & $\lesssim 5.0\,\%$ $( 90\,\%)$ & $\gtrsim 100.0\,\%$ $( 35\,\%)$ \\ \hdashline
  II & $\lesssim 2.5\,\%$ $( 96\,\%)$  & $\lesssim 2.5\,\%$ $( 97\,\%)$ & $\lesssim 2.5\,\%$ $( 82\,\%)$ & $\lesssim 2.5\,\%$ $( 94\,\%)$ & $\lesssim 2.5\,\%$ $( 95\,\%)$ & $\lesssim 17.5\,\%$ $( 50\,\%)$ \\
     & $\lesssim 5.0\,\%$ $( 99\,\%)$  & $\lesssim 5.0\,\%$ $( 99\,\%)$ & $\lesssim 7.5\,\%$ $( 93\,\%)$ & $\lesssim 2.5\,\%$ $( 98\,\%)$ & $\lesssim 5.0\,\%$ $( 98\,\%)$ & $\gtrsim 100.0\,\%$ $( 20\,\%)$ \\ \hdashline
  LS & $\lesssim 2.5\,\%$ $( 93\,\%)$  & $\lesssim 2.5\,\%$ $( 95\,\%)$ & $\lesssim 2.5\,\%$ $( 72\,\%)$ & $\lesssim 2.5\,\%$ $( 92\,\%)$ & $\lesssim 2.5\,\%$ $( 91\,\%)$ & $\lesssim 35.0\,\%$ $( 50\,\%)$ \\
     & $\lesssim 5.0\,\%$ $( 98\,\%)$  & $\lesssim 5.0\,\%$ $( 98\,\%)$ & $\lesssim 10.0\,\%$ $( 90\,\%)$ & $\lesssim 5.0\,\%$ $( 97\,\%)$ & $\lesssim 5.0\,\%$ $( 96\,\%)$ & $\gtrsim 100.0\,\%$ $( 30\,\%)$ \\ \hdashline
  FL & $\lesssim 2.5\,\%$ $( 93\,\%)$  & $\lesssim 2.5\,\%$ $( 95\,\%)$ & $\lesssim 2.5\,\%$ $( 76\,\%)$ & $\lesssim 2.5\,\%$ $( 92\,\%)$ & $\lesssim 2.5\,\%$ $( 91\,\%)$ & $\lesssim 40.0\,\%$ $( 50\,\%)$ \\  
     & $\lesssim 5.0\,\%$ $( 98\,\%)$  & $\lesssim 5.0\,\%$ $( 98\,\%)$ & $\lesssim 7.5\,\%$ $( 90\,\%)$ & $\lesssim 5.0\,\%$ $( 96\,\%)$ & $\lesssim 5.0\,\%$ $( 96\,\%)$ & $\gtrsim 100.0\,\%$ $( 31\,\%)$ \\ \hline\hline
  Type & $\Delta \text{BR}_{A\tau ^+ \tau ^-}^{\pmb{S_1}}$ & $\Delta \text{BR}_{A\tau ^+ \tau ^-}^{\pmb{S_2}}$ & $\Delta \text{BR}_{A\tau ^+ \tau ^-}^{\pmb{S_3}}$ & $\Delta \text{BR}_{A\tau ^+ \tau ^-}^{\pmb{\text{OS12}}}$ & $\Delta \text{BR}_{A\tau ^+ \tau ^-}^{\pmb{\text{BFMS}}}$ & $\Delta \text{BR}_{A\tau ^+ \tau ^-}^{\pmb{\overline{\text{MS}}}}$ \\ \hline
  I  & $\lesssim 10.0\,\%$ $( 50\,\%)$  & $\lesssim 10.0\,\%$ $( 55\,\%)$ & $\lesssim 7.5\,\%$ $(45 \,\%)$ & $\lesssim 10.0\,\%$ $( 64\,\%)$ & $\lesssim 10.0\,\%$ $( 45\,\%)$ & $\lesssim 47.5\,\%$ $( 50\,\%)$  \\  
     & $\lesssim 20.0\,\%$ $( 97\,\%)$  & $\lesssim 20.0\,\%$ $( 97\,\%)$ & $\lesssim 17.5\,\%$ $( 93\,\%)$ & $\lesssim 17.5\,\%$ $( 93\,\%)$ & $\lesssim 25.0\,\%$ $( 97\,\%)$ & $\gtrsim 100.0\,\%$ $( 36\,\%)$ \\ \hdashline
  II & $\lesssim 10.0\,\%$ $( 45\,\%)$  & $\lesssim 17.5\,\%$ $( 56\,\%)$ & $\lesssim 12.5\,\%$ $( 38\,\%)$ & $\lesssim 15.0\,\%$ $( 50\,\%)$ & $\lesssim 7.5\,\%$ $( 40\,\%)$ & $\lesssim 50.0\,\%$ $( 3\,\%)$ \\
     & $\lesssim 20.0\,\%$ $( 97\,\%)$  & $\lesssim 30.0\,\%$ $( 97\,\%)$ & $\lesssim 30.0\,\%$ $( 87\,\%)$ & $\lesssim 27.5\,\%$ $( 95\,\%)$ & $\lesssim 25.0\,\%$ $( 97\,\%)$ & $\gtrsim 100.0\,\%$ $( 95\,\%)$ \\ \hdashline
  LS & $\lesssim 10.0\,\%$ $( 53\,\%)$  & $\lesssim 15.0\,\%$ $( 55\,\%)$ & $\lesssim 10.0\,\%$ $( 40\,\%)$ & $\lesssim 17.5\,\%$ $( 46\,\%)$ & $\lesssim 7.5\,\%$ $( 55\,\%)$ & $\lesssim 50.0\,\%$ $( 5\,\%)$ \\
     & $\lesssim 20.0\,\%$ $( 97\,\%)$  & $\lesssim 27.5\,\%$ $( 98\,\%)$ & $\lesssim 27.5\,\%$ $( 93\,\%)$ & $\lesssim 30.0\,\%$ $( 96\,\%)$ & $\lesssim 20.0\,\%$ $( 98\,\%)$ & $\gtrsim 100.0\,\%$ $( 85\,\%)$ \\ \hdashline
  FL & $\lesssim 10.0\,\%$ $( 51\,\%)$  & $\lesssim 10.0\,\%$ $( 53\,\%)$ & $\lesssim 10.0\,\%$ $( 70\,\%)$ & $\lesssim 10.0\,\%$ $( 61\,\%)$ & $\lesssim 10.0\,\%$ $( 49\,\%)$ & $\lesssim 40.0\,\%$ $( 50\,\%)$ \\  
     & $\lesssim 20.0\,\%$ $( 97\,\%)$  & $\lesssim 20.0\,\%$ $( 98\,\%)$ & $\lesssim 17.5\,\%$ $( 97\,\%)$ & $\lesssim 17.5\,\%$ $( 96\,\%)$ & $\lesssim 25.0\,\%$ $( 97\,\%)$ & $\gtrsim 100.0\,\%$ $( 31\,\%)$ \\ \hline\hline
  Type & $\Delta \text{BR}_{AZh}^{\pmb{S_1}}$ & $\Delta \text{BR}_{AZh}^{\pmb{S_2}}$ & $\Delta \text{BR}_{AZh}^{\pmb{S_3}}$ & $\Delta \text{BR}_{AZh}^{\pmb{\text{OS12}}}$ & $\Delta \text{BR}_{AZh}^{\pmb{\text{BFMS}}}$ & $\Delta \text{BR}_{AZh}^{\pmb{\overline{\text{MS}}}}$ \\ \hline
  I  & $\lesssim 15.0\,\%$ $( 51\,\%)$  & $\lesssim 17.5\,\%$ $( 51\,\%)$ & $\lesssim 85.0\,\%$ $( 45\,\%)$ & $\lesssim 30.0\,\%$ $( 50\,\%)$ & $\lesssim 10.0\,\%$ $( 46\,\%)$ & $\lesssim 90.0\,\%$ $( 38\,\%)$  \\  
     & $\gtrsim 100.0\,\%$ $( 12\,\%)$  & $\gtrsim 100.0\,\%$ $( 13\,\%)$ & $\gtrsim 100.0\,\%$ $( 50\,\%)$ & $\gtrsim 100.0\,\%$ $( 22\,\%)$ & $\gtrsim 100.0\,\%$ $( 8\,\%)$ & $\gtrsim 100.0\,\%$ $( 60\,\%)$ \\ \hdashline
  II & $\lesssim 30.0\,\%$ $( 51\,\%)$  & $\lesssim 27.5\,\%$ $( 51\,\%)$ & $\lesssim 80.0\,\%$ $( 27\,\%)$ & $\lesssim 90.0\,\%$ $( 42\,\%)$ & $\lesssim 10.0\,\%$ $( 46\,\%)$ & $\lesssim 90.0\,\%$ $( 27\,\%)$ \\
     & $\gtrsim 100.0\,\%$ $( 22\,\%)$  & $\gtrsim 100.0\,\%$ $( 21\,\%)$ & $\gtrsim 100.0\,\%$ $( 67\,\%)$ & $\gtrsim 100.0\,\%$ $( 55\,\%)$ & $\gtrsim 100.0\,\%$ $( 10\,\%)$ & $\gtrsim 100.0\,\%$ $( 70\,\%)$ \\ \hdashline
  LS & $\lesssim 22.5\,\%$ $( 50\,\%)$  & $\lesssim 25.0\,\%$ $( 50\,\%)$ & $\lesssim 90.0\,\%$ $( 40\,\%)$ & $\lesssim 50.0\,\%$ $( 50\,\%)$ & $\lesssim 12.5\,\%$ $( 49\,\%)$ & $\lesssim 90.0\,\%$ $( 36\,\%)$ \\
     & $\gtrsim 100.0\,\%$ $( 17\,\%)$  & $\gtrsim 100.0\,\%$ $( 18\,\%)$ & $\gtrsim 100.0\,\%$ $( 57\,\%)$ & $\gtrsim 100.0\,\%$ $( 32\,\%)$ & $\gtrsim 100.0\,\%$ $( 10\,\%)$ & $\gtrsim 100.0\,\%$ $( 64\,\%)$ \\ \hdashline
  FL & $\lesssim 35.0\,\%$ $( 50\,\%)$  & $\lesssim 37.5\,\%$ $( 50\,\%)$ & $\lesssim 90.0\,\%$ $( 25\,\%)$ & $\lesssim 77.5\,\%$ $( 50\,\%)$ & $\lesssim 12.5\,\%$ $( 51\,\%)$ & $\lesssim 90.0\,\%$ $( 30\,\%)$ \\  
     & $\gtrsim 100.0\,\%$ $( 28\,\%)$  & $\gtrsim 100.0\,\%$ $( 28\,\%)$ & $\gtrsim 100.0\,\%$ $( 73\,\%)$ & $\gtrsim 100.0\,\%$ $( 43\,\%)$ & $\gtrsim 100.0\,\%$ $( 9\,\%)$ & $\gtrsim 100.0\,\%$ $( 67\,\%)$ \\ \hline\hline
  Type & $\Delta \text{BR}_{AZH}^{\pmb{S_1}}$ & $\Delta \text{BR}_{AZH}^{\pmb{S_2}}$ & $\Delta \text{BR}_{AZH}^{\pmb{S_3}}$ & $\Delta \text{BR}_{AZH}^{\pmb{\text{OS12}}}$ & $\Delta \text{BR}_{AZH}^{\pmb{\text{BFMS}}}$ & $\Delta \text{BR}_{AZH}^{\pmb{\overline{\text{MS}}}}$ \\ \hline
  I  & $\lesssim 2.5\,\%$ $( 61\,\%)$  & $\lesssim 2.5\,\%$ $( 69\,\%)$ & $\lesssim 7.5\,\%$ $( 55\,\%)$ & $\lesssim 2.5\,\%$ $( 69\,\%)$ & $\lesssim 2.5\,\%$ $( 49\,\%)$ & $\lesssim 90.0\,\%$ $( 37\,\%)$  \\  
     & $\lesssim 7.5\,\%$ $( 84\,\%)$  & $\lesssim 5.0\,\%$ $( 82\,\%)$ & $\lesssim 17.5\,\%$ $( 83\,\%)$ & $\lesssim 5.0\,\%$ $( 85\,\%)$ & $\lesssim 12.5\,\%$ $( 87\,\%)$ & $\gtrsim 100.0\,\%$ $( 50\,\%)$ \\ \hdashline
  II & $\lesssim 5.0\,\%$ $( 59\,\%)$  & $\lesssim 2.5\,\%$ $( 54\,\%)$ & $\lesssim 15.0\,\%$ $( 40\,\%)$ & $\lesssim 5.0\,\%$ $( 65\,\%)$ & $\lesssim 5.0\,\%$ $( 55\,\%)$ & $\lesssim 90.0\,\%$ $( 22\,\%)$ \\
     & $\lesssim 10.0\,\%$ $( 82\,\%)$  & $\lesssim 7.5\,\%$ $( 83\,\%)$ & $\lesssim 32.5\,\%$ $( 70\,\%)$ & $\lesssim 10.0\,\%$ $( 84\,\%)$ & $\lesssim 12.5\,\%$ $( 83\,\%)$ & $\gtrsim 100.0\,\%$ $( 52\,\%)$ \\ \hdashline
  LS & $\lesssim 2.5\,\%$ $( 52\,\%)$  & $\lesssim 2.5\,\%$ $( 66\,\%)$ & $\lesssim 15.0\,\%$ $( 56\,\%)$ & $\lesssim 5.0\,\%$ $( 82\,\%)$ & $\lesssim 5.0\,\%$ $( 61\,\%)$ & $\lesssim 90.0\,\%$ $( 33\,\%)$ \\
     & $\lesssim 7.5\,\%$ $( 81\,\%)$  & $\lesssim 7.5\,\%$ $( 87\,\%)$ & $\lesssim 32.5\,\%$ $( 80\,\%)$ & $\lesssim 7.5\,\%$ $( 89\,\%)$ & $\gtrsim 12.5\,\%$ $( 86\,\%)$ & $\gtrsim 100.0\,\%$ $( 49\,\%)$ \\ \hdashline
  FL & $\lesssim 5.0\,\%$ $( 62\,\%)$  & $\lesssim 2.5\,\%$ $( 57\,\%)$ & $\lesssim 10.0\,\%$ $( 48\,\%)$ & $\lesssim 2.5\,\%$ $( 56\,\%)$ & $\lesssim 5.0\,\%$ $( 50\,\%)$ & $\lesssim 90.0\,\%$ $( 23\,\%)$ \\  
     & $\lesssim 10.0\,\%$ $( 84\,\%)$  & $\lesssim 7.5\,\%$ $( 83\,\%)$ & $\lesssim 20.0\,\%$ $( 81\,\%)$ & $\lesssim 7.5\,\%$ $( 86\,\%)$ & $\lesssim 15.0\,\%$ $( 86\,\%)$ & $\gtrsim 100.0\,\%$ $( 52\,\%)$ \\ \hline
 \end{tabular}
\caption{Relative size of the EW corrections to the BRs for the
  pseudoscalar 2HDM Higgs boson $A$ decays into $b\bar{b}$, $\tau^+
  \tau^-$, $t\bar{t}$, $Zh$, and $ZH$, in the four 2HDM types I, II, LS and
FL ($h$ is SM-like). \label{tab:lighterpseudoscalar}}  
\end{center}
\vspace*{-0.5cm}
\end{table}

With $A$ being above the $t\bar{t}$ threshold, the decay $A \to
t\bar{t}$ is in general the dominant decay channel and consequently also the EW
corrections are the most moderate ones compared to the other final
states. The corrections to $b\bar{b}$ and $\tau^+\tau^-$ are larger,
but still of moderate size for the renormalization scheme sets
$\pmb{S}_1$ and $\pmb{\text{BFMS}}$. These two scheme sets show the
smallest corrections throughout all decays and 2HDM types. This implies
a good numerical stability through large parts of the parameter space
and for all considered decay channels. \s

The decay $A\,\rightarrow \, Z\,h$ 
features relatively large electroweak corrections for all 2HDM types
and scheme sets, with a lot of points leading to corrections above $\pm
100\,\%$. Again, several effects may be responsible
  for this behaviour. The decay width is proportional to
$c_{\beta-\alpha}^2$ which is very small in case that $h$ is SM-like since then,
$s_{\beta-\alpha}$ is close to 1. The tree-level BR of $A\,\rightarrow
\, Z\,h$ is hence very small, so that the relative EW corrections blow
up. Moreover, the corrections can also be parametrically enhanced in
this corner of the parameter space and the counterterms themselves can become large.
The BR of the decay $A\to ZH$ on the other hand can become
important so that here the EW corrections are of moderate
size. 

%%%%%%%%%%%%%%%%%%%%%%%%%%%%%%%%%%%%%%%%%%%%%%%%%%%%%%%%%%%%%%
\subsubsection{$H$ is the SM-like Higgs boson}
%%%%%%%%%%%%%%%%%%%%%%%%%%%%%%%%%%%%%%%%%%%%%%%%%%%%%%%%%%%%%%
\begin{table}[t!]
\begin{center}
\renewcommand{\arraystretch}{1.2}
\scriptsize
 \begin{tabular}{|c|c c c|}
  \hline
Type & $A \to t\bar{t}$ & $A \to Zh$ & $A \to ZH$ \\
   \hline
I & 291 & 558 & 124 \\
II &  39 & 14 & 12 \\
LS & 84 & 114 & 107 \\
FL & 124 & 124 & 114 \\
\hline
\end{tabular}
\caption{Number of parameter points available for the analysis in the
  individual OS decay channels of the pseudoscalar Higgs boson $A$ ($H$ is SM-like).}
\label{tab:pseudoheavier}
\vspace*{-0.4cm}
\end{center}
\end{table}
As mentioned before, the amount of parameter points available reduces considerably in case that the SM-like Higgs boson corresponds to $H$, {\it cf.}~Eq.~(\ref{eq:numbersheavier}), and hence the statistics for these scenarios is rather low. The amount of available parameter points is even further reduced by the requirement of the decays to be OS. The correponding number of
scenarios that we can use in the individual channels are listed in
Tab.~\ref{tab:pseudoheavier}. The relative corrections $\Delta \mbox{BR}$ are given in
Tab.~\ref{tab:heavierpseudoscalar}. The scheme sets shown in the
table have been combined according to: 
\begin{align}
\pmb{S_1} &\equiv \big\{ ~ \text{KOSY}^o ~,~ \text{KOSY}^c ~,~ \text{pOS}^o ~,~ \text{pOS}^c ~,~ \text{p}_*^o ~,~ \text{p}_*^c ~,~ \text{BFMS} \big\} \nonumber\\
\pmb{S_2} &\equiv \big\{ ~ \text{proc1} ~,~ \text{proc2} ~ \big\} \nonumber\\
\pmb{S_3} &\equiv \big\{ ~ \text{OS2} ~,~ \text{OS12} ~ \big\} \nonumber\\
\pmb{\text{proc3}} &\equiv \big\{ ~ \text{proc3} ~ \big\} \nonumber\\
\pmb{\text{OS1}} &\equiv \big\{ ~ \text{OS1} ~ \big\} \nonumber\\
\pmb{\overline{\text{MS}}} &\equiv \big\{ ~ \overline{\text{MS}} ~ \big\}
\nonumber
\end{align}
\begin{table}[ht!]
\begin{center}
\renewcommand{\arraystretch}{1.2}
\scriptsize
 \begin{tabular}{|c|c c c c c c|}
  \hline
 Type & $\Delta \text{BR}_{Ab\bar{b}}^{\pmb{S_1}}$ & $\Delta \text{BR}_{Ab\bar{b}}^{\pmb{S_2}}$ & $\Delta \text{BR}_{Ab\bar{b}}^{\pmb{S_3}}$ & $\Delta \text{BR}_{Ab\bar{b}}^{\pmb{\text{proc3}}}$ & $\Delta \text{BR}_{Ab\bar{b}}^{\pmb{\text{OS1}}}$ & $\Delta \text{BR}_{Ab\bar{b}}^{\pmb{\overline{\text{MS}}}}$ \\ \hline
  I  & $\lesssim 7.5\,\%$ $( 73\,\%)$  & $\lesssim 7.5\,\%$ $( 93\,\%)$ & $\lesssim 5.0\,\%$ $( 97\,\%)$ & $\lesssim 90.0\,\%$ $(39\,\%)$ & $\lesssim 50.0\,\%$ $(50\,\%)$ & $\lesssim 90.0\,\%$ $(22\,\%)$  \\  
     & $\lesssim 17.5\,\%$ $( 98\,\%)$  & $\lesssim 10.0\,\%$ $( 99\,\%)$ & $\lesssim 10.0\,\%$ $(100\,\%)$ & $\gtrsim 100.0\,\%$ $(58\,\%)$ & $\gtrsim 100.0\,\%$ $(29\,\%)$ & $\gtrsim 100.0\,\%$ $(77\,\%)$ \\ \hdashline
  II & $\lesssim 12.5\,\%$ $( 76\,\%)$  & $\lesssim 7.5\,\%$ $( 59\,\%)$ & $\lesssim 20.0\,\%$ $(46\,\%)$ & $\lesssim 2.5\,\%$ $(82\,\%)$ & $\lesssim 62.5\,\%$ $(50\,\%)$ & $\lesssim 90.0\,\%$ $(2\,\%)$  \\  
     & $\lesssim 15.0\,\%$ $( 98\,\%)$  & $\lesssim 10.0\,\%$ $( 100\,\%)$ & $\lesssim 30.0\,\%$ $(96\,\%)$ & $\lesssim 7.5\,\%$ $(100\,\%)$ & $\gtrsim 100.0\,\%$ $(31\,\%)$ & $\gtrsim 100.0\,\%$ $(97\,\%)$ \\ \hdashline
  LS & $\lesssim 7.5\,\%$ $( 57\,\%)$  & $\lesssim 20.0\,\%$ $( 58\,\%)$ & $\lesssim 5.0\,\%$ $(76\,\%)$ & $\lesssim 20.0\,\%$ $(67\,\%)$ & $\lesssim 77.5\,\%$ $(50\,\%)$ & $\lesssim 90.0\,\%$ $(2\,\%)$  \\  
     & $\lesssim 17.5\,\%$ $( 98\,\%)$  & $\lesssim 30.0\,\%$ $( 98\,\%)$ & $\lesssim 15.0\,\%$ $(100\,\%)$ & $\lesssim 25.0\,\%$ $(98\,\%)$ & $\gtrsim 100.0\,\%$ $(42\,\%)$ & $\gtrsim 100.0\,\%$ $(97\,\%)$ \\ \hdashline
  FL & $\lesssim 12.5\,\%$ $( 76\,\%)$  & $\lesssim 30.0\,\%$ $( 60\,\%)$ & $\lesssim 22.5\,\%$ $(54\,\%)$ & $\lesssim 75.0\,\%$ $(50\,\%)$ & $\lesssim 77.5\,\%$ $(50\,\%)$ & $\lesssim 90.0\,\%$ $(2\,\%)$  \\  
     & $\lesssim 15.0\,\%$ $( 98\,\%)$  & $\lesssim 35.0\,\%$ $( 98\,\%)$ & $\lesssim 32.5\,\%$ $(99\,\%)$ & $\gtrsim 100.0\,\%$ $(27\,\%)$ & $\gtrsim 100.0\,\%$ $(32\,\%)$ & $\gtrsim 100.0\,\%$ $(97\,\%)$ \\ \hline\hline
  Type & $\Delta \text{BR}_{At\bar{t}}^{\pmb{S_1}}$ & $\Delta \text{BR}_{At\bar{t}}^{\pmb{S_2}}$ & $\Delta \text{BR}_{At\bar{t}}^{\pmb{S_3}}$ & $\Delta \text{BR}_{At\bar{t}}^{\pmb{\text{proc3}}}$ & $\Delta \text{BR}_{At\bar{t}}^{\pmb{\text{OS1}}}$ & $\Delta \text{BR}_{At\bar{t}}^{\pmb{\overline{\text{MS}}}}$ \\ \hline
  I  & $\lesssim 7.5\,\%$ $( 82\,\%)$  & $\lesssim 15.0\,\%$ $(46\,\%)$ & $\lesssim 12.5\,\%$ $(54\,\%)$ & $\lesssim 90.0\,\%$ $(50\,\%)$ & $\lesssim 37.5\,\%$ $(50\,\%)$ & $\lesssim 90.0\,\%$ $(17\,\%)$  \\  
     & $\lesssim 12.5\,\%$ $( 99\,\%)$  & $\lesssim 27.5\,\%$ $(97\,\%)$ & $\lesssim 22.5\,\%$ $(97\,\%)$ & $\gtrsim 100.0\,\%$ $(47\,\%)$ & $\gtrsim 100.0\,\%$ $(24\,\%)$ & $\gtrsim 100.0\,\%$ $(82\,\%)$ \\ \hdashline
  II & $\lesssim 10.0\,\%$ $( 84\,\%)$  & $\lesssim 10.0\,\%$ $(87\,\%)$ & $\lesssim 17.5\,\%$ $(51\,\%)$ & $\lesssim 5.0\,\%$ $(82\,\%)$ & $\lesssim 60.0\,\%$ $(50\,\%)$ & $\lesssim 90.0\,\%$ $(12\,\%)$  \\  
     & $\lesssim 12.5\,\%$ $( 99\,\%)$  & $\lesssim 12.5\,\%$ $(100\,\%)$ & $\lesssim 25.0\,\%$ $(97\,\%)$ & $\lesssim 10.0\,\%$ $(100\,\%)$ & $\gtrsim 100.0\,\%$ $(31\,\%)$ & $\gtrsim 100.0\,\%$ $(87\,\%)$ \\ \hdashline
  LS & $\lesssim 10.0\,\%$ $( 91\,\%)$  & $\lesssim 10.0\,\%$ $(85\,\%)$ & $\lesssim 15.0\,\%$ $(65\,\%)$ & $\lesssim 5.0\,\%$ $(55\,\%)$ & $\lesssim 60.0\,\%$ $(50\,\%)$ & $\lesssim 95.0\,\%$ $(7\,\%)$  \\  
     & $\lesssim 12.5\,\%$ $( 99\,\%)$  & $\lesssim 15.0\,\%$ $(97\,\%)$ & $\lesssim 25.0\,\%$ $(98\,\%)$ & $\lesssim 15.0\,\%$ $(98\,\%)$ & $\gtrsim 100.0\,\%$ $(29\,\%)$ & $\gtrsim 100.0\,\%$ $(89\,\%)$ \\ \hdashline
  FL & $\lesssim 10.0\,\%$ $( 86\,\%)$  & $\lesssim 25.0\,\%$ $(49\,\%)$ & $\lesssim 20.0\,\%$ $(61\,\%)$ & $\lesssim 67.5\,\%$ $(50\,\%)$ & $\lesssim 75.0\,\%$ $(50\,\%)$ & $\lesssim 90.0\,\%$ $(7\,\%)$  \\  
     & $\lesssim 12.5\,\%$ $( 99\,\%)$  & $\lesssim 30.0\,\%$ $( 98\,\%)$ & $\lesssim 25.0\,\%$ $(97\,\%)$ & $\gtrsim 100.0\,\%$ $(25\,\%)$ & $\gtrsim 100.0\,\%$ $(31\,\%)$ & $\gtrsim 100.0\,\%$ $(92\,\%)$ \\ \hline\hline
  Type & $\Delta \text{BR}_{A\tau ^+ \tau ^-}^{\pmb{S_1}}$ & $\Delta \text{BR}_{A\tau ^+ \tau ^-}^{\pmb{S_2}}$ & $\Delta \text{BR}_{A\tau ^+ \tau ^-}^{\pmb{S_3}}$ & $\Delta \text{BR}_{A\tau ^+ \tau ^-}^{\pmb{\text{proc3}}}$ & $\Delta \text{BR}_{A\tau ^+ \tau ^-}^{\pmb{\text{OS1}}}$ & $\Delta \text{BR}_{A\tau ^+ \tau ^-}^{\pmb{\overline{\text{MS}}}}$ \\ \hline
  I  & $\lesssim 10.0\,\%$ $(62\,\%)$  & $\lesssim 5.0\,\%$ $(94\,\%)$ & $\lesssim 5.0\,\%$ $(92\,\%)$ & $\lesssim 90.0\,\%$ $(40\,\%)$ & $\lesssim 50.0\,\%$ $(50\,\%)$ & $\lesssim 90.0\,\%$ $(22\,\%)$  \\  
     & $\lesssim 25.0\,\%$ $(99\,\%)$  & $\lesssim 7.5\,\%$ $(99\,\%)$ & $\lesssim 12.5\,\%$ $(100\,\%)$ & $\gtrsim 100.0\,\%$ $(57\,\%)$ & $\gtrsim 100.0\,\%$ $(29\,\%)$ & $\gtrsim 100.0\,\%$ $(77\,\%)$ \\ \hdashline
  II & $\lesssim 12.5\,\%$ $(71\,\%)$  & $\lesssim 7.5\,\%$ $(83\,\%)$ & $\lesssim 22.5\,\%$ $(53\,\%)$ & $\lesssim 2.5\,\%$ $(82\,\%)$ & $\lesssim 62.5\,\%$ $(50\,\%)$ & $\lesssim 90.0\,\%$ $(2\,\%)$  \\  
     & $\lesssim 17.5\,\%$ $(100\,\%)$  & $\lesssim 10.0\,\%$ $(100\,\%)$ & $\lesssim 35.0\,\%$ $(100\,\%)$ & $\lesssim 5.0\,\%$ $(100\,\%)$ & $\gtrsim 100.0\,\%$ $(31\,\%)$ & $\gtrsim 100.0\,\%$ $(97\,\%)$ \\ \hdashline
  LS & $\lesssim 10.0\,\%$ $(53\,\%)$  & $\lesssim 5.0\,\%$ $(74\,\%)$ & $\lesssim 17.5\,\%$ $(50\,\%)$ & $\lesssim 2.5\,\%$ $(82\,\%)$ & $\lesssim 60.0\,\%$ $(50\,\%)$ & $\lesssim 90.0\,\%$ $(21\,\%)$  \\  
     & $\lesssim 17.5\,\%$ $(100\,\%)$  & $\lesssim 12.5\,\%$ $(100\,\%)$ & $\lesssim 30.0\,\%$ $(97\,\%)$ & $\lesssim 7.5\,\%$ $(99\,\%)$ & $\gtrsim 100.0\,\%$ $(34\,\%)$ & $\gtrsim 100.0\,\%$ $(78\,\%)$ \\ \hdashline
  FL & $\lesssim 12.5\,\%$ $(67\,\%)$  & $\lesssim 5.0\,\%$ $(50\,\%)$ & $\lesssim 5.0\,\%$ $(64\,\%)$ & $\lesssim 90.0\,\%$ $(77\,\%)$ & $\lesssim 90.0\,\%$ $(46\,\%)$ & $\lesssim 90.0\,\%$ $(8\,\%)$  \\  
     & $\lesssim 20.0\,\%$ $(98\,\%)$  & $\lesssim 12.5\,\%$ $(100\,\%)$ & $\lesssim 10.0\,\%$ $(100\,\%)$ & $\gtrsim 100.0\,\%$ $(22\,\%)$ & $\gtrsim 100.0\,\%$ $(44\,\%)$ & $\gtrsim 100.0\,\%$ $(91\,\%)$ \\ \hline\hline
  Type & $\Delta \text{BR}_{AZh}^{\pmb{S_1}}$ & $\Delta \text{BR}_{AZh}^{\pmb{S_2}}$ & $\Delta \text{BR}_{AZh}^{\pmb{S_3}}$ & $\Delta \text{BR}_{AZh}^{\pmb{\text{proc3}}}$ & $\Delta \text{BR}_{AZh}^{\pmb{\text{OS1}}}$ & $\Delta \text{BR}_{AZh}^{\pmb{\overline{\text{MS}}}}$ \\ \hline
  I  & $\lesssim 2.5\,\%$ $(100\,\%)$  & $\lesssim 2.5\,\%$ $(95\,\%)$ & $\lesssim 2.5\,\%$ $(99\,\%)$ & $\lesssim 2.5\,\%$ $(79\,\%)$ & $\lesssim 2.5\,\%$ $(95\,\%)$ & $\lesssim 90.0\,\%$ $(33\,\%)$  \\  
     & & $\lesssim 5.0\,\%$ $(97\,\%)$ & $\lesssim 5.0\,\%$ $(100\,\%)$ & $\lesssim 5.0\,\%$ $(97\,\%)$ & $\lesssim 5.0\,\%$ $(100\,\%)$ & $\gtrsim 100.0\,\%$ $(65\,\%)$ \\ \hdashline
  II & $\lesssim 2.5\,\%$ $(100\,\%)$  & $\lesssim 2.5\,\%$ $(100\,\%)$ & $\lesssim 2.5\,\%$ $(95\,\%)$ & $\lesssim 2.5\,\%$ $(100\,\%)$ & $\lesssim 2.5\,\%$ $(37\,\%)$ & $\lesssim 90.0\,\%$ $(4\,\%)$  \\  
     & & & $\lesssim 5.0\,\%$ $(100\,\%)$ & & $\lesssim 5.0\,\%$ $(100\,\%)$ & $\gtrsim 100.0\,\%$ $(95\,\%)$ \\ \hdashline
  LS & $\lesssim 2.5\,\%$ $(96\,\%)$  & $\lesssim 2.5\,\%$ $(100\,\%)$ & $\lesssim 2.5\,\%$ $(92\,\%)$ & $\lesssim 2.5\,\%$ $(100\,\%)$ & $\lesssim 2.5\,\%$ $(50\,\%)$ & $\lesssim 90.0\,\%$ $(10\,\%)$  \\  
     & $\lesssim 5.0\,\%$ $(98\,\%)$  & & $\lesssim 5.0\,\%$ $(97\,\%)$ & & $\lesssim 10.0\,\%$ $(90\,\%)$ & $\gtrsim 100.0\,\%$ $(83\,\%)$ \\ \hdashline
  FL & $\lesssim 2.5\,\%$ $(100\,\%)$  & $\lesssim 2.5\,\%$ $(92\,\%)$ & $\lesssim 2.5\,\%$ $(97\,\%)$ & $\lesssim 2.5\,\%$ $(48\,\%)$ & $\lesssim 2.5\,\%$ $(40\,\%)$ & $\lesssim 90.0\,\%$ $(6\,\%)$  \\  
     & & $\lesssim 5.0\,\%$ $(99\,\%)$ & $\lesssim 5.0\,\%$ $(100\,\%)$ & $\lesssim 5.0\,\%$ $(98\,\%)$ & $\lesssim 5.0\,\%$ $(98\,\%)$ & $\gtrsim 100.0\,\%$ $(92\,\%)$ \\ \hline\hline
  Type & $\Delta \text{BR}_{AZH}^{\pmb{S_1}}$ & $\Delta \text{BR}_{AZH}^{\pmb{S_2}}$ & $\Delta \text{BR}_{AZH}^{\pmb{S_3}}$ & $\Delta \text{BR}_{AZH}^{\pmb{\text{proc3}}}$ & $\Delta \text{BR}_{AZH}^{\pmb{\text{OS1}}}$ & $\Delta \text{BR}_{AZH}^{\pmb{\overline{\text{MS}}}}$ \\ \hline
  I  & $\lesssim 10.0\,\%$ $(53\,\%)$  & $\lesssim 35.0\,\%$ $(30\,\%)$ & $\lesssim 27.5\,\%$ $(48\,\%)$ & $\lesssim 82.5\,\%$ $(3\,\%)$ & $\lesssim 47.5\,\%$ $(20\,\%)$ & $\lesssim 90.0\,\%$ $(5\,\%)$  \\  
     & $\gtrsim 100.0\,\%$ $(6\,\%)$  & $\gtrsim 100.0\,\%$ $(54\,\%)$ & $\gtrsim 100.0\,\%$ $(15\,\%)$ & $\gtrsim 100.0\,\%$ $(96\,\%)$ & $\gtrsim 100.0\,\%$ $(68\,\%)$ & $\gtrsim 100.0\,\%$ $(94\,\%)$ \\ \hdashline
  II & $\lesssim 25.0\,\%$ $(51\,\%)$ & $\lesssim 57.5\,\%$ $(10\,\%)$ & $\lesssim 22.5\,\%$ $(10\,\%)$ & $\lesssim 72.5\,\%$ $(40\,\%)$ & $\gtrsim 100.0,\%$ $(100\,\%)$ & $\lesssim 90.0\,\%$ $(10\,\%)$  \\  
     & $\gtrsim 100.0\,\%$ $(22\,\%)$ & $\gtrsim 100.0\,\%$ $(74\,\%)$ & $\gtrsim 100.0\,\%$ $(52\,\%)$ &                               $\gtrsim 100.0\,\%$ $(48\,\%)$ & & $\gtrsim 100.0\,\%$ $(89\,\%)$ \\ \hdashline
  LS & $\lesssim 30.0\,\%$ $(36\,\%)$  & $\lesssim 67.5\,\%$ $(10\,\%)$ & $\lesssim 70.0\,\%$ $(21\,\%)$ & $\lesssim 57.5\,\%$ $(12\,\%)$ & $\lesssim 95.0\,\%$ $(5\,\%)$ & $\lesssim 90.0\,\%$ $(1\,\%)$  \\  
     & $\gtrsim 100.0\,\%$ $(34\,\%)$  & $\gtrsim 100.0\,\%$ $(79\,\%)$ & $\gtrsim 100.0\,\%$ $(69\,\%)$ & $\gtrsim 100.0\,\%$ $(61\,\%)$ & $\gtrsim 100.0\,\%$ $(94\,\%)$ & $\gtrsim 100.0\,\%$ $(98\,\%)$ \\ \hdashline
  FL & $\lesssim 27.5\,\%$ $(49\,\%)$  & $\lesssim 75.0\,\%$ $(2\,\%)$ & $\lesssim 72.5\,\%$ $(30\,\%)$ & $\gtrsim 100.0\,\%$ $(100\,\%)$ & $\gtrsim 100.0\,\%$ $(100\,\%)$ & $\lesssim 92.5\,\%$ $(7\,\%)$  \\  
     & $\gtrsim 100.0\,\%$ $(16\,\%)$ & $\gtrsim 100.0\,\%$ $(94\,\%)$ & $\gtrsim 100.0\,\%$ $(62\,\%)$ &  &  & $\gtrsim 100.0\,\%$ $(92\,\%)$ \\ \hline
 \end{tabular}
\caption{Relative size of the EW corrections to the BRs for the
  pseudoscalar 2HDM Higgs boson $A$ decays into $b\bar{b}$, $\tau^+
  \tau^-$, $t\bar{t}$, $Zh$, and $ZH$, in the four 2HDM types I, II, LS and
FL ($H$ is SM-like). \label{tab:heavierpseudoscalar}}  
\end{center}
\vspace*{-0.5cm}
\end{table}

As can be inferred from Tab.~\ref{tab:heavierpseudoscalar}, the results for the
proc3 and OS1 schemes are typically very large for all
  decays and hence indicate numerical instability of these
  renormalization schemes and for the considered decay channels.
The results computed within the other schemes are typically of medium
size in the fermionic final states. The BRs into
$Zh$ are more important now. Since $H$ is SM-like, we typically have
$c_{\beta-\alpha}$ close to 1 so that the decay into $Zh$, whose
EW corrections are proportional to $c_{\beta-\alpha}^2$, is
not suppressed, while the BR for the decay into $ZH$ is typically very
small. This is reflected in the relative corrections of the BRs which
for the $Zh$ final state are the smallest among all decays whereas for
the decay into $ZH$, they become very large. In case that the decay into
$Zh$ is the dominant one, the LO branching ratio into $t\bar{t}$ is reduced
so that the relative corrections here can become more important than
in the previous case where $h$ is SM-like. 

%%%%%%%%%%%%%%%%%%%%%%%%%%%%%%%%%%%%%%%%%%%%%%%%%%%%%%%
\subsection{N2HDM Higgs Decays}
\label{sec:N2HDMHiggsDecays}
%%%%%%%%%%%%%%%%%%%%%%%%%%%%%%%%%%%%%%%%%%%%%%%%%%%%%%%
In the N2HDM, all three CP-even Higgs bosons can in principle be
SM-like. Although the case where the heaviest one, $H_3$, is SM-like
is not completely excluded yet, it is strongly disfavored. For type I
we only found 25 valid parameter points, and for the other three types
none.\footnote{This does not necessarily mean that this case is
  excluded, but it would require a dedicated scan to find valid
  scenarios.} As the statistics would be very low, we decided not to present 
the results for this case. Instead, we focus on the scenarios where
$H_1$ or $H_2$ are SM-like. In the former case the number of valid
scenarios are as follows:
\beq
\begin{array}{llll}
\mbox{Type I:} & \; 262\,332 & \qquad \mbox{Type II:} & \; 299\,959 \\[0.2cm]
\mbox{Type LS:} & \; 283\,234 & \qquad \mbox{Type FL:} & \; 292\,634 ~.
\end{array}
\label{eq:numberslightern2hdm}
\eeq 
In case that $H_2$ is SM-like, the total amount of points that can be used for the analysis is given by:
\beq
\begin{array}{llll}
\mbox{Type I:} & \; 8540 & \qquad \mbox{Type II:} & \; 2381 \\[0.2cm]
\mbox{Type LS:} & \; 3303 & \qquad \mbox{Type FL:} & \; 1562 ~.
\end{array}
\label{eq:numbersheaviern2hdm}
\eeq 
The scheme sets shown in the following in all tables for the N2HDM are
defined as:
\begin{align}
	\pmb{S_1} &\equiv \big\{ ~ \text{KOSY}^o ~,~ \text{KOSY}^c ~,~ \text{pOS}^o ~,~ \text{pOS}^c ~ \big\} \\
	\pmb{S_2} &\equiv \big\{ ~ \text{p}_*^o ~,~ \text{p}_*^c ~ \big\} \\
	\pmb{\overline{\text{MS}}} &\equiv \big\{ ~ \overline{\text{MS}} ~ \big\}
\end{align}

%%%%%%%%%%%%%%%%%%%%%%%%%%%%%%%%%%%%%%%%%%%%%%%%%%%%%%%
\subsection{SM-like N2HDM Higgs Decays}
%%%%%%%%%%%%%%%%%%%%%%%%%%%%%%%%%%%%%%%%%%%%%%%%%%%%%%%
We again start with the presentation of the SM-like Higgs boson
results before moving on to the corrections to the BRs of the
non-SM-like CP-even and CP-odd Higgs bosons.

%%%%%%%%%%%%%%%%%%%%%%%%%%%%%%%%%%%%%%%%%%%%%%%%%%%%%%%
\subsubsection{$H_1$ is the SM-like Higgs Boson}
%%%%%%%%%%%%%%%%%%%%%%%%%%%%%%%%%%%%%%%%%%%%%%%%%%%%%%%
\begin{table}[t!]
\begin{center}
\renewcommand{\arraystretch}{1.2}
\scriptsize
\begin{tabular}{|c|c c c|}
  \hline
  Type & $\Delta \text{BR}_{H_1b\bar{b}}^{\pmb{S_1}}$ & $\Delta \text{BR}_{H_1b\bar{b}}^{\pmb{S_2}}$ & $\Delta \text{BR}_{H_1b\bar{b}}^{\pmb{\overline{\text{MS}}}}$ \\ \hline
  I  & $\lesssim 2.5\,\%$ $(99\,\%)$ & $\lesssim 2.5\,\%$ $(99\,\%)$ & $\lesssim 17.5\,\%$ $(50\,\%)$ \\  
     & & & $\gtrsim 100.0\,\%$ $(21\,\%)$ \\ \hdashline
  II & $\lesssim 2.5\,\%$ $(99\,\%)$ & $\lesssim 2.5\,\%$ $(99\,\%)$ & $\lesssim 90.0\,\%$ $(50\,\%)$ \\  
     & & & $\gtrsim 100.0\,\%$ $(42\,\%)$ \\ \hdashline
  LS & $\lesssim 2.5\,\%$ $(99\,\%)$ & $\lesssim 2.5\,\%$ $(99\,\%)$ & $\lesssim 25.0\,\%$ $(22\,\%)$ \\  
     & & & $\gtrsim 100.0\,\%$ $(\,\%)$ \\ \hdashline
  FL & $\lesssim 2.5\,\%$ $(99\,\%)$ & $\lesssim 2.5\,\%$ $(99\,\%)$ & $\lesssim 90.0\,\%$ $(45\,\%)$ \\  
     & & & $\gtrsim 100.0\,\%$ $(52\,\%)$ \\ \hline\hline
  Type & $\Delta \text{BR}_{H_1\gamma \gamma / H_1ZZ}^{\pmb{S_1}}$ & $\Delta \text{BR}_{H_1\gamma \gamma / H_1ZZ}^{\pmb{S_2}}$ & $\Delta \text{BR}_{H_1\gamma \gamma / H_1ZZ}^{\pmb{\overline{\text{MS}}}}$ \\ \hline
  I  & $\lesssim 2.5\,\%$ $(1\,\%)$  & $\lesssim 2.5\,\%$ $(3\,\%)$ & $\lesssim 35.0\,\%$ $(50\,\%)$ \\  
     & $\lesssim 5.0\,\%$ $(99\,\%)$  & $\lesssim 5.0\,\%$ $(99\,\%)$ & $\gtrsim 100.0\,\%$ $(33\,\%)$ \\ \hdashline
  II & $\lesssim 2.5\,\%$ $(66\,\%)$  & $\lesssim 2.5\,\%$ $(46\,\%)$ & $\lesssim 90.0\,\%$ $(34\,\%)$ \\  
     & $\lesssim 5.0\,\%$ $(99\,\%)$  & $\lesssim 5.0\,\%$ $(99\,\%)$ & $\gtrsim 100.0\,\%$ $(65\,\%)$ \\ \hdashline
  LS & $\lesssim 2.5\,\%$ $(1\,\%)$  & $\lesssim 2.5\,\%$ $(1\,\%)$ & $\lesssim 32.5\,\%$ $(50\,\%)$ \\  
     & $\lesssim 5.0\,\%$ $(99\,\%)$  & $\lesssim 5.0\,\%$ $(99\,\%)$ & $\gtrsim 100.0\,\%$ $(30\,\%)$ \\ \hdashline
  FL & $\lesssim 2.5\,\%$ $(60\,\%)$  & $\lesssim 2.5\,\%$ $(40\,\%)$ & $\lesssim 90.0\,\%$ $(38\,\%)$ \\  
     & $\lesssim 5.0\,\%$ $(99\,\%)$  & $\lesssim 5.0\,\%$ $(99\,\%)$ & $\gtrsim 100.0\,\%$ $(61\,\%)$ \\ \hline\hline
  Type & $\Delta \text{BR}_{H_1\tau ^+ \tau ^-}^{\pmb{S_1}}$ & $\Delta \text{BR}_{H_1\tau ^+ \tau ^-}^{\pmb{S_2}}$ & $\Delta \text{BR}_{H_1\tau ^+ \tau ^-}^{\pmb{\overline{\text{MS}}}}$ \\ \hline
  I  & $\lesssim 2.5\,\%$ $(99\,\%)$  & $\lesssim 2.5\,\%$ $(99\,\%)$ & $\lesssim 17.5\,\%$ $(50\,\%)$ \\  
     & & & $\gtrsim 100.0\,\%$ $(21\,\%)$ \\ \hdashline
  II & $\lesssim 2.5\,\%$ $(98\,\%)$  & $\lesssim 2.5\,\%$ $(93\,\%)$ & $\lesssim 90.0\,\%$ $(50\,\%)$ \\  
     & $\lesssim 5.0\,\%$ $(99\,\%)$  & $\lesssim 2.5\,\%$ $(99\,\%)$ & $\gtrsim 100.0\,\%$ $(45\,\%)$ \\ \hdashline
  LS & $\lesssim 2.5\,\%$ $(94\,\%)$  & $\lesssim 2.5\,\%$ $(73\,\%)$ & $\lesssim 90.0\,\%$ $(37\,\%)$ \\  
     & $\lesssim 5.0\,\%$ $(99\,\%)$  & $\lesssim 7.5\,\%$ $(99\,\%)$ & $\gtrsim 100.0\,\%$ $(61\,\%)$ \\ \hdashline
  FL & $\lesssim 2.5\,\%$ $(82\,\%)$  & $\lesssim 2.5\,\%$ $(93\,\%)$ & $\lesssim 90.0\,\%$ $(47\,\%)$ \\  
     & $\lesssim 5.0\,\%$ $(99\,\%)$  & $\lesssim 5.0\,\%$ $(99\,\%)$ & $\gtrsim 100.0\,\%$ $(49\,\%)$ \\ \hline
 \end{tabular}
\caption{Relative size of the EW corrections to the BRs of the N2HDM 
  SM-like $H_1$ in the four N2HDM types I, II, LS and FL. \label{tab:N2HDMH1}}
\vspace*{-0.4cm}
\end{center}
\end{table}
In Tab.~\ref{tab:N2HDMH1}, the relative corrections $\Delta \mbox{BR}$ for the decays of
the SM-like Higgs boson $H_1$ into $b\bar{b}$, $\tau^+\tau^-$,
$\gamma\gamma$ and $ZZ$ final states are shown. 
The number of points used for the analysis has been given in
Eq.~(\ref{eq:numberslightern2hdm}).
Note, however, that the decays into $ZZ$ are always
off-shell while the process $H_1 \to \gamma\gamma$ is loop-induced
already at tree level. Therefore, as in the 2HDM, no EW corrections to
$H_1\to ZZ$ and $H_1 \to \gamma\gamma$ 
are included and the relative corrections to the BRs are the same for
both decay channels. \s 

From the table we read off that the $\Delta \text{BR}$ for both scheme
sets $\pmb{S_1}$ and $\pmb{S_2}$ for all four
  N2HDM types are typically very small, with the bulk of points
  resulting in corrections below $2.5\,\%$ and $5.0\,\%$, indicating
  both small EW corrections as well as numerical stability of
  the schemes. 

%%%%%%%%%%%%%%%%%%%%%%%%%%%%%%%%%%%%%%%%%%%%%%%%%%%%%%%
\subsubsection{$H_2$ is the SM-like Higgs Boson}
%%%%%%%%%%%%%%%%%%%%%%%%%%%%%%%%%%%%%%%%%%%%%%%%%%%%%%%
\begin{table}[t!]
\begin{center}
\renewcommand{\arraystretch}{1.2}
\scriptsize
 \begin{tabular}{|c|c c c||c c c|}
  \hline
  Type & $\Delta \text{BR}_{H_2b\bar{b}}^{\pmb{S_1}}$ & $\Delta \text{BR}_{H_2b\bar{b}}^{\pmb{S_2}}$ & $\Delta \text{BR}_{H_2b\bar{b}}^{\pmb{\overline{\text{MS}}}}$ & $\Delta \text{BR}_{H_2\gamma \gamma / H_2ZZ}^{\pmb{S_1}}$ & $\Delta \text{BR}_{H_2\gamma \gamma / H_2ZZ}^{\pmb{S_2}}$ & $\Delta \text{BR}_{H_2\gamma \gamma / H_2ZZ}^{\pmb{\overline{\text{MS}}}}$ \\ \hline
  I  & $\lesssim 2.5\,\%$ $(97\,\%)$  & $\lesssim 2.5\,\%$ $(98\,\%)$ & $\lesssim 32.5\,\%$ $(50\,\%)$ & $\lesssim 5.0\,\%$ $(97\,\%)$  & $\lesssim 5.0\,\%$ $(98\,\%)$ & $\lesssim 70.0\,\%$ $(50\,\%)$ \\  
     & $\lesssim 5.0\,\%$ $(99\,\%)$  & $\lesssim 5.0\,\%$ $(99\,\%)$ & $\gtrsim 100.0\,\%$ $(7\,\%)$ & $\lesssim 7.5\,\%$ $(99\,\%)$  & $\lesssim 7.5\,\%$ $(99\,\%)$ & $\gtrsim 100.0\,\%$ $(26\,\%)$ \\ \hdashline
  II & $\lesssim 2.5\,\%$ $(97\,\%)$  & $\lesssim 2.5\,\%$ $(97\,\%)$ & $\lesssim 37.5\,\%$ $(50\,\%)$ & $\lesssim 2.5\,\%$ $(65\,\%)$  & $\lesssim 2.5\,\%$ $(42\,\%)$ & $\lesssim 75.0\,\%$ $(50\,\%)$ \\  
     & $\lesssim 7.5\,\%$ $(99\,\%)$  & $\lesssim 7.5\,\%$ $(99\,\%)$ & $\gtrsim 100.0\,\%$ $(10\,\%)$ & $\lesssim 7.5\,\%$ $(99\,\%)$  & $\lesssim 7.5\,\%$ $(99\,\%)$ & $\gtrsim 100.0\,\%$ $(14\,\%)$ \\ \hdashline
  LS & $\lesssim 2.5\,\%$ $(97\,\%)$  & $\lesssim 2.5\,\%$ $(96\,\%)$ & $\lesssim 27.5\,\%$ $(51\,\%)$ & $\lesssim 5.0\,\%$ $(96\,\%)$  & $\lesssim 5.0\,\%$ $(96\,\%)$ & $\lesssim 47.5\,\%$ $(50\,\%)$ \\  
     & $\lesssim 7.5\,\%$ $(99\,\%)$  & $\lesssim 5.0\,\%$ $(98\,\%)$ & $\gtrsim 100.0\,\%$ $(18\,\%)$ & $\lesssim 7.5\,\%$ $(98\,\%)$  & $\lesssim 10.0\,\%$ $(98\,\%)$ & $\gtrsim 100.0\,\%$ $(13\,\%)$ \\ \hdashline
  FL & $\lesssim 2.5\,\%$ $(95\,\%)$  & $\lesssim 2.5\,\%$ $(95\,\%)$ & $\lesssim 37.5\,\%$ $(51\,\%)$ & $\lesssim 2.5\,\%$ $(61\,\%)$  & $\lesssim 2.5\,\%$ $(31\,\%)$ & $\lesssim 72.5\,\%$ $(49\,\%)$ \\  
     & $\lesssim 7.5\,\%$ $(98\,\%)$  & $\lesssim 7.5\,\%$ $(98\,\%)$ & $\gtrsim 100.0\,\%$ $(10\,\%)$ & $\lesssim 5.0\,\%$ $(97\,\%)$  & $\lesssim 5.0\,\%$ $(97\,\%)$ & $\gtrsim 100.0\,\%$ $(10\,\%)$ \\ \hline\hline
  Type & $\Delta \text{BR}_{H_2\tau ^+ \tau ^-}^{\pmb{S_1}}$ & $\Delta \text{BR}_{H_2\tau ^+ \tau ^-}^{\pmb{S_2}}$ & $\Delta \text{BR}_{H_2\tau ^+ \tau ^-}^{\pmb{\overline{\text{MS}}}}$ & $\Delta \text{BR}_{H_2H_1H_1}^{\pmb{S_1}}$ & $\Delta \text{BR}_{H_2H_1H_1}^{\pmb{S_2}}$ & $\Delta \text{BR}_{H_2H_1H_1}^{\pmb{\overline{\text{MS}}}}$ \\ \hline
  I  & $\lesssim 2.5\,\%$ $(98\,\%)$  & $\lesssim 2.5\,\%$ $(98\,\%)$ & $\lesssim 32.5\,\%$ $(50\,\%)$ & $\lesssim 82.5\,\%$ $(39\,\%)$  & $\lesssim 77.5\,\%$ $(40\,\%)$ & $\lesssim 85.0\,\%$ $(4\,\%)$ \\  
     & $\lesssim 5.0\,\%$ $(99\,\%)$  & $\lesssim 5.0\,\%$ $(99\,\%)$ & $\gtrsim 100.0\,\%$ $(7\,\%)$ & $\gtrsim 100.0\,\%$ $(60\,\%)$  & $\gtrsim 100.0\,\%$ $(59\,\%)$ & $\gtrsim 100.0\,\%$ $(95\,\%)$ \\ \hdashline
  II & $\lesssim 2.5\,\%$ $(93\,\%)$  & $\lesssim 2.5\,\%$ $(83\,\%)$ & $\lesssim 37.5\,\%$ $(50\,\%)$ & $\lesssim 85.0\,\%$ $(35\,\%)$  & $\lesssim 90.0\,\%$ $(35\,\%)$ & $\lesssim 90.0\,\%$ $(1\,\%)$ \\  
     & $\lesssim 5.0\,\%$ $(98\,\%)$  & $\lesssim 5.0\,\%$ $(98\,\%)$ & $\gtrsim 100.0\,\%$ $(10\,\%)$ & $\gtrsim 100.0\,\%$ $(64\,\%)$  & $\gtrsim 100.0\,\%$ $(64\,\%)$ & $\gtrsim 100.0\,\%$ $(98\,\%)$ \\ \hdashline
  LS & $\lesssim 2.5\,\%$ $(67\,\%)$  & $\lesssim 5.0\,\%$ $(79\,\%)$ & $\lesssim 92.5\,\%$ $(50\,\%)$ & $\lesssim 87.5\,\%$ $(40\,\%)$  & $\lesssim 77.5\,\%$ $(40\,\%)$ & $\lesssim 87.5\,\%$ $(2\,\%)$ \\  
     & $\lesssim 7.5\,\%$ $(94\,\%)$  & $\lesssim 10.0\,\%$ $(94\,\%)$ & $\gtrsim 100.0\,\%$ $(48\,\%)$ & $\gtrsim 100.0\,\%$ $(57\,\%)$  & $\gtrsim 100.0\,\%$ $(54\,\%)$ & $\gtrsim 100.0\,\%$ $(97\,\%)$ \\ \hdashline
  FL & $\lesssim 2.5\,\%$ $(67\,\%)$  & $\lesssim 2.5\,\%$ $(86\,\%)$ & $\lesssim 50.0\,\%$ $(50\,\%)$ & $\lesssim 85.0\,\%$ $(35\,\%)$  & $\lesssim 87.5\,\%$ $(37\,\%)$ & $\gtrsim 100.0\,\%$ $(100\,\%)$ \\  
     & $\lesssim 7.5\,\%$ $(98\,\%)$  & $\lesssim 5.0\,\%$ $(97\,\%)$ & $\gtrsim 100.0\,\%$ $(26\,\%)$ & $\gtrsim 100.0\,\%$ $(60\,\%)$  & $\gtrsim 100.0\,\%$ $(60\,\%)$ &  \\ \hline
 \end{tabular}
\caption{Relative size of the EW corrections to the BRs of the N2HDM
  SM-like $H_2$ in the four N2HDM types I, II, LS and FL. \label{tab:N2HDMH2}}
\end{center}
\end{table}
For the case that $H_2$ is SM-like, we display in
Tab.~\ref{tab:N2HDMH2} the relative EW corrections to the BRs into
$b\bar{b}$, $\tau^+\tau^-$,  $\gamma\gamma$, $ZZ$ as well as into
$H_1H_1$ since this channel can be OS in these
scenarios. However, the additional OS condition $m_{H_2} \ge 2 m_{H_1}$ reduces the number
of parameters points available for this particular channel as follows:
\beq
\begin{array}{llll}
\mbox{Type I:} & \; 204 & \qquad \mbox{Type II:} & \; 146 \\[0.2cm]
\mbox{Type LS:} & \; 181 & \qquad \mbox{Type FL:} & \; 133  ~.
\end{array}
\label{eq:nnumberslightern2hdm}
\nonumber
\eeq 
From the table, we conclude the following:
\begin{itemize}
\item[--] The $\Delta \text{BR}$ for $H_2\gamma \gamma$ and $H_2ZZ$ are
  again the same as they do not receive EW corrections, since the
  former is loop-induced and the latter off-shell.
\item[--] The $\Delta \text{BR}$ for both scheme sets $\pmb{S_1}$ and
  $\pmb{S_2}$ for the fermionic decays for all four N2HDM types are
  typically very small, with the bulk of points resulting in
  corrections below $2.5\,\%$ and $5.0\,\%$, indicating both small
  electroweak corrections as well as numerical stability of the
  schemes. 
\item[--] For $\pmb{S_2}$, the corrections are slightly increased compared
  to $\pmb{S_1}$, \textit{i.e.} the bulk of points is slightly shifted
  towards larger corrections. Overall, the corrections are still very
  moderate, however. 
\item[--] For the decays into $\gamma \gamma $ and $ZZ$, the relative
  corrections are slightly increased compared to the case where $H_1$
  is SM-like. But since they do not receive
  electroweak corrections, this is an indirect effect through the
  corrections of the other decay channels. 
\item[--] The relative corrections to the Higgs-to-Higgs decay $H_2
  \to H_1H_1$ are very large. On the one hand this is due to
  non-decoupling effects in scenarios where the quartic couplings of
  the potential and hence the trilinear Higgs self-couplings become large. For a detailed dicussion, see
  \cite{Krause:2016xku,Krause:2017mal,Kanemura:2002vm,Kanemura:2004mg,Braathen:2019pxr,Braathen:2019zoh}. 
The other reason for the large relative corrections are very small
tree-level BRs. Due to the LHC constraints on the SM Higgs rates
$\mbox{BR}(H_2 \to H_1 H_1)$ must be small so that the $H_2$ BRs into the other SM final states
remain SM-like.
\end{itemize}

%%%%%%%%%%%%%%%%%%%%%%%%%%%%%%%%%%%%%%%%%%%%%%%%%%%%%%%
\subsection{Non-SM-Like CP-even N2HDM Higgs Decays}
%%%%%%%%%%%%%%%%%%%%%%%%%%%%%%%%%%%%%%%%%%%%%%%%%%%%%%%
We continue with the analysis of the relative EW corrections to the
BRs of the N2HDM non-SM-like CP-even Higgs bosons. Due to the extended
Higgs sector we now have two non-SM-like CP-even Higgs bosons to be
analysed in contrast to just one as in the 2HDM. 

%%%%%%%%%%%%%%%%%%%%%%%%%%%%%%%%%%%%%%%%%%%%%%%%%%%%%%%
\subsubsection{$H_1$ is the SM-like Higgs Boson}
%%%%%%%%%%%%%%%%%%%%%%%%%%%%%%%%%%%%%%%%%%%%%%%%%%%%%%%
We start with the case where $H_1$ is SM-like and investigate the
decays of the non-SM-like Higgs bosons $H_2$ and $H_3$. 

%%%%%%%%%%%%%%%%%%%%%%%%%%%%%%%%%%%%%%%%%%%%%%%%%%%%%%%
\paragraph{$H_2$ Decays:}
%%%%%%%%%%%%%%%%%%%%%%%%%%%%%%%%%%%%%%%%%%%%%%%%%%%%%%%
\begin{table}[t!]
\begin{center}
\renewcommand{\arraystretch}{1.2}
\scriptsize
 \begin{tabular}{|c|c c c c c c|}
  \hline
Type & $H_2 \to t\bar{t}$ & $H_2 \to ZA$ & $H_2 \to W^\pm H^\mp$ & $H_2 \to ZZ$ & $H_2 \to H_1 H_1$ & $H_2 \to AA$ \\
   \hline
I & $210\,527$ &  1970 & 2558 & $250\,074$ & $237\,558$ & 488 \\
II & $280\,113$ & 942 & 893 & $296\,615$ & $292\,134$ & $33\,420$ \\
LS & $255\,075$ & 498 & 592 & $277\,968$ & $271\,130$ & 134 \\
FL & $278\,944$ & 211 & 452 & $290\,381$ & $287\,446$ & 0 \\
\hline 
\end{tabular}
\caption{Number of parameter points available for the analysis in the
  individual OS decay channels of the non-SM-like Higgs boson $H_2$ ($H_1$ is SM-like).}
\label{tab:rednumbersnonsmh2}
\end{center}
\end{table}
We only show results for OS decays as only for these the EW
corrections are computed. Additional OS conditions reduce the number
of  available parameter points, \textit{cf.}
Eq.~(\ref{eq:numberslightern2hdm}), in the individual decay channels
as shown in Tab.~\ref{tab:rednumbersnonsmh2}. The relative corrections
$\Delta \mbox{BR}$ for the non-SM-like CP-even $H_2$ are displayed 
in Tab.~\ref{tab:n2hdmnonsmh2}. From the table, we observe: \s

\begin{table}[ht!]
\begin{center}
\renewcommand{\arraystretch}{1.2}
\scriptsize
 \begin{tabular}{|c|c c c||c c c|}
  \hline
  Type & $\Delta \text{BR}_{H_2b\bar{b}}^{\pmb{S_1}}$ & $\Delta \text{BR}_{H_2b\bar{b}}^{\pmb{S_2}}$ & $\Delta \text{BR}_{H_2b\bar{b}}^{\pmb{\overline{\text{MS}}}}$ & $\Delta \text{BR}_{H_2t\bar{t}}^{\pmb{S_1}}$ & $\Delta \text{BR}_{H_2t\bar{t}}^{\pmb{S_2}}$ & $\Delta \text{BR}_{H_2t\bar{t}}^{\pmb{\overline{\text{MS}}}}$ \\ \hline
  I  & $\lesssim 7.5\,\%$ $(48\,\%)$ & $\lesssim 7.5\,\%$ $(49\,\%)$ & $\lesssim 90.0\,\%$ $(50\,\%)$ & $\lesssim 5.0\,\%$ $(48\,\%)$  & $\lesssim 7.5\,\%$ $(59\,\%)$ & $\lesssim 90.0\,\%$ $(50\,\%)$ \\  
     & $\lesssim 30.0\,\%$ $(87\,\%)$ & $\lesssim 35.0\,\%$ $(89\,\%)$ & $\gtrsim 100.0\,\%$ $(48\,\%)$ & $\lesssim 25.0\,\%$ $(86\,\%)$  & $\lesssim 25.0\,\%$ $(86\,\%)$ & $\gtrsim 100.0\,\%$ $(48\,\%)$ \\ \hdashline
  II & $\lesssim 7.5\,\%$ $(46\,\%)$ & $\lesssim 12.5\,\%$ $(51\,\%)$ & $\lesssim 90.0\,\%$ $(29\,\%)$ & $\lesssim 5.0\,\%$ $(57\,\%)$  & $\lesssim 5.0\,\%$ $(57\,\%)$ & $\lesssim 95.0\,\%$ $(50\,\%)$ \\  
     & $\lesssim 25.0\,\%$ $(90\,\%)$ & $\lesssim 27.5\,\%$ $(90\,\%)$ & $\gtrsim 100.0\,\%$ $(69\,\%)$ & $\lesssim 25.0\,\%$ $(89\,\%)$  & $\lesssim 27.5\,\%$ $(90\,\%)$ & $\gtrsim 100.0\,\%$ $(49\,\%)$ \\ \hdashline
  LS & $\lesssim 5.0\,\%$ $(45\,\%)$ & $\lesssim 7.5\,\%$ $(59\,\%)$ & $\lesssim 80.0\,\%$ $(50\,\%)$ & $\lesssim 5.0\,\%$ $(57\,\%)$  & $\lesssim 5.0\,\%$ $(57\,\%)$ & $\lesssim 85.0\,\%$ $(50\,\%)$ \\  
     & $\lesssim 22.5\,\%$ $(90\,\%)$ & $\lesssim 22.5\,\%$ $(90\,\%)$ & $\gtrsim 100.0\,\%$ $(46\,\%)$ & $\lesssim 20.0\,\%$ $(90\,\%)$  & $\lesssim 20.0\,\%$ $(90\,\%)$ & $\gtrsim 100.0\,\%$ $(47\,\%)$ \\ \hdashline
  FL & $\lesssim 7.5\,\%$ $(50\,\%)$ & $\lesssim 12.5\,\%$ $(60\,\%)$ & $\lesssim 90.0\,\%$ $(25\,\%)$ & $\lesssim 5.0\,\%$ $(52\,\%)$  & $\lesssim 5.0\,\%$ $(52\,\%)$ & $\lesssim 95.0\,\%$ $(50\,\%)$ \\  
     & $\lesssim 30.0\,\%$ $(90\,\%)$ & $\lesssim 30.0\,\%$ $(90\,\%)$ & $\gtrsim 100.0\,\%$ $(72\,\%)$ & $\lesssim 30.0\,\%$ $(89\,\%)$  & $\lesssim 25.0\,\%$ $(87\,\%)$ & $\gtrsim 100.0\,\%$ $(49\,\%)$ \\ \hline\hline
  Type & $\Delta \text{BR}_{H_2\tau ^+ \tau ^-}^{\pmb{S_1}}$ & $\Delta \text{BR}_{H_2\tau ^+ \tau ^-}^{\pmb{S_2}}$ & $\Delta \text{BR}_{H_2\tau ^+ \tau ^-}^{\pmb{\overline{\text{MS}}}}$ & $\Delta \text{BR}_{H_2ZA}^{\pmb{S_1}}$ & $\Delta \text{BR}_{H_2ZA}^{\pmb{S_2}}$ & $\Delta \text{BR}_{H_2ZA}^{\pmb{\overline{\text{MS}}}}$ \\ \hline
  I  & $\lesssim 10.0\,\%$ $(51\,\%)$ & $\lesssim 10.0\,\%$ $(52\,\%)$ & $\lesssim 92.5\,\%$ $(50\,\%)$ & $\lesssim 12.5\,\%$ $(50\,\%)$  & $\lesssim 12.5\,\%$ $(50\,\%)$ & $\lesssim 90.0\,\%$ $(41\,\%)$ \\  
     & $\lesssim 35.0\,\%$ $(87\,\%)$ & $\lesssim 32.5\,\%$ $(86\,\%)$ & $\gtrsim 100.0\,\%$ $(48\,\%)$ & $\gtrsim 100.0\,\%$ $(25\,\%)$  & $\gtrsim 100.0\,\%$ $(25\,\%)$ & $\gtrsim 100.0\,\%$ $(47\,\%)$ \\ \hdashline
  II & $\lesssim 7.5\,\%$ $(50\,\%)$ & $\lesssim 12.5\,\%$ $(55\,\%)$ & $\lesssim 90.0\,\%$ $(28\,\%)$ & $\lesssim 5.0\,\%$ $(71\,\%)$  & $\lesssim 5.0\,\%$ $(46\,\%)$ & $\lesssim 90.0\,\%$ $(20\,\%)$ \\  
     & $\lesssim 25.0\,\%$ $(90\,\%)$ & $\lesssim 27.5\,\%$ $(90\,\%)$ & $\gtrsim 100.0\,\%$ $(69\,\%)$ & $\lesssim 15.0\,\%$ $(94\,\%)$  & $\lesssim 17.5\,\%$ $(95\,\%)$ & $\gtrsim 100.0\,\%$ $(57\,\%)$ \\ \hdashline
  LS & $\lesssim 10.0\,\%$ $(62\,\%)$ & $\lesssim 10.0\,\%$ $(44\,\%)$ & $\lesssim 90.0\,\%$ $(31\,\%)$ & $\lesssim 5.0\,\%$ $(57\,\%)$  & $\lesssim 5.0\,\%$ $(54\,\%)$ & $\lesssim 90.0\,\%$ $(38\,\%)$ \\  
     & $\lesssim 22.5\,\%$ $(90\,\%)$ & $\lesssim 25.0\,\%$ $(90\,\%)$ & $\gtrsim 100.0\,\%$ $(65\,\%)$ & $\lesssim 20.0\,\%$ $(83\,\%)$  & $\lesssim 25.0\,\%$ $(85\,\%)$ & $\gtrsim 100.0\,\%$ $(47\,\%)$ \\ \hdashline
  FL & $\lesssim 10.0\,\%$ $(58\,\%)$ & $\lesssim 7.5\,\%$ $(47\,\%)$ & $\lesssim 92.5\,\%$ $(50\,\%)$ & $\lesssim 10.0\,\%$ $(55\,\%)$  & $\lesssim 7.5\,\%$ $(52\,\%)$ & $\lesssim 90.0\,\%$ $(20\,\%)$ \\  
     & $\lesssim 30.0\,\%$ $(86\,\%)$ & $\lesssim 30.0\,\%$ $(86\,\%)$ & $\gtrsim 100.0\,\%$ $(48\,\%)$ & $\gtrsim 100.0\,\%$ $(16\,\%)$  & $\gtrsim 100.0\,\%$ $(16\,\%)$ & $\gtrsim 100.0\,\%$ $(45\,\%)$ \\ \hline\hline
  Type & $\Delta \text{BR}_{H_2W^\pm H^\mp}^{\pmb{S_1}}$ & $\Delta \text{BR}_{H_2W^\pm H^\mp}^{\pmb{S_2}}$ & $\Delta \text{BR}_{H_2W^\pm H^\mp}^{\pmb{\overline{\text{MS}}}}$ & $\Delta \text{BR}_{H_2ZZ}^{\pmb{S_1}}$ & $\Delta \text{BR}_{H_2ZZ}^{\pmb{S_2}}$ & $\Delta \text{BR}_{H_2ZZ}^{\pmb{\overline{\text{MS}}}}$ \\ \hline
  I  & $\lesssim 12.5\,\%$ $(41\,\%)$ & $\lesssim 10.0\,\%$ $(38\,\%)$ & $\lesssim 90.0\,\%$ $(34\,\%)$ & $\lesssim 7.5\,\%$ $(42\,\%)$  & $\lesssim 10.0\,\%$ $(43\,\%)$ & $\lesssim 50.0\,\%$ $(62.5\,\%)$ \\  
     & $\gtrsim 100.0\,\%$ $(33\,\%)$ & $\gtrsim 100.0\,\%$ $(33\,\%)$ & $\gtrsim 100.0\,\%$ $(57\,\%)$ & $\lesssim 22.5\,\%$ $(71\,\%)$  & $\lesssim 30.0\,\%$ $(72\,\%)$ & $\gtrsim 100.0\,\%$ $(42\,\%)$ \\ \hdashline
  II & $\lesssim 7.5\,\%$ $(59\,\%)$ & $\lesssim 5.0\,\%$ $(58\,\%)$ & $\lesssim 90.0\,\%$ $(27\,\%)$ & $\lesssim 7.5\,\%$ $(47\,\%)$  & $\lesssim 10.0\,\%$ $(48\,\%)$ & $\lesssim 75.0\,\%$ $(50\,\%)$ \\  
     & $\gtrsim 100.0\,\%$ $(19\,\%)$ & $\gtrsim 100.0\,\%$ $(18\,\%)$ & $\gtrsim 100.0\,\%$ $(49\,\%)$ & $\lesssim 25.0\,\%$ $(81\,\%)$  & $\lesssim 30.0\,\%$ $(80\,\%)$ & $\gtrsim 100.0\,\%$ $(45\,\%)$ \\ \hdashline
  LS & $\lesssim 5.0\,\%$ $(51\,\%)$ & $\lesssim 5.0\,\%$ $(57\,\%)$ & $\lesssim 90.0\,\%$ $(41\,\%)$ & $\lesssim 7.5\,\%$ $(47\,\%)$  & $\lesssim 10.0\,\%$ $(48\,\%)$ & $\lesssim 70.0\,\%$ $(50\,\%)$ \\  
     & $\lesssim 20.0\,\%$ $(83\,\%)$ & $\lesssim 15.0\,\%$ $(80\,\%)$ & $\gtrsim 100.0\,\%$ $(50\,\%)$ & $\lesssim 27.5\,\%$ $(81\,\%)$  & $\lesssim 35.0\,\%$ $(79\,\%)$ & $\gtrsim 100.0\,\%$ $(44\,\%)$ \\ \hdashline
  FL & $\lesssim 15.0\,\%$ $(21\,\%)$ & $\lesssim 15.0\,\%$ $(23\,\%)$ & $\lesssim 90.0\,\%$ $(18\,\%)$ & $\lesssim 7.5\,\%$ $(48\,\%)$  & $\lesssim 7.5\,\%$ $(44\,\%)$ & $\lesssim 57.5\,\%$ $(50\,\%)$ \\  
     & $\gtrsim 100.0\,\%$ $(55\,\%)$ & $\gtrsim 100.0\,\%$ $(54\,\%)$ & $\gtrsim 100.0\,\%$ $(63\,\%)$ & $\lesssim 25.0\,\%$ $(81\,\%)$  & $\lesssim 30.0\,\%$ $(81\,\%)$ & $\gtrsim 100.0\,\%$ $(40\,\%)$ \\ \hline\hline
  Type & $\Delta \text{BR}_{H_2H_1H_1}^{\pmb{S_1}}$ & $\Delta \text{BR}_{H_2H_1H_1}^{\pmb{S_2}}$ & $\Delta \text{BR}_{H_2H_1H_1}^{\pmb{\overline{\text{MS}}}}$ & $\Delta \text{BR}_{H_2AA}^{\pmb{S_1}}$ & $\Delta \text{BR}_{H_2AA}^{\pmb{S_2}}$ & $\Delta \text{BR}_{H_2AA}^{\pmb{\overline{\text{MS}}}}$ \\ \hline
  I  & $\lesssim 50.0\,\%$ $(50\,\%)$ & $\lesssim 47.5\,\%$ $(50\,\%)$ & $\lesssim 90.0\,\%$ $(41\,\%)$ & $\lesssim 12.5\,\%$ $(52\,\%)$ & $\lesssim 12.5\,\%$ $(52\,\%)$ & $\lesssim 25.0\,\%$ $(50\,\%)$ \\  
     & $\gtrsim 100.0\,\%$ $(37\,\%)$ & $\gtrsim 100.0\,\%$ $(34\,\%)$ & $\gtrsim 100.0\,\%$ $(57\,\%)$ & $\gtrsim 100.0\,\%$ $(25\,\%)$ & $\gtrsim 100.0\,\%$ $(25\,\%)$ & $\gtrsim 100.0\,\%$ $(28\,\%)$ \\ \hdashline
  II & $\lesssim 40.0\,\%$ $(50\,\%)$ & $\lesssim 40.0\,\%$ $(51\,\%)$ & $\lesssim 90.0\,\%$ $(31\,\%)$ & $\lesssim 5.0\,\%$ $(64\,\%)$ & $\lesssim 5.0\,\%$ $(67\,\%)$ & $\lesssim 90.0\,\%$ $(21\,\%)$ \\  
     & $\gtrsim 100.0\,\%$ $(31\,\%)$ & $\gtrsim 100.0\,\%$ $(30\,\%)$ & $\gtrsim 100.0\,\%$ $(67\,\%)$ & $\lesssim 20.0\,\%$ $(87\,\%)$ & $\lesssim 20.0\,\%$ $(88\,\%)$ & $\gtrsim 100.0\,\%$ $(40\,\%)$ \\ \hdashline
  LS & $\lesssim 45.0\,\%$ $(50\,\%)$ & $\lesssim 45.0\,\%$ $(50\,\%)$ & $\lesssim 90.0\,\%$ $(35\,\%)$ & $\lesssim 7.5\,\%$ $(56\,\%)$ & $\lesssim 7.5\,\%$ $(58\,\%)$ & $\lesssim 15.0\,\%$ $(52\,\%)$ \\  
     & $\gtrsim 100.0\,\%$ $(32\,\%)$ & $\gtrsim 100.0\,\%$ $(32\,\%)$ & $\gtrsim 100.0\,\%$ $(63\,\%)$ & $\gtrsim 100.0\,\%$ $(17\,\%)$ & $\gtrsim 100.0\,\%$ $(17\,\%)$ & $\gtrsim 100.0\,\%$ $(19\,\%)$ \\ \hdashline
  FL & $\lesssim 37.5\,\%$ $(50\,\%)$ & $\lesssim 37.5\,\%$ $(50\,\%)$ & $\lesssim 90.0\,\%$ $(34\,\%)$ & - & - & - \\  
     & $\gtrsim 100.0\,\%$ $(29\,\%)$ & $\gtrsim 100.0\,\%$ $(29\,\%)$ & $\gtrsim 100.0\,\%$ $(64\,\%)$ & - & - & - \\  \hline
 \end{tabular}
\caption{Relative size of the EW corrections to the BRs of the
  non-SM-like N2HDM Higgs boson $H_2$ into
  $b\bar{b}$, $\tau^+ \tau^-$, $ZA$, $W^\pm H^\mp$, $ZZ$, $H_1 H_1$,
  and $AA$ for the four N2HDM types I, II, LS and FL ($H_1$ is SM-like).}
\label{tab:n2hdmnonsmh2}
\end{center}
\end{table}
\begin{itemize}
\item[--] The $\Delta \text{BR}$ for the scheme sets $\pmb{S_1}$ and
  $\pmb{S_2}$ for all four N2HDM types typically range between
  moderate and significant corrections of up to 35\%. However, they can
  also become very large in the Higgs plus gauge boson and di-Higgs final
  states. For most of the channels and types, the scheme set
  $\pmb{S_1}$ produces smaller corrections than $\pmb{S_2}$. 
\item[--] The enhanced relative EW corrections to the BRs of the decay
  channels $H_2 \to ZA$ and $H_2\to W^\pm H^\mp$
  computed with the scheme sets $\pmb{S_1}$ and $\pmb{S_2}$ are due to
  a very small BR at leading order which is due to a
    small phase space.
  Hence, this does not indicate the
  numerical instability of the two scheme sets but rather demonstrates artificially
  large electroweak corrections due to the small LO
  BRs. 
%These large corrections can also be due to
%    parametrically enhanced counterterm contributions or large
%    counterterms themselves that blow up in this corner of the
%    parameter space.
\item[--] For the Higgs-to-Higgs decays $H_2 \to H_1H_1$
  and $H_2 \to AA$, the corrections are typically
  large for all schemes. This is either due to very small LO BRs or due
  to parametrically enhanced Higgs-to-Higgs decays in certain regions
  of the N2HDM parameter space.
\item[--] We remark that the huge corrections indirectly affect
  through the total width the 
  other decay channels into fermion and
  $ZZ$\footnote{Note, that 
    although $H_1$ behaves SM-like, the couplings of $H_2$ to $ZZ$
    need not necessarily be suppressed, as we would expect from sum
    rules. The N2HDM with its larger number of parameters allows for
    a Higgs boson that behaves SM-like although its individual
    couplings to SM particles are not very close to the SM values.
    Since it is the rates that decide about the SM-like behaviour, the
  interplay of production and decay can still lead to very SM-like
  signatures. This leaves room for the couplings of the other Higgs
  bosons to the $ZZ$ bosons to be non-zero. Additionally, the singlet
  admixture influences the Higgs couplings. \label{foot:zzdecays}} final states so that
  their corrections can become significant in particular if the LO BR 
  is not large.
\end{itemize}

%%%%%%%%%%%%%%%%%%%%%%%%%%%%%%%%%%%%%%%%%%%%%%%%%%%%%%%
\paragraph{$H_3$ Decays:}
%%%%%%%%%%%%%%%%%%%%%%%%%%%%%%%%%%%%%%%%%%%%%%%%%%%%%%%
\begin{table}[t!]
\begin{center}
\renewcommand{\arraystretch}{1.2}
\scriptsize
 \begin{tabular}{|c|c c c c c c c c|}
  \hline
Type & $H_3 \to t\bar{t}$ & $H_3 \to ZA$ & $H_3 \to W^\pm H^\mp$ & $H_3 \to ZZ$ & $H_3 \to H_1 H_1$ & $H_3 \to H_1 H_2$ & $H_3 \to H_2 H_2$ & $H_3 \to AA$ \\
   \hline
I & $254\,113$ & $120\,280$ & $137\,260$ & $262\,051$ & $260\,531$ & $185\,847$ & $53\,558$ & $25\,033$ \\
II & $299\,959$ & $118\,359$ & $133\,277$ & $299\,959$ & $299\,959$ & $215\,367$ & $31\,552$ & 4055 \\
LS & $281\, 975$ & $121\,501$ & $138\,894$ & $283\,221$ & $283\,037$ & $202\,202$ & $44\,160$ & $15\,533$ \\
FL & $292\,634$ & $116\,923$ & $133\,182$ & $292\,634$ & $292\,634$ & $208\,005$ & $24\,927$ & 4519  \\
\hline 
\end{tabular}
\caption{Number of parameter points available for the analysis in the
  individual OS decay channels of the non-SM-like Higgs boson $H_3$ ($H_1$ is SM-like).}
\label{tab:rednumbersnonsmh3}
\end{center}
\end{table}
As before, we restrict the analysis to OS decays which reduces the number of available
parameter points in the individual channels as shown in
Tab.~\ref{tab:rednumbersnonsmh3}\footnote{A comparison with
  Eq.~(\ref{eq:numberslightern2hdm}) shows that the amount of
  available parameter points is not reduced for all N2HDM types and
  final states, actually.}. The relative corrections $\Delta \mbox{BR}$ are given in Tab.~\ref{tab:n2hdmnonsmh3}
from which we conclude the following:
\begin{table}[t!]
\begin{center}
\renewcommand{\arraystretch}{1.2}
\scriptsize
 \begin{tabular}{|c|c c c||c c c|}
  \hline
  Type & $\Delta \text{BR}_{H_3b\bar{b}}^{\pmb{S_1}}$ & $\Delta \text{BR}_{H_3b\bar{b}}^{\pmb{S_2}}$ & $\Delta \text{BR}_{H_3b\bar{b}}^{\pmb{\overline{\text{MS}}}}$ & $\Delta \text{BR}_{H_3t\bar{t}}^{\pmb{S_1}}$ & $\Delta \text{BR}_{H_3t\bar{t}}^{\pmb{S_2}}$ & $\Delta \text{BR}_{H_3t\bar{t}}^{\pmb{\overline{\text{MS}}}}$ \\ \hline
  I  & $\lesssim 12.5\,\%$ $(53\,\%)$ & $\lesssim 12.5\,\%$ $(53\,\%)$ & $\lesssim 70.0\,\%$ $(50\,\%)$ & $\lesssim 7.5\,\%$ $(45\,\%)$  & $\lesssim 7.5\,\%$ $(45\,\%)$ & $\lesssim 67.5\,\%$ $(50\,\%)$ \\  
     & $\lesssim 40.0\,\%$ $(80\,\%)$ & $\lesssim 40.0\,\%$ $(80\,\%)$ & $\gtrsim 100.0\,\%$ $(42\,\%)$ & $\lesssim 25.0\,\%$ $(72\,\%)$  & $\lesssim 30.0\,\%$ $(75\,\%)$ & $\gtrsim 100.0\,\%$ $(41\,\%)$ \\ \hdashline
  II & $\lesssim 12.5\,\%$ $(51\,\%)$ & $\lesssim 17.5\,\%$ $(55\,\%)$ & $\lesssim 90.0\,\%$ $(28\,\%)$ & $\lesssim 5.0\,\%$ $(46\,\%)$  & $\lesssim 7.5\,\%$ $(54\,\%)$ & $\lesssim 75.0\,\%$ $(50\,\%)$ \\  
     & $\lesssim 35.0\,\%$ $(81\,\%)$ & $\lesssim 32.5\,\%$ $(80\,\%)$ & $\gtrsim 100.0\,\%$ $(69\,\%)$ & $\lesssim 30.0\,\%$ $(78\,\%)$  & $\lesssim 35.0\,\%$ $(79\,\%)$ & $\gtrsim 100.0\,\%$ $(44\,\%)$ \\ \hdashline
  LS & $\lesssim 10.0\,\%$ $(49\,\%)$ & $\lesssim 10.0\,\%$ $(49\,\%)$ & $\lesssim 65.0\,\%$ $(50\,\%)$ & $\lesssim 7.5\,\%$ $(49\,\%)$  & $\lesssim 7.5\,\%$ $(49\,\%)$ & $\lesssim 65.0\,\%$ $(50\,\%)$ \\  
     & $\lesssim 37.5\,\%$ $(80\,\%)$ & $\lesssim 37.5\,\%$ $(80\,\%)$ & $\gtrsim 100.0\,\%$ $(41\,\%)$ & $\lesssim 35.0\,\%$ $(78\,\%)$  & $\lesssim 30.0\,\%$ $(76\,\%)$ & $\gtrsim 100.0\,\%$ $(41\,\%)$ \\ \hdashline
  FL & $\lesssim 12.5\,\%$ $(50\,\%)$ & $\lesssim 15.0\,\%$ $(50\,\%)$ & $\lesssim 90.0\,\%$ $(27\,\%)$ & $\lesssim 7.5\,\%$ $(51\,\%)$  & $\lesssim 7.5\,\%$ $(50\,\%)$ & $\lesssim 77.5\,\%$ $(50\,\%)$ \\  
     & $\lesssim 42.5\,\%$ $(80\,\%)$ & $\lesssim 40.0\,\%$ $(80\,\%)$ & $\gtrsim 100.0\,\%$ $(70\,\%)$ & $\lesssim 35.0\,\%$ $(78\,\%)$  & $\lesssim 30.0\,\%$ $(75\,\%)$ & $\gtrsim 100.0\,\%$ $(44\,\%)$ \\ \hline\hline
  Type & $\Delta \text{BR}_{H_3\tau ^+ \tau ^-}^{\pmb{S_1}}$ & $\Delta \text{BR}_{H_3\tau ^+ \tau ^-}^{\pmb{S_2}}$ & $\Delta \text{BR}_{H_3\tau ^+ \tau ^-}^{\pmb{\overline{\text{MS}}}}$ & $\Delta \text{BR}_{H_3ZA}^{\pmb{S_1}}$ & $\Delta \text{BR}_{H_3ZA}^{\pmb{S_2}}$ & $\Delta \text{BR}_{H_3ZA}^{\pmb{\overline{\text{MS}}}}$ \\ \hline
  I  & $\lesssim 12.5\,\%$ $(50\,\%)$ & $\lesssim 12.5\,\%$ $(51\,\%)$ & $\lesssim 70.0\,\%$ $(50\,\%)$ & $\lesssim 7.5\,\%$ $(35\,\%)$  & $\lesssim 7.5\,\%$ $(34\,\%)$ & $\lesssim 90.0\,\%$ $(44\,\%)$ \\  
     & $\lesssim 40.0\,\%$ $(80\,\%)$ & $\lesssim 32.5\,\%$ $(77\,\%)$ & $\gtrsim 100.0\,\%$ $(42\,\%)$ & $\lesssim 25.0\,\%$ $(65\,\%)$  & $\lesssim 32.5\,\%$ $(70\,\%)$ & $\gtrsim 100.0\,\%$ $(52\,\%)$ \\ \hdashline
  II & $\lesssim 10.0\,\%$ $(49\,\%)$ & $\lesssim 12.5\,\%$ $(41\,\%)$ & $\lesssim 90.0\,\%$ $(28\,\%)$ & $\lesssim 10.0\,\%$ $(36\,\%)$  & $\lesssim 10.0\,\%$ $(39\,\%)$ & $\lesssim 90.0\,\%$ $(39\,\%)$ \\  
     & $\lesssim 27.5\,\%$ $(80\,\%)$ & $\lesssim 25.0\,\%$ $(77\,\%)$ & $\gtrsim 100.0\,\%$ $(69\,\%)$ & $\lesssim 30.0\,\%$ $(64\,\%)$  & $\lesssim 35.0\,\%$ $(70\,\%)$ & $\gtrsim 100.0\,\%$ $(55\,\%)$ \\ \hdashline
  LS & $\lesssim 12.5\,\%$ $(55\,\%)$ & $\lesssim 10.0\,\%$ $(33\,\%)$ & $\lesssim 80.0\,\%$ $(35\,\%)$ & $\lesssim 10.0\,\%$ $(41\,\%)$  & $\lesssim 10.0\,\%$ $(41\,\%)$ & $\lesssim 90.0\,\%$ $(47\,\%)$ \\  
     & $\lesssim 32.5\,\%$ $(80\,\%)$ & $\lesssim 30.0\,\%$ $(79\,\%)$ & $\gtrsim 100.0\,\%$ $(60\,\%)$ & $\lesssim 35.0\,\%$ $(70\,\%)$  & $\lesssim 35.0\,\%$ $(70\,\%)$ & $\gtrsim 100.0\,\%$ $(49\,\%)$ \\ \hdashline
  FL & $\lesssim 10.0\,\%$ $(48\,\%)$ & $\lesssim 10.0\,\%$ $(48\,\%)$ & $\lesssim 77.5\,\%$ $(50\,\%)$ & $\lesssim 10.0\,\%$ $(36\,\%)$  & $\lesssim 10.0\,\%$ $(39\,\%)$ & $\lesssim 90.0\,\%$ $(38\,\%)$ \\  
     & $\lesssim 40.0\,\%$ $(80\,\%)$ & $\lesssim 35.0\,\%$ $(79\,\%)$ & $\gtrsim 100.0\,\%$ $(44\,\%)$ & $\lesssim 35.0\,\%$ $(66\,\%)$  & $\lesssim 35.0\,\%$ $(67\,\%)$ & $\gtrsim 100.0\,\%$ $(57\,\%)$ \\ \hline\hline
  Type & $\Delta \text{BR}_{H_3W^\pm H^\mp}^{\pmb{S_1}}$ & $\Delta \text{BR}_{H_3W^\pm H^\mp}^{\pmb{S_2}}$ & $\Delta \text{BR}_{H_3W^\pm H^\mp}^{\pmb{\overline{\text{MS}}}}$ & $\Delta \text{BR}_{H_3ZZ}^{\pmb{S_1}}$ & $\Delta \text{BR}_{H_3ZZ}^{\pmb{S_2}}$ & $\Delta \text{BR}_{H_3ZZ}^{\pmb{\overline{\text{MS}}}}$ \\ \hline
  I  & $\lesssim 7.5\,\%$ $(36\,\%)$ & $\lesssim 7.5\,\%$ $(35\,\%)$ & $\lesssim 90.0\,\%$ $(44\,\%)$ & $\lesssim 7.5\,\%$ $(43\,\%)$  & $\lesssim 7.5\,\%$ $(39\,\%)$ & $\lesssim 52.5\,\%$ $(50\,\%)$ \\  
     & $\lesssim 30.0\,\%$ $(69\,\%)$ & $\lesssim 30.0\,\%$ $(68\,\%)$ & $\gtrsim 100.0\,\%$ $(52\,\%)$ & $\lesssim 30.0\,\%$ $(73\,\%)$  & $\lesssim 30.0\,\%$ $(70\,\%)$ & $\gtrsim 100.0\,\%$ $(38\,\%)$ \\ \hdashline
  II & $\lesssim 7.5\,\%$ $(30\,\%)$ & $\lesssim 10.0\,\%$ $(40\,\%)$ & $\lesssim 90.0\,\%$ $(38\,\%)$ & $\lesssim 7.5\,\%$ $(41\,\%)$  & $\lesssim 10.0\,\%$ $(43\,\%)$ & $\lesssim 70.0\,\%$ $(50\,\%)$ \\  
     & $\lesssim 35.0\,\%$ $(68\,\%)$ & $\lesssim 35.0\,\%$ $(69\,\%)$ & $\gtrsim 100.0\,\%$ $(56\,\%)$ & $\lesssim 30.0\,\%$ $(72\,\%)$  & $\lesssim 32.5\,\%$ $(70\,\%)$ & $\gtrsim 100.0\,\%$ $(42\,\%)$ \\ \hdashline
  LS & $\lesssim 7.5\,\%$ $(34\,\%)$ & $\lesssim 7.5\,\%$ $(34\,\%)$ & $\lesssim 90.0\,\%$ $(47\,\%)$ & $\lesssim 7.5\,\%$ $(43\,\%)$  & $\lesssim 10.0\,\%$ $(46\,\%)$ & $\lesssim 47.5\,\%$ $(50\,\%)$ \\  
     & $\lesssim 35.0\,\%$ $(70\,\%)$ & $\lesssim 35.0\,\%$ $(70\,\%)$ & $\gtrsim 100.0\,\%$ $(51\,\%)$ & $\lesssim 32.5\,\%$ $(74\,\%)$  & $\lesssim 30.0\,\%$ $(70\,\%)$ & $\gtrsim 100.0\,\%$ $(36\,\%)$ \\ \hdashline
  FL & $\lesssim 10.0\,\%$ $(36\,\%)$ & $\lesssim 10.0\,\%$ $(38\,\%)$ & $\lesssim 95.0\,\%$ $(38\,\%)$ & $\lesssim 7.5\,\%$ $(44\,\%)$  & $\lesssim 7.5\,\%$ $(41\,\%)$ & $\lesssim 42.5\,\%$ $(50\,\%)$ \\  
     & $\lesssim 40.0\,\%$ $(69\,\%)$ & $\lesssim 40.0\,\%$ $(70\,\%)$ & $\gtrsim 100.0\,\%$ $(58\,\%)$ & $\lesssim 25.0\,\%$ $(70\,\%)$  & $\lesssim 27.5\,\%$ $(70\,\%)$ & $\gtrsim 100.0\,\%$ $(34\,\%)$ \\ \hline\hline
  Type & $\Delta \text{BR}_{H_3H_1H_1}^{\pmb{S_1}}$ & $\Delta \text{BR}_{H_3H_1H_1}^{\pmb{S_2}}$ & $\Delta \text{BR}_{H_3H_1H_1}^{\pmb{\overline{\text{MS}}}}$ & $\Delta \text{BR}_{H_3H_1H_2}^{\pmb{S_1}}$ & $\Delta \text{BR}_{H_3H_1H_2}^{\pmb{S_2}}$ & $\Delta \text{BR}_{H_3H_1H_2}^{\pmb{\overline{\text{MS}}}}$ \\ \hline
  I  & $\lesssim 35.0\,\%$ $(50\,\%)$ & $\lesssim 37.5\,\%$ $(50\,\%)$ & $\lesssim 85.0\,\%$ $(45\,\%)$ & $\lesssim 35.0\,\%$ $(21\,\%)$  & $\lesssim 32.5\,\%$ $(20\,\%)$ & $\lesssim 47.5\,\%$ $(20\,\%)$ \\  
     & $\gtrsim 100.0\,\%$ $(32\,\%)$ & $\gtrsim 100.0\,\%$ $(32\,\%)$ & $\gtrsim 100.0\,\%$ $(52\,\%)$ & $\gtrsim 100.0\,\%$ $(62\,\%)$  & $\gtrsim 100.0\,\%$ $(62\,\%)$ & $\gtrsim 100.0\,\%$ $(68\,\%)$ \\ \hdashline
  II & $\lesssim 57.5\,\%$ $(50\,\%)$ & $\lesssim 57.5\,\%$ $(50\,\%)$ & $\lesssim 87.5\,\%$ $(37\,\%)$ & $\lesssim 50.0\,\%$ $(10\,\%)$  & $\lesssim 50.0\,\%$ $(10\,\%)$ & $\lesssim 35.0\,\%$ $(10\,\%)$ \\  
     & $\gtrsim 100.0\,\%$ $(41\,\%)$ & $\gtrsim 100.0\,\%$ $(41\,\%)$ & $\gtrsim 100.0\,\%$ $(60\,\%)$ & $\gtrsim 100.0\,\%$ $(81\,\%)$  & $\gtrsim 100.0\,\%$ $(81\,\%)$ & $\gtrsim 100.0\,\%$ $(76\,\%)$ \\ \hdashline
  LS & $\lesssim 37.5\,\%$ $(50\,\%)$ & $\lesssim 37.5\,\%$ $(50\,\%)$ & $\lesssim 85.0\,\%$ $(44\,\%)$ & $\lesssim 50.0\,\%$ $(20\,\%)$  & $\lesssim 50.0\,\%$ $(20\,\%)$ & $\lesssim 10.0\,\%$ $(20\,\%)$ \\  
     & $\gtrsim 100.0\,\%$ $(33\,\%)$ & $\gtrsim 100.0\,\%$ $(33\,\%)$ & $\gtrsim 100.0\,\%$ $(53\,\%)$ & $\gtrsim 100.0\,\%$ $(69\,\%)$  & $\gtrsim 100.0\,\%$ $(68\,\%)$ & $\gtrsim 100.0\,\%$ $(69\,\%)$ \\ \hdashline
  FL & $\lesssim 42.5\,\%$ $(50\,\%)$ & $\lesssim 42.5\,\%$ $(50\,\%)$ & $\lesssim 80.0\,\%$ $(40\,\%)$ & $\lesssim 10.0\,\%$ $(10\,\%)$  & $\lesssim 47.5\,\%$ $(10\,\%)$ & $\lesssim 30.0\,\%$ $(11\,\%)$ \\  
     & $\gtrsim 100.0\,\%$ $(36\,\%)$ & $\gtrsim 100.0\,\%$ $(36\,\%)$ & $\gtrsim 100.0\,\%$ $(56\,\%)$ & $\gtrsim 100.0\,\%$ $(81\,\%)$  & $\gtrsim 100.0\,\%$ $(81\,\%)$ & $\gtrsim 100.0\,\%$ $(73\,\%)$ \\ \hline\hline
  Type & $\Delta \text{BR}_{H_3H_2H_2}^{\pmb{S_1}}$ & $\Delta \text{BR}_{H_3H_2H_2}^{\pmb{S_2}}$ & $\Delta \text{BR}_{H_3H_2H_2}^{\pmb{\overline{\text{MS}}}}$ & $\Delta \text{BR}_{H_3AA}^{\pmb{S_1}}$ & $\Delta \text{BR}_{H_3AA}^{\pmb{S_2}}$ & $\Delta \text{BR}_{H_3AA}^{\pmb{\overline{\text{MS}}}}$ \\ \hline
  I  & $\lesssim 35.0\,\%$ $(40\,\%)$ & $\lesssim 20.0\,\%$ $(31\,\%)$ & $\lesssim 27.5\,\%$ $(20\,\%)$ & $\lesssim 7.5\,\%$ $(43\,\%)$ & $\lesssim 7.5\,\%$ $(43\,\%)$ & $\lesssim 27.5\,\%$ $(30\,\%)$ \\  
     & $\gtrsim 100.0\,\%$ $(42\,\%)$ & $\gtrsim 100.0\,\%$ $(43\,\%)$ & $\gtrsim 100.0\,\%$ $(63\,\%)$ & $\lesssim 30.0\,\%$ $(75\,\%)$ & $\lesssim 30.0\,\%$ $(76\,\%)$ & $\gtrsim 100.0\,\%$ $(41\,\%)$ \\ \hdashline
  II & $\lesssim 37.5\,\%$ $(10\,\%)$ & $\lesssim 35.0\,\%$ $(10\,\%)$ & $\lesssim 32.5\,\%$ $(10\,\%)$ & $\lesssim 25.0\,\%$ $(42\,\%)$ & $\lesssim 25.0\,\%$ $(46\,\%)$ & $\lesssim 20.0\,\%$ $(30\,\%)$ \\  
     & $\gtrsim 100.0\,\%$ $(81\,\%)$ & $\gtrsim 100.0\,\%$ $(80\,\%)$ & $\gtrsim 100.0\,\%$ $(79\,\%)$ & $\gtrsim 100.0\,\%$ $(21\,\%)$ & $\gtrsim 100.0\,\%$ $(19\,\%)$ & $\gtrsim 100.0\,\%$ $(44\,\%)$ \\ \hdashline
  LS & $\lesssim 37.5\,\%$ $(30\,\%)$ & $\lesssim 20.0\,\%$ $(21\,\%)$ & $\lesssim 25.0\,\%$ $(20\,\%)$ & $\lesssim 10.0\,\%$ $(40\,\%)$ & $\lesssim 10.0\,\%$ $(43\,\%)$ & $\lesssim 25.0\,\%$ $(37\,\%)$ \\  
     & $\gtrsim 100.0\,\%$ $(52\,\%)$ & $\gtrsim 100.0\,\%$ $(53\,\%)$ & $\gtrsim 100.0\,\%$ $(62\,\%)$ & $\lesssim 30.0\,\%$ $(70\,\%)$ & $\lesssim 32.5\,\%$ $(74\,\%)$ & $\gtrsim 100.0\,\%$ $(36\,\%)$ \\ \hdashline
  FL & $\lesssim 30.0\,\%$ $(10\,\%)$ & $\lesssim 30.0\,\%$ $(10\,\%)$ & $\lesssim 22.5\,\%$ $(10\,\%)$ & $\lesssim 30.0\,\%$ $(42\,\%)$ & $\lesssim 22.5\,\%$ $(40\,\%)$ & $\lesssim 35.0\,\%$ $(37\,\%)$ \\  
     & $\gtrsim 100.0\,\%$ $(77\,\%)$ & $\gtrsim 100.0\,\%$ $(77\,\%)$ & $\gtrsim 100.0\,\%$ $(72\,\%)$ & $\gtrsim 100.0\,\%$ $(27\,\%)$ & $\gtrsim 100.0\,\%$ $(26\,\%)$ & $\gtrsim 100.0\,\%$ $(45\,\%)$ \\ \hline
 \end{tabular}
\caption{Relative size of the EW corrections to the BRs of the
  non-SM-like N2HDM Higgs boson $H_3$ ($H_1$ is SM-like) into
  $b\bar{b}$, $t\bar{t}$, $\tau^+ \tau^-$, $ZA$, $W^\pm H^\mp$, $ZZ$,
  $H_1 H_1$, $H_1 H_2$, $H_2 H_2$, 
  and $AA$ for the four N2HDM types I, II, LS and FL.}
\label{tab:n2hdmnonsmh3}
\end{center}
\vspace*{-0.5cm}
\end{table}
\begin{itemize}
\item[--] The corrections for the renormalization
  scheme sets $\pmb{S_1}$ and $\pmb{S_2}$ are of similar size. 
\item[--] Considering the decay channels, the smallest corrections are found for the decays into
  $t\bar{t}$, $W^\pm H^\mp$, $ZA$, and $ZZ$\footnote{The same argument
    as given in footnote \ref{foot:zzdecays} for $H_2$ also applies
    here for $H_3$.} and range from below
  7.5\% up to 40\% at most. Note that a heavy non-SM-like Higgs boson $H_3$ can
  have important BRs not only into $t\bar{t}$ but also
  into gauge plus Higgs final states. 
  The relative corrections to the
  BRs into $b\bar{b}$ and $\tau^+ \tau^-$ are slightly larger.
\item[--] The relative corrections to the Higgs-to-Higgs decays $H_3
  \to  H_iH_j$ ($i,j=1,2$) and $H_3 \to AA$
  on the other hand can become again very large which is due to
  either small LO BRs (see next item) or
    parametrically enhanced corrections. This again indirectly affects
    through the total width the decay channels with typically more
    moderate corrections, inducing more significant $\Delta \mbox{BR}$ then.
\item[--] We remark that the scheme sets $\pmb{S_1}$ and $\pmb{S_2}$,
  \textit{independently of the N2HDM type}, typically feature
  corrections larger than $100\,\%$ for roughly $10\,\%$ to $13\,\%$
  of the parameter points for \textit{all decay channels} (of course,
  for the Higgs-to-Higgs decays, the amount of parameter points
  leading to large corrections is even larger)\footnote{This is not shown
  in the table in order not to blow up the presentation.}. This is not
a sign of numerical instability of the renormalization schemes though
but instead this stems from small LO BRs and hence from a large
sensitivity on the higher-order corrections. Moreover, BRs containing
mixing angle CTs and off-diagonal scalar WFRCs become numerically
enhanced in certain corners of parameter space and hence, $\Delta
\text{BR}$ blows up there. Additionally, the counterterms are
  multiplied with coefficients containing couplings in the denominator
  that become small here and thereby lead to parametrically enhanced
  counterterm contributions.
\end{itemize}

%%%%%%%%%%%%%%%%%%%%%%%%%%%%%%%%%%%%%%%%%%%%%%%%%%%%%%%
\subsubsection{$H_2$ is the SM-like Higgs Boson}
%%%%%%%%%%%%%%%%%%%%%%%%%%%%%%%%%%%%%%%%%%%%%%%%%%%%%%%
We turn to the case where $H_2$ is SM-like so that we have an
additional light CP-even Higgs boson in the spectrum, and investigate the
decays of the non-SM-like Higgs bosons $H_1$ and $H_3$. For this
scenario the number of parameter points compatible with all
constraints, and hence the statistics of the analysis, is significantly reduced, {\it
  cf.}~Eq.~(\ref{eq:numbersheaviern2hdm}). 

%%%%%%%%%%%%%%%%%%%%%%%%%%%%%%%%%%%%%%%%%%%%%%%%%%%%%%%
\paragraph{$H_1$ Decays:}
%%%%%%%%%%%%%%%%%%%%%%%%%%%%%%%%%%%%%%%%%%%%%%%%%%%%%%%
\begin{table}[t!]
\begin{center}
\renewcommand{\arraystretch}{1.2}
\scriptsize
 \begin{tabular}{|c|c c c||c c c|}
  \hline
  Type & $\Delta \text{BR}_{H_1b\bar{b}}^{\pmb{S_1}}$ & $\Delta \text{BR}_{H_1b\bar{b}}^{\pmb{S_2}}$ & $\Delta \text{BR}_{H_1b\bar{b}}^{\pmb{\overline{\text{MS}}}}$ & $\Delta \text{BR}_{H_1\tau ^+ \tau ^-}^{\pmb{S_1}}$ & $\Delta \text{BR}_{H_1\tau ^+ \tau ^-}^{\pmb{S_2}}$ & $\Delta \text{BR}_{H_1\tau ^+ \tau ^-}^{\pmb{\overline{\text{MS}}}}$ \\ \hline
  I  & $\lesssim 2.5\,\%$ $(85\,\%)$  & $\lesssim 2.5\,\%$ $(85\,\%)$ & $\lesssim 15.0\,\%$ $(71\,\%)$ & $\lesssim 2.5\,\%$ $(73\,\%)$  & $\lesssim 2.5\,\%$ $(71\,\%)$ & $\lesssim 15.0\,\%$ $(70\,\%)$ \\  
     & $\lesssim 10.0\,\%$ $(94\,\%)$  & $\lesssim 10.0\,\%$ $(93\,\%)$ & $\gtrsim 100.0\,\%$ $(6\,\%)$ & $\lesssim 10.0\,\%$ $(92\,\%)$  & $\lesssim 10.0\,\%$ $(91\,\%)$ & $\gtrsim 100.0\,\%$ $(6\,\%)$ \\ \hdashline
  II & $\lesssim 2.5\,\%$ $(95\,\%)$  & $\lesssim 2.5\,\%$ $(93\,\%)$ & $\lesssim 20.0\,\%$ $(60\,\%)$ & $\lesssim 2.5\,\%$ $(93\,\%)$  & $\lesssim 2.5\,\%$ $(94\,\%)$ & $\lesssim 37.5\,\%$ $(70\,\%)$ \\  
     & $\lesssim 7.5\,\%$ $(97\,\%)$  & $\lesssim 7.5\,\%$ $(96\,\%)$ & $\gtrsim 100.0\,\%$ $(18\,\%)$ & $\lesssim 7.5\,\%$ $(98\,\%)$  & $\lesssim 7.5\,\%$ $(97\,\%)$ & $\gtrsim 100.0\,\%$ $(18\,\%)$ \\ \hdashline
  LS & $\lesssim 2.5\,\%$ $(77\,\%)$  & $\lesssim 2.5\,\%$ $(74\,\%)$ & $\lesssim 45.0\,\%$ $(50\,\%)$ & $\lesssim 2.5\,\%$ $(70\,\%)$  & $\lesssim 2.5\,\%$ $(60\,\%)$ & $\lesssim 85.0\,\%$ $(50\,\%)$ \\  
     & $\lesssim 10.0\,\%$ $(92\,\%)$  & $\lesssim 10.0\,\%$ $(92\,\%)$ & $\gtrsim 100.0\,\%$ $(35\,\%)$ & $\lesssim 10.0\,\%$ $(92\,\%)$  & $\lesssim 10.0\,\%$ $(92\,\%)$ & $\gtrsim 100.0\,\%$ $(47\,\%)$ \\ \hdashline
  FL & $\lesssim 2.5\,\%$ $(93\,\%)$  & $\lesssim 2.5\,\%$ $(92\,\%)$ & $\lesssim 17.5\,\%$ $(60\,\%)$ & $\lesssim 2.5\,\%$ $(76\,\%)$  & $\lesssim 2.5\,\%$ $(67\,\%)$ & $\lesssim 80.0\,\%$ $(35\,\%)$ \\  
     & $\lesssim 10.0\,\%$ $(97\,\%)$  & $\lesssim 10.0\,\%$ $(96\,\%)$ & $\gtrsim 100.0\,\%$ $(20\,\%)$ & $\lesssim 7.5\,\%$ $(98\,\%)$  & $\lesssim 7.5\,\%$ $(97\,\%)$ & $\gtrsim 100.0\,\%$ $(61\,\%)$ \\ \hline
 \end{tabular}
\caption{Relative size of the EW corrections to the BRs of the
  non-SM-like N2HDM Higgs boson $H_1$ into
  $b\bar{b}$ and $\tau^+\tau^-$, for the four N2HDM types I, II, LS and
  FL ($H_2$ is SM-like).} 
\label{tab:h1forh2smlike}
\end{center}
\end{table}
Since the Higgs boson $H_1$ is rather light for this mass hierarchy, the only decay
channels with important BRs (which are not
loop-induced) are those into $b\bar{b}$ and $\tau^+\tau^-$ for which
we show the relative EW corrections to the BRs in
Tab.~\ref{tab:h1forh2smlike}. The table shows that the relative corrections are small to moderate
for the majority of all input parameter points for the two scheme sets
$\pmb{S_1}$ and $\pmb{S_2}$ in all four N2HDM types. 

%%%%%%%%%%%%%%%%%%%%%%%%%%%%%%%%%%%%%%%%%%%%%%%%%%%%%%%
\paragraph{$H_3$ Decays:}
%%%%%%%%%%%%%%%%%%%%%%%%%%%%%%%%%%%%%%%%%%%%%%%%%%%%%%%
\begin{table}[t!]
\begin{center}
\renewcommand{\arraystretch}{1.2}
\scriptsize
 \begin{tabular}{|c|c c c c|}
  \hline
Type & $H_3 \to t\bar{t}$ & $H_3 \to ZA$ & $H_3 \to W^\pm H^\mp$ & $H\to ZZ$\\
   \hline
I & 5856 & 2612 & 3330 & 8302 \\
II &  2381 & 74 & 28 & 2381 \\
LS & 2923 & 157 & 310 & 3203 \\
FL & 1562 & 12 & 8 & 1562 \\
\hline \hline
Type & $H_3 \to H_1 H_1$ & $H_3 \to H_1 H_2$ & $H_3 \to H_2 H_2$ & $H_3 \to AA$ \\ \hline
I & 8205 & 7976 & 7609 & 1448 \\
II & 2381 & 2381 & 2381 & 0 \\
LS & 3065 & 2807 & 2173 & 66 \\
FL & 1562 & 1562 & 1562 & 0 \\
\hline
\end{tabular}
\caption{Number of parameter points available for the analysis in the
  individual OS decay channels of the non-SM-like Higgs boson $H_3$
  ($H_2$ is SM-like).} 
\label{tab:rednumbersnonsmh3smh2}
\vspace*{-0.5cm}
\end{center}
\end{table}
\begin{table}[t!]
\begin{center}
\renewcommand{\arraystretch}{1.2}
\scriptsize
 \begin{tabular}{|c|c c c||c c c|}
  \hline
  Type & $\Delta \text{BR}_{H_3b\bar{b}}^{\pmb{S_1}}$ & $\Delta \text{BR}_{H_3b\bar{b}}^{\pmb{S_2}}$ & $\Delta \text{BR}_{H_3b\bar{b}}^{\pmb{\overline{\text{MS}}}}$ & $\Delta \text{BR}_{H_3t\bar{t}}^{\pmb{S_1}}$ & $\Delta \text{BR}_{H_3t\bar{t}}^{\pmb{S_2}}$ & $\Delta \text{BR}_{H_3t\bar{t}}^{\pmb{\overline{\text{MS}}}}$ \\ \hline
  I  & $\lesssim 15.0\,\%$ $(40\,\%)$ & $\lesssim 17.5\,\%$ $(43\,\%)$ & $\lesssim 80.0\,\%$ $(35\,\%)$ & $\lesssim 15.0\,\%$ $(43\,\%)$  & $\lesssim 15.0\,\%$ $(42\,\%)$ & $\lesssim 82.5\,\%$ $(40\,\%)$ \\  
     & $\gtrsim 100.0\,\%$ $(21\,\%)$ & $\gtrsim 100.0\,\%$ $(22\,\%)$ & $\gtrsim 100.0\,\%$ $(59\,\%)$ & $\gtrsim 100.0\,\%$ $(23\,\%)$  & $\gtrsim 100.0\,\%$ $(24\,\%)$ & $\gtrsim 100.0\,\%$ $(55\,\%)$ \\ \hdashline
  II & $\lesssim 12.5\,\%$ $(57\,\%)$ & $\lesssim 17.5\,\%$ $(54\,\%)$ & $\lesssim 85.0\,\%$ $(15\,\%)$ & $\lesssim 2.5\,\%$ $(54\,\%)$  & $\lesssim 2.5\,\%$ $(54\,\%)$ & $\lesssim 75.0\,\%$ $(40\,\%)$ \\  
     & $\lesssim 20.0\,\%$ $(91\,\%)$ & $\lesssim 25.0\,\%$ $(91\,\%)$ & $\gtrsim 100.0\,\%$ $(83\,\%)$ & $\lesssim 12.5\,\%$ $(92\,\%)$  & $\lesssim 10.0\,\%$ $(90\,\%)$ & $\gtrsim 100.0\,\%$ $(55\,\%)$ \\ \hdashline
  LS & $\lesssim 10.0\,\%$ $(51\,\%)$ & $\lesssim 10.0\,\%$ $(51\,\%)$ & $\lesssim 77.5\,\%$ $(30\,\%)$ & $\lesssim 12.5\,\%$ $(70\,\%)$  & $\lesssim 12.5\,\%$ $(70\,\%)$ & $\lesssim 80.0\,\%$ $(33\,\%)$ \\  
     & $\gtrsim 100.0\,\%$ $(9\,\%)$ & $\gtrsim 100.0\,\%$ $(9\,\%)$ & $\gtrsim 100.0\,\%$ $(65\,\%)$ & $\gtrsim 100.0\,\%$ $(8\,\%)$  & $\gtrsim 100.0\,\%$ $(8\,\%)$ & $\gtrsim 100.0\,\%$ $(62\,\%)$ \\ \hdashline
  FL & $\lesssim 10.0\,\%$ $(45\,\%)$ & $\lesssim 15.0\,\%$ $(53\,\%)$ & $\lesssim 85.0\,\%$ $(22\,\%)$ & $\lesssim 2.5\,\%$ $(44\,\%)$  & $\lesssim 2.5\,\%$ $(44\,\%)$ & $\lesssim 87.5\,\%$ $(30\,\%)$ \\  
     & $\lesssim 20.0\,\%$ $(86\,\%)$ & $\lesssim 27.5\,\%$ $(88\,\%)$ & $\gtrsim 100.0\,\%$ $(75\,\%)$ & $\lesssim 20.0\,\%$ $(87\,\%)$  & $\lesssim 15.0\,\%$ $(85\,\%)$ & $\gtrsim 100.0\,\%$ $(67\,\%)$ \\ \hline\hline
  Type & $\Delta \text{BR}_{H_3\tau ^+ \tau ^-}^{\pmb{S_1}}$ & $\Delta \text{BR}_{H_3\tau ^+ \tau ^-}^{\pmb{S_2}}$ & $\Delta \text{BR}_{H_3\tau ^+ \tau ^-}^{\pmb{\overline{\text{MS}}}}$ & $\Delta \text{BR}_{H_3ZA}^{\pmb{S_1}}$ & $\Delta \text{BR}_{H_3ZA}^{\pmb{S_2}}$ & $\Delta \text{BR}_{H_3ZA}^{\pmb{\overline{\text{MS}}}}$ \\ \hline
  I  & $\lesssim 17.5\,\%$ $(42\,\%)$ & $\lesssim 20.0\,\%$ $(45\,\%)$ & $\lesssim 80.0\,\%$ $(35\,\%)$ & $\lesssim 20.0\,\%$ $(42\,\%)$  & $\lesssim 15.0\,\%$ $(35\,\%)$ & $\lesssim 77.5\,\%$ $(30\,\%)$ \\  
     & $\gtrsim 100.0\,\%$ $(20\,\%)$ & $\gtrsim 100.0\,\%$ $(21\,\%)$ & $\gtrsim 100.0\,\%$ $(59\,\%)$ & $\gtrsim 100.0\,\%$ $(31\,\%)$  & $\gtrsim 100.0\,\%$ $(31\,\%)$ & $\gtrsim 100.0\,\%$ $(62\,\%)$ \\ \hdashline
  II & $\lesssim 10.0\,\%$ $(45\,\%)$ & $\lesssim 15.0\,\%$ $(42\,\%)$ & $\lesssim 85.0\,\%$ $(15\,\%)$ & $\lesssim 5.0\,\%$ $(80\,\%)$  & $\lesssim 5.0\,\%$ $(49\,\%)$ & $\lesssim 90.0\,\%$ $(34\,\%)$ \\  
     & $\lesssim 17.5\,\%$ $(89\,\%)$ & $\lesssim 22.5\,\%$ $(89\,\%)$ & $\gtrsim 100.0\,\%$ $(83\,\%)$ & $\lesssim 10.0\,\%$ $(93\,\%)$  & $\lesssim 12.5\,\%$ $(93\,\%)$ & $\gtrsim 100.0\,\%$ $(38\,\%)$ \\ \hdashline
  LS & $\lesssim 15.0\,\%$ $(58\,\%)$ & $\lesssim 15.0\,\%$ $(47\,\%)$ & $\lesssim 80.0\,\%$ $(30\,\%)$ & $\lesssim 12.5\,\%$ $(30\,\%)$  & $\lesssim 15.0\,\%$ $(30\,\%)$ & $\lesssim 90.0\,\%$ $(32\,\%)$ \\  
     & $\gtrsim 100.0\,\%$ $(9\,\%)$ & $\gtrsim 100.0\,\%$ $(9\,\%)$ & $\gtrsim 100.0\,\%$ $(63\,\%)$ & $\gtrsim 100.0\,\%$ $(47\,\%)$  & $\gtrsim 100.0\,\%$ $(43\,\%)$ & $\gtrsim 100.0\,\%$ $(61\,\%)$ \\ \hdashline
  FL & $\lesssim 10.0\,\%$ $(47\,\%)$ & $\lesssim 10.0\,\%$ $(49\,\%)$ & $\lesssim 87.5\,\%$ $(30\,\%)$ & $\lesssim 5.0\,\%$ $(58\,\%)$  & $\lesssim 7.5\,\%$ $(58\,\%)$ & $\lesssim 82.5\,\%$ $(25\,\%)$ \\  
     & $\lesssim 25.0\,\%$ $(88\,\%)$ & $\lesssim 20.0\,\%$ $(86\,\%)$ & $\gtrsim 100.0\,\%$ $(67\,\%)$ & $\gtrsim 100.0\,\%$ $(33\,\%)$  & $\gtrsim 100.0\,\%$ $(33\,\%)$ & $\gtrsim 100.0\,\%$ $(58\,\%)$ \\ \hline\hline
  Type & $\Delta \text{BR}_{H_3W^\pm H^\mp}^{\pmb{S_1}}$ & $\Delta \text{BR}_{H_3W^\pm H^\mp}^{\pmb{S_2}}$ & $\Delta \text{BR}_{H_3W^\pm H^\mp}^{\pmb{\overline{\text{MS}}}}$ & $\Delta \text{BR}_{H_3ZZ}^{\pmb{S_1}}$ & $\Delta \text{BR}_{H_3ZZ}^{\pmb{S_2}}$ & $\Delta \text{BR}_{H_3ZZ}^{\pmb{\overline{\text{MS}}}}$ \\ \hline
  I  & $\lesssim 17.5\,\%$ $(37\,\%)$ & $\lesssim 15.0\,\%$ $(33\,\%)$ & $\lesssim 75.0\,\%$ $(30\,\%)$ & $\lesssim 15.0\,\%$ $(30\,\%)$  & $\lesssim 20.0\,\%$ $(30\,\%)$ & $\lesssim 60.0\,\%$ $(30\,\%)$ \\  
     & $\gtrsim 100.0\,\%$ $(32\,\%)$ & $\gtrsim 100.0\,\%$ $(33\,\%)$ & $\gtrsim 100.0\,\%$ $(61\,\%)$ & $\gtrsim 100.0\,\%$ $(26\,\%)$  & $\gtrsim 100.0\,\%$ $(29\,\%)$ & $\gtrsim 100.0\,\%$ $(58\,\%)$ \\ \hdashline
  II & $\lesssim 2.5\,\%$ $(41\,\%)$ & $\lesssim 2.5\,\%$ $(50\,\%)$ & $\lesssim 80.0\,\%$ $(29\,\%)$ & $\lesssim 12.5\,\%$ $(39\,\%)$  & $\lesssim 12.5\,\%$ $(29\,\%)$ & $\lesssim 65.0\,\%$ $(30\,\%)$ \\  
     & $\lesssim 10.0\,\%$ $(93\,\%)$ & $\lesssim 12.5\,\%$ $(93\,\%)$ & $\gtrsim 100.0\,\%$ $(36\,\%)$ & $\gtrsim 100.0\,\%$ $(12\,\%)$  & $\gtrsim 100.0\,\%$ $(19\,\%)$ & $\gtrsim 100.0\,\%$ $(60\,\%)$ \\ \hdashline
  LS & $\lesssim 22.5\,\%$ $(29\,\%)$ & $\lesssim 22.5\,\%$ $(30\,\%)$ & $\lesssim 72.5\,\%$ $(30\,\%)$ & $\lesssim 15.0\,\%$ $(31\,\%)$  & $\lesssim 20.0\,\%$ $(28\,\%)$ & $\lesssim 70.0\,\%$ $(30\,\%)$ \\  
     & $\gtrsim 100.0\,\%$ $(41\,\%)$ & $\gtrsim 100.0\,\%$ $(40\,\%)$ & $\gtrsim 100.0\,\%$ $(58\,\%)$ & $\gtrsim 100.0\,\%$ $(18\,\%)$  & $\gtrsim 100.0\,\%$ $(25\,\%)$ & $\gtrsim 100.0\,\%$ $(60\,\%)$ \\ \hdashline
  FL & $\lesssim 5.0\,\%$ $(13\,\%)$ & $\lesssim 2.5\,\%$ $(12.5\,\%)$ & $\lesssim 95.0\,\%$ $(25\,\%)$ & $\lesssim 12.5\,\%$ $(32\,\%)$  & $\lesssim 20.0\,\%$ $(37\,\%)$ & $\lesssim 67.5\,\%$ $(30\,\%)$ \\  
     & $\gtrsim 100.0\,\%$ $(50\,\%)$ & $\gtrsim 100.0\,\%$ $(50\,\%)$ & $\gtrsim 100.0\,\%$ $(63\,\%)$ & $\gtrsim 100.0\,\%$ $(17\,\%)$  & $\gtrsim 100.0\,\%$ $(20\,\%)$ & $\gtrsim 100.0\,\%$ $(57\,\%)$ \\ \hline\hline
  Type & $\Delta \text{BR}_{H_3H_1H_1}^{\pmb{S_1}}$ & $\Delta \text{BR}_{H_3H_1H_1}^{\pmb{S_2}}$ & $\Delta \text{BR}_{H_3H_1H_1}^{\pmb{\overline{\text{MS}}}}$ & $\Delta \text{BR}_{H_3H_1H_2}^{\pmb{S_1}}$ & $\Delta \text{BR}_{H_3H_1H_2}^{\pmb{S_2}}$ & $\Delta \text{BR}_{H_3H_1H_2}^{\pmb{\overline{\text{MS}}}}$ \\ \hline
  I  & $\lesssim 17.5\,\%$ $(30\,\%)$ & $\lesssim 20.0\,\%$ $(31\,\%)$ & $\lesssim 57.5\,\%$ $(10\,\%)$ & $\lesssim 37.5\,\%$ $(30\,\%)$  & $\lesssim 37.5\,\%$ $(30\,\%)$ & $\lesssim 85.0\,\%$ $(8\,\%)$ \\  
     & $\gtrsim 100.0\,\%$ $(44\,\%)$ & $\gtrsim 100.0\,\%$ $(44\,\%)$ & $\gtrsim 100.0\,\%$ $(85\,\%)$ & $\gtrsim 100.0\,\%$ $(50\,\%)$  & $\gtrsim 100.0\,\%$ $(50\,\%)$ & $\gtrsim 100.0\,\%$ $(91\,\%)$ \\ \hdashline
  II & $\lesssim 40.0\,\%$ $(10\,\%)$ & $\lesssim 40.0\,\%$ $(10\,\%)$ & $\lesssim 95.0\,\%$ $(2\,\%)$ & $\lesssim 62.5\,\%$ $(10\,\%)$  & $\lesssim 60.0\,\%$ $(10\,\%)$ & $\lesssim 90.0\,\%$ $(1\,\%)$ \\  
     & $\gtrsim 100.0\,\%$ $(78\,\%)$ & $\gtrsim 100.0\,\%$ $(77\,\%)$ & $\gtrsim 100.0\,\%$ $(97\,\%)$ & $\gtrsim 100.0\,\%$ $(85\,\%)$  & $\gtrsim 100.0\,\%$ $(84\,\%)$ & $\gtrsim 100.0\,\%$ $(98\,\%)$ \\ \hdashline
  LS & $\lesssim 27.5\,\%$ $(25\,\%)$ & $\lesssim 37.5\,\%$ $(30\,\%)$ & $\lesssim 80.0\,\%$ $(8\,\%)$ & $\lesssim 67.5\,\%$ $(30\,\%)$  & $\lesssim 40.0\,\%$ $(20\,\%)$ & $\lesssim 87.5\,\%$ $(6\,\%)$ \\  
     & $\gtrsim 100.0\,\%$ $(53\,\%)$ & $\gtrsim 100.0\,\%$ $(52\,\%)$ & $\gtrsim 100.0\,\%$ $(90\,\%)$ & $\gtrsim 100.0\,\%$ $(62\,\%)$  & $\gtrsim 100.0\,\%$ $(62\,\%)$ & $\gtrsim 100.0\,\%$ $(93\,\%)$ \\ \hdashline
  FL & $\lesssim 32.5\,\%$ $(10\,\%)$ & $\lesssim 32.5\,\%$ $(10\,\%)$ & $\lesssim 90.0\,\%$ $(3\,\%)$ & $\lesssim 47.5\,\%$ $(10\,\%)$  & $\lesssim 45.0\,\%$ $(10\,\%)$ & $\lesssim 82.5\,\%$ $(2\,\%)$ \\  
     & $\gtrsim 100.0\,\%$ $(75\,\%)$ & $\gtrsim 100.0\,\%$ $(74\,\%)$ & $\gtrsim 100.0\,\%$ $(96\,\%)$ & $\gtrsim 100.0\,\%$ $(81\,\%)$  & $\gtrsim 100.0\,\%$ $(80\,\%)$ & $\gtrsim 100.0\,\%$ $(97\,\%)$ \\ \hline\hline
  Type & $\Delta \text{BR}_{H_3H_2H_2}^{\pmb{S_1}}$ & $\Delta \text{BR}_{H_3H_2H_2}^{\pmb{S_2}}$ & $\Delta \text{BR}_{H_3H_2H_2}^{\pmb{\overline{\text{MS}}}}$ & $\Delta \text{BR}_{H_3AA}^{\pmb{S_1}}$ & $\Delta \text{BR}_{H_3AA}^{\pmb{S_2}}$ & $\Delta \text{BR}_{H_3AA}^{\pmb{\overline{\text{MS}}}}$ \\ \hline
  I  & $\lesssim 32.5\,\%$ $(30\,\%)$ & $\lesssim 20.0\,\%$ $(20\,\%)$ & $\lesssim 67.5\,\%$ $(10\,\%)$ & $\lesssim 10.0\,\%$ $(43\,\%)$  & $\lesssim 12.5\,\%$ $(45\,\%)$ & $\lesssim 30.0\,\%$ $(20\,\%)$ \\  
     & $\gtrsim 100.0\,\%$ $(43\,\%)$ & $\gtrsim 100.0\,\%$ $(44\,\%)$ & $\gtrsim 100.0\,\%$ $(87\,\%)$ & $\gtrsim 100.0\,\%$ $(22\,\%)$  & $\gtrsim 100.0\,\%$ $(23\,\%)$ & $\gtrsim 100.0\,\%$ $(56\,\%)$ \\ \hdashline
  II & $\lesssim 45.0\,\%$ $(10\,\%)$ & $\lesssim 42.5\,\%$ $(10\,\%)$ & $\lesssim 90.0\,\%$ $(4\,\%)$ & - & - & - \\  
     & $\gtrsim 100.0\,\%$ $(79\,\%)$ & $\gtrsim 100.0\,\%$ $(80\,\%)$ & $\gtrsim 100.0\,\%$ $(95\,\%)$ & - & - & - \\ \hdashline
  LS & $\lesssim 32.5\,\%$ $(21\,\%)$ & $\lesssim 37.5\,\%$ $(21\,\%)$ & $\lesssim 92.5\,\%$ $(10\,\%)$ & $\lesssim 7.5\,\%$ $(44\,\%)$  & $\lesssim 10.0\,\%$ $(44\,\%)$ & $\lesssim 32.5\,\%$ $(30\,\%)$ \\  
     & $\gtrsim 100.0\,\%$ $(52\,\%)$ & $\gtrsim 100.0\,\%$ $(54\,\%)$ & $\gtrsim 100.0\,\%$ $(89\,\%)$ & $\gtrsim 100.0\,\%$ $(30\,\%)$  & $\gtrsim 100.0\,\%$ $(30\,\%)$ & $\gtrsim 100.0\,\%$ $(47\,\%)$ \\ \hdashline
  FL & $\lesssim 37.5\,\%$ $(10\,\%)$ & $\lesssim 40.0\,\%$ $(10\,\%)$ & $\lesssim 90.0\,\%$ $(4\,\%)$ & - & - & - \\  
     & $\gtrsim 100.0\,\%$ $(71\,\%)$ & $\gtrsim 100.0\,\%$ $(72\,\%)$ & $\gtrsim 100.0\,\%$ $(95\,\%)$ & - & - & - \\ \hline
 \end{tabular}
\caption{Relative size of the EW corrections to the BRs of the
  non-SM-like N2HDM Higgs boson $H_3$ into
$b\bar{b}$, $t\bar{t}$, $\tau^+ \tau^-$, $ZA$, $W^\pm H^\mp$, $ZZ$,
  $H_1 H_1$, $H_1 H_2$, $H_2 H_2$, 
  and $AA$, for the four N2HDM types I, II, LS and FL ($H_2$ is SM-like).}
\label{tab:h3forh2smlike}
\end{center}
\end{table}
For the heaviest Higgs boson $H_3$, many more decay channels are kinematically open. Still, the requirement of
OS decays reduces the original number of parameter points available,
({\it cf.}~Eq.~(\ref{eq:numbersheaviern2hdm})), for the individual OS
channels to the values listed in
Tab.~\ref{tab:rednumbersnonsmh3smh2}\footnote{Comparison with
  Eq.~(\ref{eq:numbersheaviern2hdm}) shows that again some of the
  numbers are actually not reduced.}. Note, in particular, that the
statistics for the FL type in the $ZA$ and $W^\pm H^\mp$ final states
becomes very low in this scenario. The relative corrections $\Delta
\mbox{BR}$ for the decays of $H_3$ 
are summarized in Tab.~\ref{tab:h3forh2smlike}. We observe:
\begin{itemize}
\item[--] The relative corrections for the Higgs-to-Higgs decays are very
  large for all N2HDM types and scheme sets. The reasons are 
  small LO BRs and/or non-decoupling effects. 
\item[--] For the scheme sets $\pmb{S_1}$ and $\pmb{S_2}$, the relative
  corrections to the other decays of $H_3$ are more moderate, but
  larger than for the scenarios where $H_1$ is the SM-like Higgs
  boson. 
\item[--] For the N2HDM types I and LS, all decay channels typically
  feature huge corrections for a considerable amount of input
  parameters. This might stem from 
\begin{itemize}
\item[$\ast$] small LO BRs;
\item[$\ast$] relatively large EW corrections (also
  indirectly) in these two N2HDM types; 
\item[$\ast$] parametrical enhancement of the mixing angle CTs and
  off-diagonal WFRCs for these two N2HDM types (since the CTs and
  WFRCs come in combination with the Yukawa couplings, their
  contribution strongly depends on the N2HDM type); 
\item[$\ast$] enhanced uncanceled contributions from the singlet
  (through the off-diagonal WFRCs connected to $H_3$ and the mixing
  angle CTs $\delta \alpha _2$ and $\delta \alpha _3$). 
\end{itemize}
\item[--] For the decays $H_3\to ZA$, $W^\pm H^\mp$ and $ZZ$, the
  corrections can become very large. The reasons are small LO branching
    ratios due to suppressed couplings, counterterm
    contributions that are parametrically enhanced or large
    counterterms themselves.
\end{itemize} 

%%%%%%%%%%%%%%%%%%%%%%%%%%%%%%%%%%%%%%%%%%%%%%%%%%%%%%%
\subsection{Pseudoscalar N2HDM Decays}
\label{sec:pseudoscalarn2hdm}
%%%%%%%%%%%%%%%%%%%%%%%%%%%%%%%%%%%%%%%%%%%%%%%%%%%%%%%
We finally turn to the decays of the pseudoscalar Higgs boson $A$ in
the N2HDM and again investigate the two cases where either $H_1$ or
$H_2$ is the SM-like Higgs boson. 

%%%%%%%%%%%%%%%%%%%%%%%%%%%%%%%%%%%%%%%%%%%%%%%%%%%%%%%
\subsubsection{$H_1$ is the SM-like Higgs Boson}
%%%%%%%%%%%%%%%%%%%%%%%%%%%%%%%%%%%%%%%%%%%%%%%%%%%%%%%
\begin{table}[t!]
\begin{center}
\renewcommand{\arraystretch}{1.2}
\scriptsize
 \begin{tabular}{|c|c c c c|}
  \hline
Type & $A \to t\bar{t}$ & $A \to ZH_1$ & $A \to ZH_2$ & $A\to ZH_3$\\
   \hline
I & $245\,233$ & $257\,182$ & $127\,677$ & $17\,786$ \\
II & $299\,959$ & $299\,959$ & $155\,270$ & $14\,149$ \\
LS & $281\,965$ & $282\,998$ & $145\,633$ & $18\,419$ \\
FL & $292\,634$ & $292\,634$ & $150\,051$ & $14\,481$ \\
\hline 
\end{tabular}
\caption{Number of parameter points available for the analysis in the
  individual OS decay channels of $A$ ($H_1$ is SM-like).}
\label{tab:rednumberspseudosmh1}
\end{center}
\end{table}
\begin{table}[t!]
\begin{center}
\renewcommand{\arraystretch}{1.2}
\scriptsize
 \begin{tabular}{|c|c c c||c c c|}
  \hline
  Type & $\Delta \text{BR}_{Ab\bar{b}}^{\pmb{S_1}}$ & $\Delta \text{BR}_{Ab\bar{b}}^{\pmb{S_2}}$ & $\Delta \text{BR}_{Ab\bar{b}}^{\pmb{\overline{\text{MS}}}}$ & $\Delta \text{BR}_{At\bar{t}}^{\pmb{S_1}}$ & $\Delta \text{BR}_{At\bar{t}}^{\pmb{S_2}}$ & $\Delta \text{BR}_{At\bar{t}}^{\pmb{\overline{\text{MS}}}}$ \\ \hline
  I  & $\lesssim 5.0\,\%$ $(52\,\%)$ & $\lesssim 5.0\,\%$ $(96\,\%)$ & $\lesssim 35.0\,\%$ $(50\,\%)$ & $\lesssim 2.5\,\%$ $(94\,\%)$  & $\lesssim 2.5\,\%$ $(95\,\%)$ & $\lesssim 32.5\,\%$ $(50\,\%)$ \\  
     & $\lesssim 12.5\,\%$ $(96\,\%)$ & $\lesssim 12.5\,\%$ $(97\,\%)$ & $\gtrsim 100.0\,\%$ $(29\,\%)$ & $\lesssim 7.5\,\%$ $(99\,\%)$  & $\lesssim 7.5\,\%$ $(99\,\%)$ & $\gtrsim 100.0\,\%$ $(27\,\%)$ \\ \hdashline
  II & $\lesssim 7.5\,\%$ $(49\,\%)$ & $\lesssim 12.5\,\%$ $(43\,\%)$ & $\lesssim 90.0\,\%$ $(8\,\%)$ & $\lesssim 2.5\,\%$ $(98\,\%)$  & $\lesssim 2.5\,\%$ $(98\,\%)$ & $\lesssim 12.5\,\%$ $(50\,\%)$ \\  
     & $\lesssim 17.5\,\%$ $(94\,\%)$ & $\lesssim 22.5\,\%$ $(94\,\%)$ & $\gtrsim 100.0\,\%$ $(90\,\%)$ & $\lesssim 5.0\,\%$ $(99\,\%)$  & $\lesssim 5.0\,\%$ $(99\,\%)$ & $\gtrsim 100.0\,\%$ $(12\,\%)$ \\ \hdashline
  LS & $\lesssim 5.0\,\%$ $(52\,\%)$ & $\lesssim 5.0\,\%$ $(56\,\%)$ & $\lesssim 27.5\,\%$ $(50\,\%)$ & $\lesssim 2.5\,\%$ $(98\,\%)$  & $\lesssim 2.5\,\%$ $(98\,\%)$ & $\lesssim 25.0\,\%$ $(50\,\%)$ \\  
     & $\lesssim 12.5\,\%$ $(97\,\%)$ & $\lesssim 10.0\,\%$ $(93\,\%)$ & $\gtrsim 100.0\,\%$ $(24\,\%)$ & $\lesssim 5.0\,\%$ $(99\,\%)$  & $\lesssim 5.0\,\%$ $(99\,\%)$ & $\gtrsim 100.0\,\%$ $(24\,\%)$ \\ \hdashline
  FL & $\lesssim 7.5\,\%$ $(56\,\%)$ & $\lesssim 12.5\,\%$ $(60\,\%)$ & $\lesssim 90.0\,\%$ $(10\,\%)$ & $\lesssim 2.5\,\%$ $(98\,\%)$  & $\lesssim 2.5\,\%$ $(98\,\%)$ & $\lesssim 22.5\,\%$ $(50\,\%)$ \\  
     & $\lesssim 17.5\,\%$ $(96\,\%)$ & $\lesssim 20.0\,\%$ $(94\,\%)$ & $\gtrsim 100.0\,\%$ $(89\,\%)$ & $\lesssim 5.0\,\%$ $(99\,\%)$  & $\lesssim 5.0\,\%$ $(99\,\%)$ & $\gtrsim 100.0\,\%$ $(19\,\%)$ \\ \hline\hline
  Type & $\Delta \text{BR}_{A\tau ^+\tau ^-}^{\pmb{S_1}}$ & $\Delta \text{BR}_{A\tau ^+\tau ^-}^{\pmb{S_2}}$ & $\Delta \text{BR}_{A\tau ^+\tau ^-}^{\pmb{\overline{\text{MS}}}}$ & $\Delta \text{BR}_{AZH_1}^{\pmb{S_1}}$ & $\Delta \text{BR}_{AZH_1}^{\pmb{S_2}}$ & $\Delta \text{BR}_{AZH_1}^{\pmb{\overline{\text{MS}}}}$ \\ \hline
  I  & $\lesssim 7.5\,\%$ $(49\,\%)$ & $\lesssim 7.5\,\%$ $(54\,\%)$ & $\lesssim 37.5\,\%$ $(50\,\%)$ & $\lesssim 12.5\,\%$ $(48\,\%)$  & $\lesssim 12.5\,\%$ $(51\,\%)$ & $\lesssim 85.0\,\%$ $(25\,\%)$ \\  
     & $\lesssim 15.0\,\%$ $(98\,\%)$ & $\lesssim 12.5\,\%$ $(95\,\%)$ & $\gtrsim 100.0\,\%$ $(29\,\%)$ & $\gtrsim 100.0\,\%$ $(10\,\%)$  & $\gtrsim 100.0\,\%$ $(9\,\%)$ & $\gtrsim 100.0\,\%$ $(73\,\%)$ \\ \hdashline
  II & $\lesssim 7.5\,\%$ $(57\,\%)$ & $\lesssim 12.5\,\%$ $(49\,\%)$ & $\lesssim 90.0\,\%$ $(8\,\%)$ & $\lesssim 17.5\,\%$ $(48\,\%)$  & $\lesssim 12.5\,\%$ $(47\,\%)$ & $\lesssim 82.5\,\%$ $(16\,\%)$ \\  
     & $\lesssim 15.0\,\%$ $(97\,\%)$ & $\lesssim 20.0\,\%$ $(96\,\%)$ & $\gtrsim 100.0\,\%$ $(90\,\%)$ & $\gtrsim 100.0\,\%$ $(14\,\%)$  & $\gtrsim 100.0\,\%$ $(10\,\%)$ & $\gtrsim 100.0\,\%$ $(81\,\%)$ \\ \hdashline
  LS & $\lesssim 7.5\,\%$ $(55\,\%)$ & $\lesssim 12.5\,\%$ $(61\,\%)$ & $\lesssim 90.0\,\%$ $(15\,\%)$ & $\lesssim 17.5\,\%$ $(47\,\%)$  & $\lesssim 15.0\,\%$ $(49\,\%)$ & $\lesssim 77.5\,\%$ $(18\,\%)$ \\  
     & $\lesssim 15.0\,\%$ $(98\,\%)$ & $\lesssim 17.5\,\%$ $(95\,\%)$ & $\gtrsim 100.0\,\%$ $(82\,\%)$ & $\gtrsim 100.0\,\%$ $(13\,\%)$  & $\gtrsim 100.0\,\%$ $(11\,\%)$ & $\gtrsim 100.0\,\%$ $(79\,\%)$ \\ \hdashline
  FL & $\lesssim 7.5\,\%$ $(61\,\%)$ & $\lesssim 7.5\,\%$ $(66\,\%)$ & $\lesssim 25.0\,\%$ $(50\,\%)$ & $\lesssim 15.0\,\%$ $(47\,\%)$  & $\lesssim 15.0\,\%$ $(52\,\%)$ & $\lesssim 90.0\,\%$ $(19\,\%)$ \\  
     & $\lesssim 12.5\,\%$ $(94\,\%)$ & $\lesssim 12.5\,\%$ $(96\,\%)$ & $\gtrsim 100.0\,\%$ $(19\,\%)$ & $\gtrsim 100.0\,\%$ $(13\,\%)$  & $\gtrsim 100.0\,\%$ $(12\,\%)$ & $\gtrsim 100.0\,\%$ $(80\,\%)$ \\ \hline\hline  
  Type & $\Delta \text{BR}_{AZH_2}^{\pmb{S_1}}$ & $\Delta \text{BR}_{AZH_2}^{\pmb{S_2}}$ & $\Delta \text{BR}_{AZH_2}^{\pmb{\overline{\text{MS}}}}$ & $\Delta \text{BR}_{AZH_3}^{\pmb{S_1}}$ & $\Delta \text{BR}_{AZH_3}^{\pmb{S_2}}$ & $\Delta \text{BR}_{AZH_3}^{\pmb{\overline{\text{MS}}}}$ \\ \hline
  I  & $\lesssim 2.5\,\%$ $(60\,\%)$ & $\lesssim 2.5\,\%$ $(70\,\%)$ & $\lesssim 92.5\,\%$ $(30\,\%)$ & $\lesssim 20.0\,\%$ $(7\,\%)$  & $\lesssim 27.5\,\%$ $(11\,\%)$ & $\lesssim 75.0\,\%$ $(10\,\%)$ \\  
     & $\lesssim 10.0\,\%$ $(97\,\%)$ & $\lesssim 7.5\,\%$ $(95\,\%)$ & $\gtrsim 100.0\,\%$ $(67\,\%)$ & $\lesssim 70.0\,\%$ $(42\,\%)$  & $\lesssim 60.0\,\%$ $(33\,\%)$ & $\gtrsim 100.0\,\%$ $(53\,\%)$ \\ \hdashline
  II & $\lesssim 2.5\,\%$ $(52\,\%)$ & $\lesssim 2.5\,\%$ $(55\,\%)$ & $\lesssim 90.0\,\%$ $(23\,\%)$ & $\lesssim 22.5\,\%$ $(10\,\%)$  & $\lesssim 25.0\,\%$ $(11\,\%)$ & $\lesssim 75.0\,\%$ $(10\,\%)$ \\  
     & $\lesssim 10.0\,\%$ $(96\,\%)$ & $\lesssim 10.0\,\%$ $(97\,\%)$ & $\gtrsim 100.0\,\%$ $(72\,\%)$ & $\lesssim 50.0\,\%$ $(63\,\%)$  & $\lesssim 55.0\,\%$ $(77\,\%)$ & $\gtrsim 100.0\,\%$ $(70\,\%)$ \\ \hdashline
  LS & $\lesssim 2.5\,\%$ $(58\,\%)$ & $\lesssim 2.5\,\%$ $(69\,\%)$ & $\lesssim 85.0\,\%$ $(25\,\%)$ & $\lesssim 22.5\,\%$ $(9\,\%)$  & $\lesssim 25.0\,\%$ $(10\,\%)$ & $\lesssim 75.0\,\%$ $(10\,\%)$ \\  
     & $\lesssim 10.0\,\%$ $(96\,\%)$ & $\lesssim 7.5\,\%$ $(96\,\%)$ & $\gtrsim 100.0\,\%$ $(69\,\%)$ & $\lesssim 67.5\,\%$ $(50\,\%)$  & $\lesssim 67.5\,\%$ $(50\,\%)$ & $\gtrsim 100.0\,\%$ $(57\,\%)$ \\ \hdashline
  FL & $\lesssim 2.5\,\%$ $(56\,\%)$ & $\lesssim 2.5\,\%$ $(67\,\%)$ & $\lesssim 87.5\,\%$ $(25\,\%)$ & $\lesssim 22.5\,\%$ $(12\,\%)$  & $\lesssim 22.5\,\%$ $(10\,\%)$ & $\lesssim 80.0\,\%$ $(10\,\%)$ \\  
     & $\lesssim 10.0\,\%$ $(96\,\%)$ & $\lesssim 7.5\,\%$ $(96\,\%)$ & $\gtrsim 100.0\,\%$ $(69\,\%)$ & $\lesssim 60.0\,\%$ $(70\,\%)$  & $\lesssim 60.0\,\%$ $(69\,\%)$ & $\gtrsim 100.0\,\%$ $(73\,\%)$ \\ \hline
 \end{tabular}
\caption{Relative size of the EW corrections to the BRs of the
  pseudoscalar $A$ into
$b\bar{b}$, $t\bar{t}$, $\tau^+ \tau^-$, $ZH_1$, $ZH_2$, and $ZH_3$,
for the four N2HDM types I, II, LS and FL ($H_1$ is SM-like).} 
\label{tab:aforh1smlike}
\end{center}
\end{table}
The requirement of the pseudoscalar Higgs boson decays to be OS
reduces the originally available parameter points to 
those given in Tab.~\ref{tab:rednumberspseudosmh1} for the individual
channels\footnote{As before, the amount of available input parameters for some N2HDM types is actually not reduced.}. In
Tab.~\ref{tab:aforh1smlike}, we display the relative EW corrections 
to the BRs of the pseudoscalar Higgs boson for the case where $H_1$ is
SM-like. The table allows for the following observations:  
\begin{itemize}
\item[--] The corrections for the decay $A \to  t\bar{t}$
  (which is typically the dominant one if it is kinematically allowed)
  are typically very small for the scheme sets $\pmb{S_1}$ and
  $\pmb{S_2}$, indicating both low corrections for this decay in these
  schemes and numerical stability of the two scheme sets. 
\item[--] For the other fermionic decay channels as well as for $A
  \to ZH_2$, the corrections in the two schemesets
  $\pmb{S_1}$ and $\pmb{S_2}$ are small to moderate for all four types
  of the N2HDM. 
\item[--] On the other hand, the decay channels $A \to
  ZH_1$ and $A \to ZH_3$ typically
  feature very large EW corrections for all N2HDM types and
  for all scheme sets. This is due to suppressed BRs
    at leading order. For an SM-like $H_1$ the coupling 
    $AZH_1$ is very suppressed due to sum rules. The
  coupling of the $H_3$, which is rather singlet-like, is also
  suppressed but less strongly, so that the resulting decays widths and BRs for $A \to
  ZH_1$ become very small and also those for $A\to ZH_3$, which are less
  suppressed however. Apart from small coupling constants, also
  parametrically enhanced 
  counterterm contributions or uncanceled counterterms that blow up in this
  parameter region can be responsible for the large corrections. The coupling $AZH_2$ on the other hand is close
to its maximum value so that the decay $A\to ZH_2$ can become
significant, resulting in small relative
corrections.
\end{itemize}

%%%%%%%%%%%%%%%%%%%%%%%%%%%%%%%%%%%%%%%%%%%%%%%%%%%%%%%
\subsubsection{$H_2$ is the SM-like Higgs Boson}
%%%%%%%%%%%%%%%%%%%%%%%%%%%%%%%%%%%%%%%%%%%%%%%%%%%%%%%
\begin{table}[t!]
\begin{center}
\renewcommand{\arraystretch}{1.2}
\scriptsize
 \begin{tabular}{|c|c c c c|}
  \hline
Type & $A \to t\bar{t}$ & $A \to ZH_1$ & $A \to ZH_2$ & $A\to ZH_3$\\
   \hline
I & 4737 & 7566 & 7040 & 1632 \\
II &  2381 & 2381 & 2381 & 342 \\
LS & 3122 & 3303 & 3203 & 961 \\
FL & 1562 & 1562 & 1562 & 304 \\
\hline 
\end{tabular}
\caption{Number of parameter points available for the analysis in the
  individual OS decay channels of $A$ ($H_2$ is SM-like).}
\label{tab:rednumberspseudosmh2}
\end{center}
\end{table}
In case that $H_2$ is the SM-like Higgs boson, the additional OS requirement for all pseudoscalar Higgs decays to be OS reduces the amount of available
parameter points in some channels as shown in
Tab.~\ref{tab:rednumberspseudosmh2}\footnote{As before, some of the
  numbers are actually not reduced.}. The results for the relative corrections $\Delta \mbox{BR}$ of the
pseuoscalar decays for the scenarios where $H_2$ is SM-like are displayed in
Tab.~\ref{tab:aforh2smlike}. We make the following observations:
\begin{table}[t!]
\begin{center}
\renewcommand{\arraystretch}{1.2}
\scriptsize
 \begin{tabular}{|c|c c c||c c c|}
  \hline
  Type & $\Delta \text{BR}_{Ab\bar{b}}^{\pmb{S_1}}$ & $\Delta \text{BR}_{Ab\bar{b}}^{\pmb{S_2}}$ & $\Delta \text{BR}_{Ab\bar{b}}^{\pmb{\overline{\text{MS}}}}$ & $\Delta \text{BR}_{At\bar{t}}^{\pmb{S_1}}$ & $\Delta \text{BR}_{At\bar{t}}^{\pmb{S_2}}$ & $\Delta \text{BR}_{At\bar{t}}^{\pmb{\overline{\text{MS}}}}$ \\ \hline
  I  & $\lesssim 5.0\,\%$ $(47\,\%)$ & $\lesssim 5.0\,\%$ $(51\,\%)$ & $\lesssim 80.0\,\%$ $(50\,\%)$ & $\lesssim 2.5\,\%$ $(74\,\%)$  & $\lesssim 2.5\,\%$ $(73\,\%)$ & $\lesssim 67.5\,\%$ $(50\,\%)$ \\  
     & $\lesssim 15.0\,\%$ $(90\,\%)$ & $\lesssim 15.0\,\%$ $(90\,\%)$ & $\gtrsim 100.0\,\%$ $(43\,\%)$ & $\lesssim 15.0\,\%$ $(91\,\%)$  & $\lesssim 12.5\,\%$ $(91\,\%)$ & $\gtrsim 100.0\,\%$ $(41\,\%)$ \\ \hdashline
  II & $\lesssim 10.0\,\%$ $(64\,\%)$ & $\lesssim 15.0\,\%$ $(53\,\%)$ & $\lesssim 82.5\,\%$ $(5\,\%)$ & $\lesssim 2.5\,\%$ $(98\,\%)$  & $\lesssim 2.5\,\%$ $(98\,\%)$ & $\lesssim 77.5\,\%$ $(50\,\%)$ \\  
     & $\lesssim 17.5\,\%$ $(92\,\%)$ & $\lesssim 22.5\,\%$ $(90\,\%)$ & $\gtrsim 100.0\,\%$ $(94\,\%)$ & $\lesssim 5.0\,\%$ $(99\,\%)$  & $\lesssim 5.0\,\%$ $(99\,\%)$ & $\gtrsim 100.0\,\%$ $(44\,\%)$ \\ \hdashline
  LS & $\lesssim 7.5\,\%$ $(75\,\%)$ & $\lesssim 5.0\,\%$ $(46\,\%)$ & $\lesssim 75.0\,\%$ $(50\,\%)$ & $\lesssim 2.5\,\%$ $(90\,\%)$  & $\lesssim 2.5\,\%$ $(90\,\%)$ & $\lesssim 70.0\,\%$ $(50\,\%)$ \\  
     & $\lesssim 12.5\,\%$ $(95\,\%)$ & $\lesssim 12.5\,\%$ $(96\,\%)$ & $\gtrsim 100.0\,\%$ $(44\,\%)$ & $\lesssim 7.5\,\%$ $(96\,\%)$  & $\lesssim 5.0\,\%$ $(95\,\%)$ & $\gtrsim 100.0\,\%$ $(58\,\%)$ \\ \hdashline
  FL & $\lesssim 7.5\,\%$ $(48\,\%)$ & $\lesssim 12.5\,\%$ $(51\,\%)$ & $\lesssim 92.5\,\%$ $(8\,\%)$ & $\lesssim 2.5\,\%$ $(97\,\%)$  & $\lesssim 2.5\,\%$ $(97\,\%)$ & $\lesssim 77.5\,\%$ $(40\,\%)$ \\  
     & $\lesssim 17.5\,\%$ $(94\,\%)$ & $\lesssim 20.0\,\%$ $(91\,\%)$ & $\gtrsim 100.0\,\%$ $(91\,\%)$ & $\lesssim 7.5\,\%$ $(99\,\%)$  & $\lesssim 5.0\,\%$ $(98\,\%)$ & $\gtrsim 100.0\,\%$ $(55\,\%)$ \\ \hline\hline
  Type & $\Delta \text{BR}_{A\tau ^+\tau ^-}^{\pmb{S_1}}$ & $\Delta \text{BR}_{A\tau ^+\tau ^-}^{\pmb{S_2}}$ & $\Delta \text{BR}_{A\tau ^+\tau ^-}^{\pmb{\overline{\text{MS}}}}$ & $\Delta \text{BR}_{AZH_1}^{\pmb{S_1}}$ & $\Delta \text{BR}_{AZH_1}^{\pmb{S_2}}$ & $\Delta \text{BR}_{AZH_1}^{\pmb{\overline{\text{MS}}}}$ \\ \hline
  I  & $\lesssim 7.5\,\%$ $(43\,\%)$ & $\lesssim 7.5\,\%$ $(48\,\%)$ & $\lesssim 80.0\,\%$ $(50\,\%)$ & $\lesssim 2.5\,\%$ $(73\,\%)$  & $\lesssim 2.5\,\%$ $(71\,\%)$ & $\lesssim 92.5\,\%$ $(25\,\%)$ \\  
     & $\lesssim 15.0\,\%$ $(89\,\%)$ & $\lesssim 15.0\,\%$ $(90\,\%)$ & $\gtrsim 100.0\,\%$ $(43\,\%)$ & $\lesssim 10.0\,\%$ $(93\,\%)$  & $\lesssim 7.5\,\%$ $(91\,\%)$ & $\gtrsim 100.0\,\%$ $(74\,\%)$ \\ \hdashline
  II & $\lesssim 7.5\,\%$ $(45\,\%)$ & $\lesssim 15.0\,\%$ $(60\,\%)$ & $\lesssim 80.0\,\%$ $(5\,\%)$ & $\lesssim 5.0\,\%$ $(56\,\%)$  & $\lesssim 5.0\,\%$ $(47\,\%)$ & $\lesssim 90.0\,\%$ $(3\,\%)$ \\  
     & $\lesssim 15.0\,\%$ $(95\,\%)$ & $\lesssim 20.0\,\%$ $(93\,\%)$ & $\gtrsim 100.0\,\%$ $(94\,\%)$ & $\lesssim 17.5\,\%$ $(89\,\%)$  & $\lesssim 15.0\,\%$ $(91\,\%)$ & $\gtrsim 100.0\,\%$ $(96\,\%)$ \\ \hdashline
  LS & $\lesssim 10.0\,\%$ $(67\,\%)$ & $\lesssim 12.5\,\%$ $(61\,\%)$ & $\lesssim 87.5\,\%$ $(16\,\%)$ & $\lesssim 2.5\,\%$ $(53\,\%)$  & $\lesssim 2.5\,\%$ $(47\,\%)$ & $\lesssim 87.5\,\%$ $(11\,\%)$ \\  
     & $\lesssim 15.0\,\%$ $(95\,\%)$ & $\lesssim 17.5\,\%$ $(93\,\%)$ & $\gtrsim 100.0\,\%$ $(82\,\%)$ & $\lesssim 15.0\,\%$ $(92\,\%)$  & $\lesssim 12.5\,\%$ $(92\,\%)$ & $\gtrsim 100.0\,\%$ $(88\,\%)$ \\ \hdashline
  FL & $\lesssim 7.5\,\%$ $(46\,\%)$ & $\lesssim 7.5\,\%$ $(52\,\%)$ & $\lesssim 77.5\,\%$ $(40\,\%)$ & $\lesssim 2.5\,\%$ $(42\,\%)$  & $\lesssim 5.0\,\%$ $(57\,\%)$ & $\lesssim 90.0\,\%$ $(3\,\%)$ \\  
     & $\lesssim 15.0\,\%$ $(97\,\%)$ & $\lesssim 15.0\,\%$ $(98\,\%)$ & $\gtrsim 100.0\,\%$ $(55\,\%)$ & $\lesssim 15.0\,\%$ $(89\,\%)$  & $\lesssim 15.0\,\%$ $(91\,\%)$ & $\gtrsim 100.0\,\%$ $(96\,\%)$ \\ \hline\hline  
  Type & $\Delta \text{BR}_{AZH_2}^{\pmb{S_1}}$ & $\Delta \text{BR}_{AZH_2}^{\pmb{S_2}}$ & $\Delta \text{BR}_{AZH_2}^{\pmb{\overline{\text{MS}}}}$ & $\Delta \text{BR}_{AZH_3}^{\pmb{S_1}}$ & $\Delta \text{BR}_{AZH_3}^{\pmb{S_2}}$ & $\Delta \text{BR}_{AZH_3}^{\pmb{\overline{\text{MS}}}}$ \\ \hline
  I  & $\lesssim 15.0\,\%$ $(49\,\%)$ & $\lesssim 12.5\,\%$ $(47\,\%)$ & $\lesssim 85.0\,\%$ $(10\,\%)$ & $\lesssim 50.0\,\%$ $(10\,\%)$  & $\lesssim 50.0\,\%$ $(10\,\%)$ & $\lesssim 82.5\,\%$ $(10\,\%)$ \\  
     & $\gtrsim 100.0\,\%$ $(13\,\%)$ & $\gtrsim 100.0\,\%$ $(13\,\%)$ & $\gtrsim 100.0\,\%$ $(89\,\%)$ & $\lesssim 95.0\,\%$ $(53\,\%)$  & $\lesssim 90.0\,\%$ $(40\,\%)$ & $\gtrsim 100.0\,\%$ $(49\,\%)$ \\ \hdashline
  II & $\lesssim 17.5\,\%$ $(41\,\%)$ & $\lesssim 15.0\,\%$ $(50\,\%)$ & $\lesssim 90.0\,\%$ $(3\,\%)$ & $\lesssim 27.5\,\%$ $(20\,\%)$  & $\lesssim 35.0\,\%$ $(30\,\%)$ & $\lesssim 92.5\,\%$ $(20\,\%)$ \\  
     & $\gtrsim 100.0\,\%$ $(15\,\%)$ & $\gtrsim 100.0\,\%$ $(10\,\%)$ & $\gtrsim 100.0\,\%$ $(96\,\%)$ & $\lesssim 55.0\,\%$ $(78\,\%)$  & $\lesssim 57.5\,\%$ $(81\,\%)$ & $\gtrsim 100.0\,\%$ $(65\,\%)$ \\ \hdashline
  LS & $\lesssim 20.0\,\%$ $(38\,\%)$ & $\lesssim 15.0\,\%$ $(38\,\%)$ & $\lesssim 87.5\,\%$ $(7\,\%)$ & $\lesssim 35.0\,\%$ $(10\,\%)$  & $\lesssim 60.0\,\%$ $(21\,\%)$ & $\lesssim 85.0\,\%$ $(10\,\%)$ \\  
     & $\gtrsim 100.0\,\%$ $(20\,\%)$ & $\gtrsim 100.0\,\%$ $(18\,\%)$ & $\gtrsim 100.0\,\%$ $(92\,\%)$ & $\lesssim 80.0\,\%$ $(39\,\%)$  & $\lesssim 90.0\,\%$ $(51\,\%)$ & $\gtrsim 100.0\,\%$ $(58\,\%)$ \\ \hdashline
  FL & $\lesssim 20.0\,\%$ $(49\,\%)$ & $\lesssim 15.0\,\%$ $(48\,\%)$ & $\lesssim 85.0\,\%$ $(3\,\%)$ & $\lesssim 37.5\,\%$ $(32\,\%)$  & $\lesssim 37.5\,\%$ $(29\,\%)$ & $\lesssim 85.0\,\%$ $(10\,\%)$ \\  
     & $\gtrsim 100.0\,\%$ $(15\,\%)$ & $\gtrsim 100.0\,\%$ $(13\,\%)$ & $\gtrsim 100.0\,\%$ $(96\,\%)$ & $\lesssim 67.5\,\%$ $(81\,\%)$  & $\lesssim 72.5\,\%$ $(88\,\%)$ & $\gtrsim 100.0\,\%$ $(77\,\%)$ \\ \hline
 \end{tabular}
\caption{Relative size of the EW corrections to the BRs of the
  pseudoscalar $A$ into
$b\bar{b}$, $t\bar{t}$, $\tau^+ \tau^-$, $ZH_1$, $ZH_2$, and $ZH_3$,
for the four N2HDM types I, II, LS and FL ($H_2$ is SM-like).} 
\label{tab:aforh2smlike}
\end{center}
\vspace*{-0.5cm}
\end{table}
\begin{itemize}
\item[--] The corrections for the fermionic decays and for $A
  \to ZH_1$ are typically small to moderate for the
  scheme sets $\pmb{S_1}$ and $\pmb{S_2}$ and for all N2HDM types,
  indicating again both moderate corrections for these channels in these
  schemes and numerical stability of the two scheme sets. Overall the
  corrections are slightly larger than in the case where $H_1$ is
  SM-like. 
\item[--] On the other hand, the decay channels $A \to
  ZH_2$ and $A \to ZH_3$ typically
  feature very large EW corrections for all N2HDM types and
  for all scheme sets. Again the reason are suppressed tree-level
  decays and BRs. With $H_2$ being SM-like, the sum rules lead to
  suppressed $AZH_2$ and $AZH_3$ couplings with the former being more
  suppressed so that the LO decays for $A\to ZH_{2,3}$ are very small,
  with $A\to ZH_3$ in general being a bit less suppressed. Again, 
    also parametrically enhanced counterterm contributions or large
    uncanceled counterterms themselves can lead to large EW corrections in this corner or the parameter space. The
  coupling involving the singlet-like $H_1$, $AZH_1$, on the other
  hand is not suppressed, apart from the singlet admixture
  suppression, so that the LO BR for the decay $A\to ZH_1$ can be moderate to
  large, inducing small relative corrections. 
\end{itemize}

%%%%%%%%%%%%%%%%%%%%%%%%%%%%%%%%%%%%%%%%%%%%%%%%%%%%%%%
\section{Conclusions}
\label{sec:conclusions}
In this paper, we presented an overview over the size of the EW corrections to
the decays of the neutral Higgs bosons of the 2HDM and the N2HDM for different
renormalization schemes. Our aim was to quantify the EW corrections
that typically appear in beyond-the-SM models with non-minimal Higgs sectors, to
identify decays and parameter regions that lead to large corrections and require further
treatment, and finally to classify renormalization schemes with
respect to the size of EW corrections they produce in order to filter
for suitable renormalization schemes that do not induce unnaturally
large corrections. For our analysis, we only considered 
parameter scenarios that fulfill theoretical and experimental
constraints and that are obtained from a scan in the parameter ranges
of the two considered 
models. Furthermore, EW corrections were only computed for
OS and non-loop induced decays. \s

Our thus obtained results show that the corrections are in general
well-behaving for renormalization schemes that
are not process-dependent. For these schemes, the relative corrections
to the SM-like Higgs bosons into SM final states are typically small,
with the bulk of the corrections being situated below 5\%. For some
scenarios, they can go up to 7.5\%. For the case in the N2HDM where
the second-lightest Higgs boson $H_2$ is SM-like, also decays 
into a lighter $H_1 H_1$ pair are possible, and the
corrections to these decays can become very large for some parameter
scenarios. This is due to small tree-level BRs or non-decoupling
effects. For the decays of the non-SM-like 
Higgs bosons, the corrections are in general more important 
and can become significant. Also here, some parameter scenarios
feature very large corrections which can be traced back to suppressed
tree-level 
decays, parametrically enhanced corrections or uncancelled large 
counterterm contributions. Large counterterms can appear {\it e.g.}~in
certain regions of the parameter space with small parameter
values (namely small couplings because of sum rule constraints) in the
denominator. The large corrections require further
investigation and call for an improvement of the fixed-order
calculation, such as inclusion of higher-orders or the resummation of
the corrections to all orders. \s 

Concerning the renormalization schemes, we found that the
process-dependent schemes typically lead to larger corrections that
can even become unphysically large with corrections beyond 100\% that
may also be negative. The $\overline{\mbox{MS}}$ scheme throughout
leads to huge corrections in the decays and hence turns out to be
unsuitable. 

%%%%%%%%%%%%%%%%%%%%%%%%%%%%%%%%%%%%%%%%%%%%%%%%%%%%%%%
\subsection*{Acknowledgments}
We thank Ansgar Denner and Stefan Dittmaier for fruitful discussion on
renormalization issues, we thank Michael Spira for discussions on
combining electroweak and QCD corrections. We are grateful to  Philipp
Basler for providing us with valid data samples. MK and MM acknowledge
financial support from the DFG project “Precision Calculations in the
Higgs Sector - Paving the Way to the New Physics Landscape” (ID: MU
3138/1-1).  

%%%%%%%%%%%%%%%%%%%%%%%%%%%%%%%%%%%%%%%%%%%%%%%%%%%%%%%%%%
%\bibliographystyle{h-physrev}
%%\bibliographystyle{plain}
%\bibliography{refs}
%%%%%%%%%%%%%%%%%%%%%%%%%%%%%%%%%%%%%%%%%%%%%%%%%%%%%%%%%%

\end{document}